\documentclass[fleqn,usenatbib]{mnras}

\usepackage[T1]{fontenc}
\usepackage{ae,aecompl}
\usepackage{pdflscape}
\usepackage{graphicx}	
\usepackage{amsmath}	
\usepackage{amssymb}	
\newcommand{\Ha}{H$\alpha$}

\newcommand{\Hb}{H$\beta$}

\newcommand{\htwo}{H$_{2}$}

\newcommand{\kms}{km~s{$^{-1}$}}

\newcommand{\cmq}{cm$^{-3}$}

\newcommand{\tC}{{$ \theta^1$~Ori-C}}
\newcommand{\tB}{{$ \theta^1$~Ori-B}}

\newcommand{\tA}{{$\theta^2$~Ori-A}}

\newcommand{\hii}{{H}{\sc ii}}
\newcommand{\Hi}{{H}{\sc i}}
\newcommand{\Hei}{{He}{\sc i}}
\newcommand{\Caii}{{Ca}{\sc ii}}
\newcommand{\Caiii}{{Ca}{\sc iii}}
\newcommand{\Nai}{{Na}{\sc i}}
\newcommand{\Naii}{{Na}{\sc ii}}
\newcommand{\cii}{[\ion{C}{\sc ii}]}
\newcommand{\sii}{[\ion{S}{\sc ii}]}
\newcommand{\siii}{[\ion{S}{\sc iii}]}
\newcommand{\Siii}{\ion{S}{\sc iii}}
\newcommand{\Piii}{\ion{P}{\sc iii}}
\newcommand{\oiii}{[\ion{O}{\sc iii}]}
\newcommand{\oii}{[\ion{O}{\sc ii}]}
\newcommand{\oi}{[\ion{O}{\sc i}]}
\newcommand{\nii}{[\ion{N}{\sc ii}]}

\newcommand{\Ne}{n$\rm_{e}$}
\newcommand{\Te}{T$\rm_{e}$}

\newcommand{\ozone}{He$^{+}$+H$^{+}$}
\newcommand{\nzone}{He$\rm ^{o}$+H$^{+}$}

\newcommand{\Alow}{A$\rm _{low}$}

\newcommand{\Ascat}{A$\rm _{scat}$}

\newcommand{\Vomc}{V$\rm_{OMC}$}

\newcommand{\Vobs}{V$\rm_{obs}$}
\newcommand{\Vobsnii}{V$\rm_{obs,[N~II]}$}
\newcommand{\Vobsoiii}{V$\rm_{obs,[O~III]}$}

 \newcommand{\Vevap}{V$\rm_{evap}$}

\newcommand{\Vevapnii}{V$\rm_{evap,[N~II]}$}

\newcommand{\Vpdr}{V$\rm_{PDR}$}
\newcommand{\Vmif}{V$\rm_{MIF}$}
\newcommand{\Vmifnii}{V$\rm_{MIF,[N~II]}$}
\newcommand{\Vmifoiii}{V$\rm_{MIF,[O~III]}$}
\newcommand{\Vscat}{V$\rm_{scat}$}
\newcommand{\Vscatoiii}{V$\rm_{scat,[O~III]}$}
\newcommand{\Vwide}{V$\rm_{wide}$}
\newcommand{\Vwideoiii}{V$\rm_{wide,[O~III]}$}
\newcommand{\Vscatnii}{V$\rm_{scat,[N~II]}$}
\newcommand{\Vwidenii}{V$\rm_{wide,[N~II]}$}
\newcommand{\Vblue}{V$\rm_{blue}$}
\newcommand{\Vbluenii}{V$\rm_{blue,[N~II]}$}
\newcommand{\Vlow}{V$\rm _{low}$}
\newcommand{\Vlownii}{V$\rm_{low,[N~II]}$}
\newcommand{\Vlowoiii}{V$\rm_{low,[O~III]}$}
\newcommand{\Vblueoiii}{V$\rm_{blue,[O~III]}$}

\newcommand{\Vnewoiii}{V$\rm_{new,[O~III]}$}

\newcommand{\Vfront}{V$\rm_{front}$}

\newcommand{\Vnew}{V$\rm _{new,[O~III]}$}

\newcommand{\Soiii}{S$\rm _{[O~III]}$}

\title{Layers in the Central Orion Nebula}

\author[C. R. O'Dell]{
C. R. O'Dell\thanks{E-mail: cr.odell@vanderbilt.edu}\\
Department of Physics and Astronomy, Vanderbilt University, Box 1807-B, Nashville, TN 37235 USA}
\date{Accepted 2018 April 6. Received 2018 March 25; in original form 2018 February 27}
\pubyear{2018}
\begin{document}
\label{firstpage}
\pagerange{\pageref{firstpage}--\pageref{lastpage}}
\maketitle

\begin{abstract}
The existence of multiple layers in the inner Orion Nebula has been revealed using data
from an Atlas of spectra at 2\arcsec\ and 12 \kms\ resolution. 
These data were sometimes grouped over Samples of 10\arcsec $\times$10\arcsec\ to produce high Signal to Noise spectra and sometimes  grouped into sequences of pseudo-slit Spectra of 12\farcs8~--~39\arcsec width for high spatial resolution studies. Multiple velocity systems were found: \Vmif\ traces the Main Ionization Front (MIF), \Vscat\ arises from back-scattering of \Vmif\ emission by particles in the background Photon Dissociation
Region (PDR), \Vlow\ is an ionized layer in front of the MIF and if it is the source of the stellar absorption lines seen in the Trapezium stars, it must 
lie between the foreground Veil and those stars, \Vnew\ may represent ionized gas evaporating from the Veil away from the observer.
There are features such as the Bright Bar where variations of velocities are due to changing tilts of the MIF, but velocity changes above about
25\arcsec\ arise from variations in velocity of the background PDR. In a region 25\arcsec\ ENE of the  Orion-S Cloud one finds dramatic
changes in the \oiii\ components, including the signals from the \Vlowoiii\ and \Vmifoiii\ becoming equal, indicating shadowing of gas
from stellar photons of >24.6 eV. This feature is also seen in areas to the west and south of the Orion-S Cloud.

\end{abstract}
\begin{keywords}
\hii\ regions -- ISM:atoms -- ISM:dust,extinction --  atomic processes -- radiation mechanisms:general -- ISM:individual objects:Orion Nebula (NGC 1976)
\end{keywords}

\section{Introduction}
\label{sec:intro}

The general nature of the Orion Nebula and its associated Orion Nebula Cluster (ONC) of young stars is now well established \citep{mue08,ode08}. The visual wavelength images are brightest within a region designated as the Huygens Region of about 5\arcmin\ diameter lying in the NE\footnote{Throughout this paper we frequently use capital letters for the abbreviations for common directions, such as northeast. When the full spelling is used and capitalized, it indicates a region, when it is not, it indicates a direction.}corner of a 24\arcmin\ $\times$ 33 \arcmin\ region designated as the Extended Orion Nebula (EON), \citep{gud08}. The goal of 
the study reported on here is to determine the nature of large-scale features that appear only in high velocity resolution spectroscopy
and establish what these features tell us about the true nature of the Orion Nebula.

\subsection{The Appearance and the basic 3-D model for the Huygens Region}
\label{sec:appearance}

The Huygens Region is dominated by emission from a thin layer of ionized gas on the facing surface of the host Orion Molecular Cloud (OMC). The dominant emitting volume is designated as the Main Ionization Front (MIF) and it is stratified in ionization, become more highly ionized  and of lower density away from the actual ionization boundary. The boundary between the MIF and the OMC  is the Main Ionization Boundary (MIB), the region where ionization of hydrogen stops. On the other side of the MIB lies a dense layer of gas, dust, and molecules known as the Photon Dissociation Region (PDR) recently imaged at 11\farcs4 resolution in \cii\ 
emission \citep{goi15} that arises slightly within the PDR.  Variations in surface brightness occur because of increasing distance from the dominant ionizing star \tC, which lies between the MIF and the observer, and limb-brightening effects in tilted regions of the MIF.
 
In the foreground of the ionized volume there are two irregular layers of primarily neutral gas \citep{vdw89,vdw13,abel16} known as the Veil. The dust component of the Veil \citep{ode00} (henceforth OY-Z) accounts for most of the optical extinction. This extinction further modifies the appearance of the Huygens Region. The region of highest extinction lies to the east of \tC\ and is commonly called the Dark Bay.  

About 60\arcsec\ southwest\footnote{Henceforth in this paper directions such as southwest will be abbreviated to use combinations of upper case letters, e.g. SW.}of \tC\ there is an imbedded group of stars within a dense region designated as Orion-S \citep{ode08} that is the source of multiple collimated outflows from young stars \citep{ode15} (henceforth O15). The nature of the Orion-S feature is not exactly understood but certainly includes a neutral cloud of gas that we refer to in this paper as the Orion-S Cloud. The presence of molecular and neutral hydrogen lines in absorption means that there is an ionized region beyond it. There are no optical features that
can be attributed to the Orion-S Cloud, but about 25\arcsec\ ENE from the centre of this cloud is the brightest part of the Huygens 
Region, having multiple structures and rapid velocity and brightness changes. We designate this as the Orion-S Crossing. This Crossing
is near the centre of imbedded young stars and the numerous collimated outflows from them.

\citet{ode09b} and \citet{ode10} argue that Orion-S is a free-floating cloud within the cavity of the concave MIF. This would be an isolated remnant of dense gas and dust within the OMC. An isolated dense cloud would cast a radiation shadow in ionizing Lyman Continuum (LyC) photons  and thus could represent the tip of a ionized pillar such as seen in NGC~6611. Isolated cloud or pillar will depend upon
the strength of the LyC radiation field in the shadowed region. If the diffuse LyC radiation field that is formed by recombining hydrogen ions is weak, then the shadowed region will have an ionization boundary and we
would see a pillar. If this diffuse LyC field, supplemented by photons from \tA, is strong, then we will see
an isolated cloud. There are no optical features that indicate the boundaries of a pillar seen edge-on and the isolated
cloud model is the more likely. More creative but probably less likely \citet{tom16} argue that it possible that the Orion-S feature actually lies beyond the MIF and its background radio continuum arises in an otherwise unobserved ionized region beyond the PDR. In any event, its NE  boundary forms the brightest feature within the Huygens Region through proximity to \tC\ and limb-brightening. Henceforth we will refer to this bright region as the Orion-S Region. 

About 110\arcsec\ to the SE of \tC\ there is a long linear feature known as the Bright Bar, again being bright because of limb-brightening. This was most recently studied in the optical by O'Dell, Ferland, \&\ Peimbert (2017a) (henceforth O17a) and in the radio by \citet{goi16}. 

\subsection{Dynamics of the Huygens Region}
\label{sec:Dynamics}

In the current study we investigate the detailed structure of regions of the Huygens Region, the Veil, and a recently established \citep{abel16} layer of ionized gas lying between \tC\ and the Veil that we call, the \Vlow\ component. 
\footnote{In this paper all velocities are in \kms\ unless otherwise stated and all radial velocities are in the Heliocentric velocity system. To convert Heliocentric radial velocities to the Local Standard of Rest (LSR) system, subtract 18.1 \kms.
 When angular distances are converted to linear distances, we have adopted a distance of 
388$\pm$8 pc \citep{mk17},
 although we recognize that 414 pc, based on the study of \citet{men07} has been used frequently. The \citet{mk17} distance gives a scale of 0.01 pc = 5\farcs32.} 
 
 Our approach is one of trying to explain variations in velocity according to their individual features and regions. This is in contrast with the earliest studies that sought to characterize variations in velocities as being due to features of turbulence. The most recent and arguably best study using the statistical approach is that of Arthur, Medina, and Henney (2016), where they conclude that turbulence dominates the velocities between scales of 8\arcsec\ and 22\arcsec\ and that the emitting gas is confined to a thick shell.  The idea of the gas being primarily in a layer is confirmed in the study reported 
 here, but we establish that important variations in velocity occur locally within 25\arcsec\ due to variations in tilt of the emitting layer. 
 
The region closest to the MIB is a thin layer of  H$^+$+He$\rm ^o$ that gives rise to the \nii\ emission and 
has a characteristic e$^{-1}$ thickness of 0.0012 pc, corresponding to 1.0\arcsec, from the Cloudy models for the central region of the Huygens Region (O17a). 
The same set of calculations give an \oiii\ e$^{-1}$ thickness of 0.026 pc, corresponding to 14\arcsec.
In the seminal Orion study of \citet{bal91}, they derived the thickness of the ionized hydrogen zone as 0.08 pc, corresponding to
43\arcsec, from their extinction corrected Pa11 surface brightness. Given that the hydrogen emission arises from both the \nzone\ and \ozone\ zones and the method of calculation does not account for the increase in density within both zones, the models and the derived values are compatible.

The surface brightness of the MIF will decrease with increasing spatial separation from \tC, but limb brightening will enhance the local brightness of a tilted region. For a fixed size of the tilted region, more of the \nii\ emitting region will be seen edge-on and the  \nii/\oiii\ surface brightness ratio will be increased, being a maximum near the \nii\ boundary. The thinner nature of the \nii\ layer makes it the more useful measure of what is happening in a tilted region and \nii\ is usually preferred in tracing conditions within the MIF.  Outside of a tilted region the \nii/\oiii\ ratio depends on many local factors, in particular, the illumination by \tC.

In addition to high velocity features arising from outflows from young stars within the ONC, variations of the observed radial velocity (\Vobs) across the face of the Huygens Region are well known. 
In the case of material flowing away from an ionization front (Henney, Arthur, and Garc\'ia-D\'iaz 2005) the gas will have a characteristic flow velocity away from the underlying PDR; where, in the case of the \nii\ emitting layer, the material receded about 300 years before.  If the underlying OMC velocity was constant across the nebula and the MIF lay in the plane of the sky (henceforth simply the sky), then the observed radial velocities would be the velocity of the OMC blue shifted by the flow velocity for each ion (the material is accelerated during the flow, so that the evaporative flow velocity for the \oiii\ emission is greater than that for \nii\ emission).  

If there were no significant differences of velocities of the PDR, the differences in the observed optical line velocities at different lines-of-sight  will reflect the tilt of the surface of the MIF. This means that a MIF surface tilted perpendicular to the sky will have no component of radial velocity due to photo-evaporation flow and the observed velocity will be that of the PDR. However, there may be large variations in the radial velocity of the OMC gas, which means that the PDR velocity is not constant. 
Evidence for this is given in the statistical study of the radial velocities by \citet{art16}, who conclude that density variations within the OMC lead to much of the velocity and brightness variations seen in ionized gas. One goal of the present study is to determine where the radial velocities change because of differences of tilt and the velocity of the underlying gas.

In this analysis we often need to refer to a characteristic value of \Vpdr . When this is necessary we will use the results for the entire Huygens Region.
The recent study of \cii\  by \citet{goi15} gave an average velocity of 27.5 \kms\ with a characteristic Full Width at Half Maximum (FWHM)  line width of about 5 \kms.
 Examination of their velocity channel images indicates no radial velocity changes within the line's FWHM that cannot be ascribed to tilted or similarly peculiar regions, therefore our assumption of a constant \Vpdr\ is useful at the level of the \cii\ study's angular resolution and the radial velocity is V(\cii) = 27.5$\pm$1.5 \kms. 
 
 The average velocity of  molecules more massive than H$\rm _{2}$O (but excluding CO) as compiled in Table 3.3.VII of \citet{gou82}, is V$\rm _{ave}$ = 25.9$\pm$1.5 \kms, while his tabulation for the bright CO lines (which must arise further into the PDR than the \cii\ emitting layer) yields V$_{CO}$ = 27.3$\pm$0.3 \kms. In this study we will adopt \Vpdr\ = Vco = 27.3$\pm$0.3 \kms, which is consistent with the other PDR values and is similar to the velocities of the ONC stars of 25$\pm$2 \kms\ \citep{sic05} and 26.1$\pm$3.1 \kms\ \citep{fur08}. In our modeling we assume for the reference value of \Vpdr\  27.3$\pm$0.3 \kms, which is within the probable error of   
 the molecular cloud velocity \Vomc\ of 25.9$\pm$1.5 \kms\ and the \cii\ velocity of 27.5$\pm$1.5 \kms.

\subsection{Outline of This Paper}
\label{sec:outline}

In Section~\ref{sec:data} we describe the observational data used. The visual images were all from the Hubble Space Telescope (HST), the \cii\ images and spectra were from the Herschel/HIFI instrument observations \citep{goi15}, and the visual range spectra from a Spectral Atlas \citep{gar08} (henceforth the Atlas). In Section~\ref{sec:NE} we use spectroscopic data of large areas and high signal to noise ratio (S/N) to demonstrate the nature of red and blue velocity components on the shoulders of the MIF emission and establish their origins. The related Appendix \ref{app:Accuracy} establishes how well one can identify these components.  Regions where variations in velocity are primarily produced by changes in the tilt of the MIF are discussed in Section~\ref{sec:Inclined}. The significant variations
in the vicinity of the Orion-S Cloud are discussed in Section~\ref{sec:OriS}. Section~\ref{sec:RedFan} treats a region near the outer Bright Bar. Section~\ref{sec:SWcloud} discusses a region of high localized extinction. Section~\ref{sec:Minus90} considers a profile of spectra that cross both the Bright Bar and the Dark Bay. The interpretation 
of the velocity systems is then discussed in Section~\ref{sec:discussion}.

\subsection{A Glossary of Terms}
\label{glossary}

This paper addresses many features in the Huygens Region, some of which have been noted before and named. Sometimes
the names of individual features have evolved as different investigators emphasized different aspects of the objects. Some of
the features are newly recognized and we have tried to use simple but descriptive names for them. In order to help the reader
make their way through this paper, we give below a simple glossary of terms.

{\bf Atlas.} The compilation of high resolution spectra prepared by \citet{gar07}.

{\bf Boldface numbers.} These are to designate Slit Spectra in a profile, e.g. {\bf 1}, {\bf 2}~--~{\bf 7}.

{\bf Dark Arc.} An arcuate feature within the Orion-S Crossing that appears as darker than its surroundings.

{\bf \nzone\ zone.} The \nzone\ layer that is closest to the Main Ionization Front and produces the \nii\ emission.

{\bf \ozone\ zone.} The \ozone\ layer that overlies the \nzone\ zone and produces the \oiii\ emission.

{\bf Line.} A series of 10\farcs0$\times$10\farcs2\ Samples having the same Declination.

{\bf MIF.} The boundary between gas ionized by high energy stellar photons and the Photon Dominated Region.

{\bf NE-Region.} A grouping of 75 10\farcs0$\times$10\farcs2\ Samples selected to avoid the Dark Bay, the Bright Bar and the Orion-S Crossing, thereby forming a good representation of the quiescent parts of the Huygens Region.

{\bf Orion-S Cloud.} A molecule rich region seen in absorption lines in the radio continuum of the nebula.

{\bf Orion-S Crossing.} A region of rapid changes 
lying about 25\arcsec\ ENE of the Orion-S Cloud.

{\bf Orion-S region.} A broader region including both the Orion-S Cloud and the Orion-S Crossing.

{\bf OriS-IF.} The ionized region facing the observer on the surface of the Orion-S Cloud.

{\bf Profile.} A series of Slit Spectra chosen to cross a feature at a diagnostically useful angle.

{\bf Sample.} An area over which spectra in the Atlas have been averaged.

{\bf Slit Spectrum or Spectrum.} A spectrum composed of Atlas spectra selected to lie closest to a line on the sky. This simulates
the results from a long-slit spectrum.

{\bf Southwest Spoked Feature.} An unusual structure that appears to be moving past the Bright Bar to the NNW.

{\bf SW Cloud.} A discrete cloud of foreground material, associated with the Red Fan feature pointed out by \citet{gar07}.

{\bf Velocity Components.} Discrete features identified by the deconvolution of a spectrum. 

{\bf \Vblue .} Infrequent velocity components lying to the blue of the \Vlow\ components

{\bf \Vlow .} A system of velocity components blue shifted with respect to \Vmif .

{\bf \Vmif .} Usually the strongest component of a spectrum and attributed to emission by a layer of material close to the Main Ionization Front.

{\bf \Vnew .} A newly recognized but difficult to detect \oiii\ velocity component lying between \Vmif\ and \Vscat .

{\bf \Vscat .} A weak component redshifted with respect to \Vmif\ and attributed to dust back-scattering.

  \begin{figure}
	\includegraphics
	[width=3.5in]
	{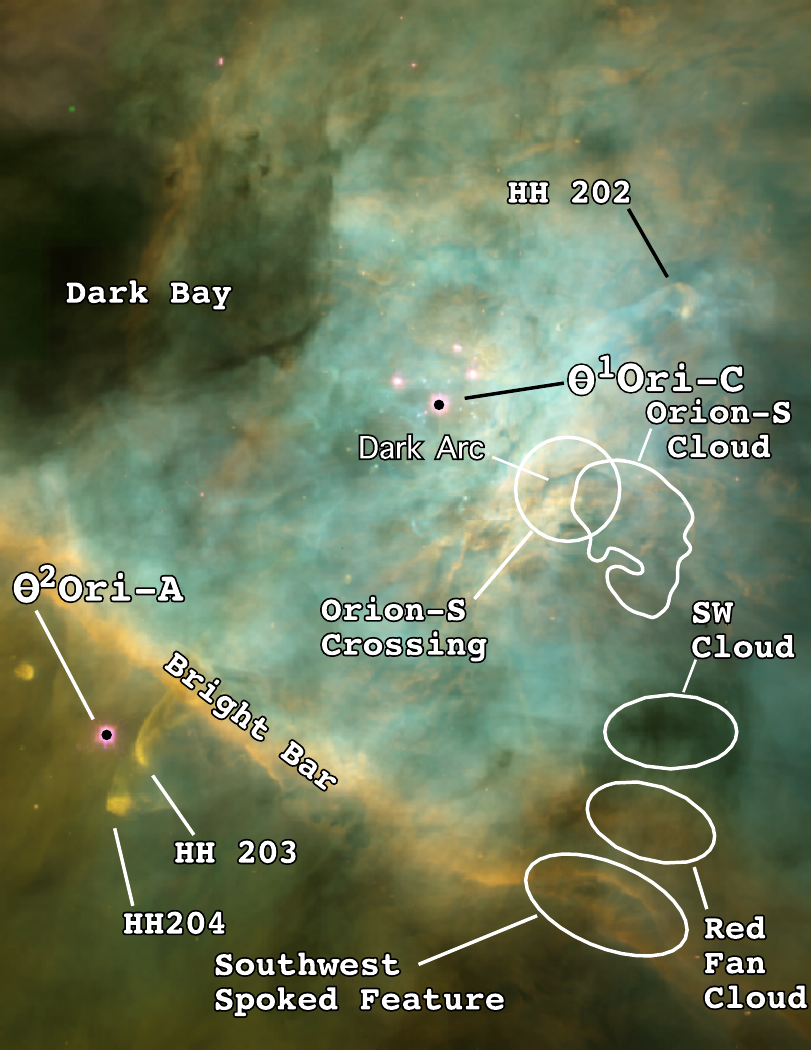}
    \caption{This 233\arcsec $\times$302\arcsec\ image of the Huygens Region is a reduced resolution sample from a larger mosaic \citep{ode96}. Blue indicates \oiii , green indicates \Ha , and red indicates \nii\ emission.
Major features of the nebula plus objects and regions discussed in this article are labelled. 
    As in all images in this paper, north is at the top and west to the right.The irregular form labelled as the Orion-S Cloud is the outer
    contour of the 21-cm absorption line of \Hi\  \citep{vdw13}.}
\label{fig:HRone}
\end{figure}

\section{Observational Data}
\label{sec:data}
We have been able to draw upon published  images and spectra of the Huygens region in this study. The images were made with the HST and are better than 0\farcs 1\ spatial resolution. The spectra were made with several ground-based telescopes and typically have resolutions of about 2\arcsec. Fortunately, the small velocity shifts caused by strong tilts in the MIF are well defined in the spectra.

\subsection{Images}
\label{sec:images}
We have used a wide variety of Hubble Space Telescope images made in narrow-band filters that isolate individual emission lines. We present in Fig.~\ref{fig:HRone} a low resolution replication of an early \citep{ode96} mosaic of WFPC2 camera images. This illustrates quite well both the large and small scale variations in the ionization in the nebula. The brightest part of the Huygens Region lies in the Orion-S Crossing and in the \nii\ line is rivaled in surface brightness by the Bright Bar. In this figure we
show the position of the two stars discussed below (\tC\ and \tA), the former being the dominant source of ionizing photons within the Bright Bar and the latter dominant outside the Bright Bar (O'Dell, Kollatschny, and Ferland 2017b).

\subsection{Spectra}
\label{sec:spectra}
The spectra we use are from the compilation of \citet{gar08}, where a compilation of north-south oriented slits at spacings of 2\arcsec\ in Right Ascension  covering much of the Huygens Region are given. This Spectroscopic Atlas (henceforth the Atlas) is made from spectra of about 10 \kms\ spectral resolution and is calibrated to 2 \kms\ accuracy and presented in steps pf 4 \kms. It is quite complete in the \Ha\ 656.3 nm, \oiii\ 500.7 nm, and \nii\ 658.4 nm lines, but has less complete coverage 
in \oi\ 630.0 nm, \sii\ 671.6 nm+673.1 nm, and  \siii\ 631.2 nm. The spectra  were sampled along the original north-south slits in steps of 0\farcs53.  We have used only the \nii\ and \oiii\ emission lines because the large width of the \Ha\ line caused by thermal broadening precluded study of small velocity differences.

In order to characterize the spectra, we performed deconvolution of each using
the IRAF task `splot'. A discussion of the accuracy of the results of using `splot' is discussed in Appendix~\ref{app:Accuracy}.

We made several types of samples of the spectra in this study, some of large areas and others that simulate subject-specific slit Spectra. The idea in each case was to increase the S/N above that in individual Atlas spectra.
This enhanced the visibility of faint features on the shoulder of the strong emission lines. 
When we created a pseudo-slit Spectrum along a non-north-south angle, we call that a Spectrum.

\subsubsection{Surface Brightness Calibration of the Spectra}
\label{sec:calibration}

The spectra were calibrated by taking the average signal of spectra in a 10\arcsec$\times$30\arcsec\  region beginning 5\arcsec\ west of \tC\ and comparing the same region with Hubble Space Telescope filter images that had been calibrated using the technique and reference numbers in \citet{ode09}. For convenience, we work in units of 100,000 original Atlas units. Conversion of our Atlas units to the surface
brightness units of ergs cm$^{-2}$ sec$^{-1}$\ steradian$^{-1}$ is found by multiplying the 500.7 nm instrumental units by 0.0614 and multiplying the 
658.4 nm instrumental units by 0.00782.

\section{Identification of Velocity Components using averaged lower spatial resolution spectra}
\label{sec:NE} 

In order to derive the velocity components in spectra from a relatively simple region of the nebula, which is also a region used in many other spectroscopic studies of the Huygens Region, we first derived 
high S/N spectra. These spectra were then studied for patterns in their velocity components, with multiple components being identified.

\subsection{Creation of high S/N spectra of a central region of the nebula}
\label{sec:makingNE}

\begin{figure}
	\includegraphics
	[width=3.5in]
	{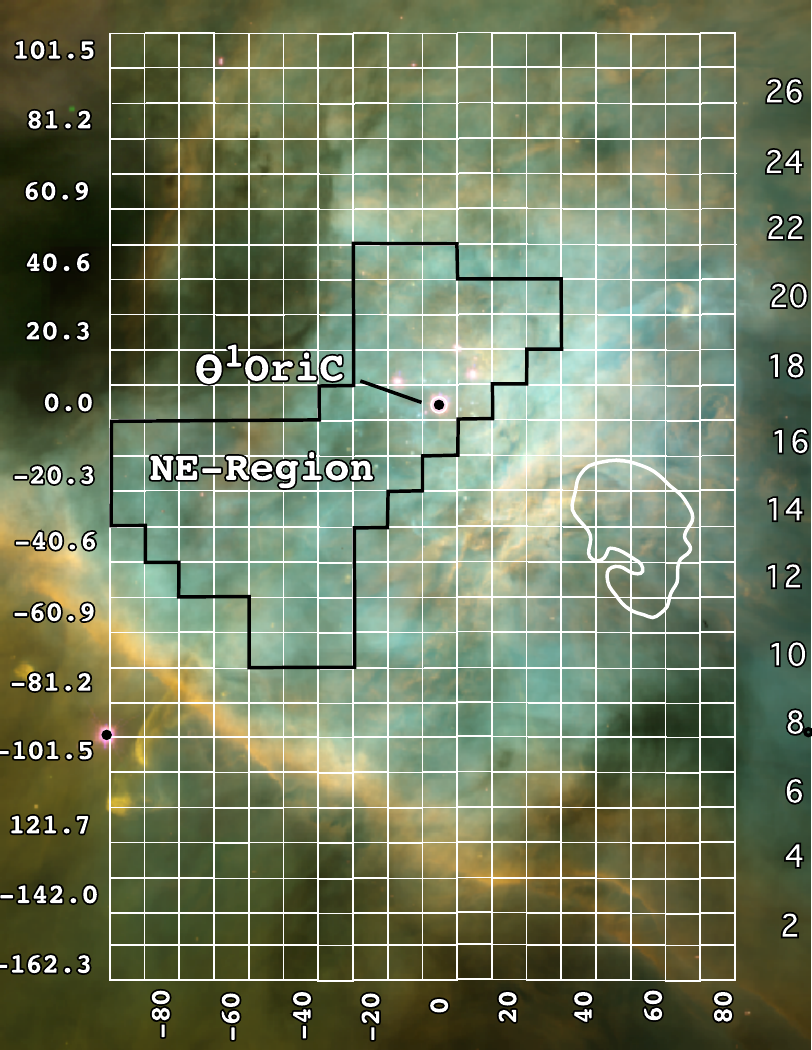}
    \caption{Like Fig.~\ref{fig:HRone} but excluding most of the labels and showing the white line orthogonal grid indicating the positions of 10\farcs0$\times$10\farcs2 boxes, whose spectra were averaged to produce lower spatial resolution but high S/N Samples, as described in Section~\ref{sec:NE}. The left hand labels indicate the displacements in seconds of arc from the box with \tC\ at its centre. For convenience, the corresponding box numbers along the north-south  axis are given on the right side. These are called Lines in the text. The irregular black line indicates the Samples included in the NE-Region that is used to characterize the different velocity components (Section~\ref{sec:NE})}
\label{fig:HRtwo}
\end{figure}

We averaged the spectra in the Spectroscopic Atlas in boxes of 10\farcs0$\times$10\farcs15 with the reference Sample (X=0, Y=0) centred on \tC. These are shown in Fig.~\ref{fig:HRtwo}. Within this array we identify a region designated as the NE-Region. This region represents a less complex area within the Huygens Region as it does not include features like the Dark Bay, the Orion-S Cloud, and the Bright Bar.

\subsection{Identification of the Velocity Components}
\label{sec:identification}

Study of the deconvolution of the spectra of the boxes within the NE-Region indicated that there were patterns in the velocity components. Analysis of these patterns indicate
that there are multiple components, each having similar characteristics of radial velocity and signal. 
The criteria we used in the assignment of velocity components are given in Table~\ref{tab:criteria} and the frequency of their occurrence is shown in Fig.~\ref{fig:HistogramsBoth}.  Not all Samples showed every component.

Usually the \Vmif\ component was strongest and is associated with emission from the evaporating gas near the MIF.  The largest velocity \Vscat\ component was discovered in earlier studies and is discussed in Section~\ref{sec:backscat}. The \Vnew\ component lying between the \Vscat\ and \Vmif\ components is present in only the \oiii\ spectra and is significantly stronger than the \Vscat\ component, although difficult to deconvolute because of the small
separation in velocity from the \Vmif\ component. Although seen in the high resolution \oiii\ study of \citet{hoc88}, it is elaborated in this study and is discussed in Section~\ref{sec:VnewProfiles}. 
The \Vlow\ component has a lower velocity than the \Vmif\ component and is also a discovery of this study. In earlier studies the term \Vblue\ was often used, but now we see that there are two low velocity systems, the frequent \Vlow\ and the more rare \Vblue . 
\Vlow\ is sometimes stronger that the \Vmif\ emission and is discussed in Section~\ref{sec:Vlow}. The lowest velocity and weak \Vblue\ components have probably been discovered before and are discussed in Sections~\ref{sec:VlowProfiles} and \ref{sec:VblueProfiles} .
All of the components are present in both \nii\ and \oiii\ with the exception of \Vnew , as noted.

\begin{table*}
\centering
\caption{Criteria for Identifying Velocity Components}
\label{tab:criteria}
\begin{tabular}{lll}
\hline  
Ion &Component&Criteria \\ 
   \hline
   \nii & \Vscat & \Vobsnii $\geq$30.0 \kms, the single red component.\\
\nii& \Vmifnii &The strongest component with \Vobsnii $\geq$15 and FWHM$\leq$18.00 \kms .\\
 \nii &\Vwidenii &  A \Vmifnii\ component with FWHM$\geq$18.00 \kms .\\
\nii &\Vlow & Velocity range 0 -- 15.00 \kms. The longer of two blue components when there are\\
 &  &two or S(obs,\nii)/S(mif,\nii)$\geq$0.05 when there is one. \\
\nii &\Vblue  &Velocity range -10  -- 1  \kms. The shorter of two blue components when there are\\
 & & two and S(obs,\nii)/S(mif,\nii)$\leq$0.05 when there is one. \\
 \oiii & \Vscat & Longer of two red components, when there are two or S(obs,oiii)/S(mif,\oiii)$\leq$0.1\\
  & &            when there is one red component.\\
\oiii & \Vnew  & Shorter of two red components or S(obs,oiii)/S(mif,\oiii)$\geq$0.1\\
 & & when there is one red component.\\
\oiii & \Vmifoiii & The strongest component with \Vobsoiii $\geq$15.00 \kms\ and FWHM$\leq$16.00 \kms .\\
\oiii &\Vwideoiii &  A \Vmifoiii\ component with  FWHM $\geq$16.00 \kms .\\
\oiii &\Vlow &Velocity range 0 -- 15.00 \kms. Longer of two blue components when there are two\\
 & & or S(\oiii)/S(mif,\oiii)$\geq$0.07 when there is one.\\
\oiii  &\Vblue\ &Velocity range -10 -- 6 \kms. Shorter of two blue components when there are two \\
 & &or S(\oiii)/S(mif,\oiii)$\leq$0.07 when there is one. \\
\hline 
\end{tabular}\\
\end{table*}

\subsection{Characteristics of the Velocity Components}
\label{sec:characteristics}

  \begin{figure}
	\includegraphics
	[width=3.5in]
	{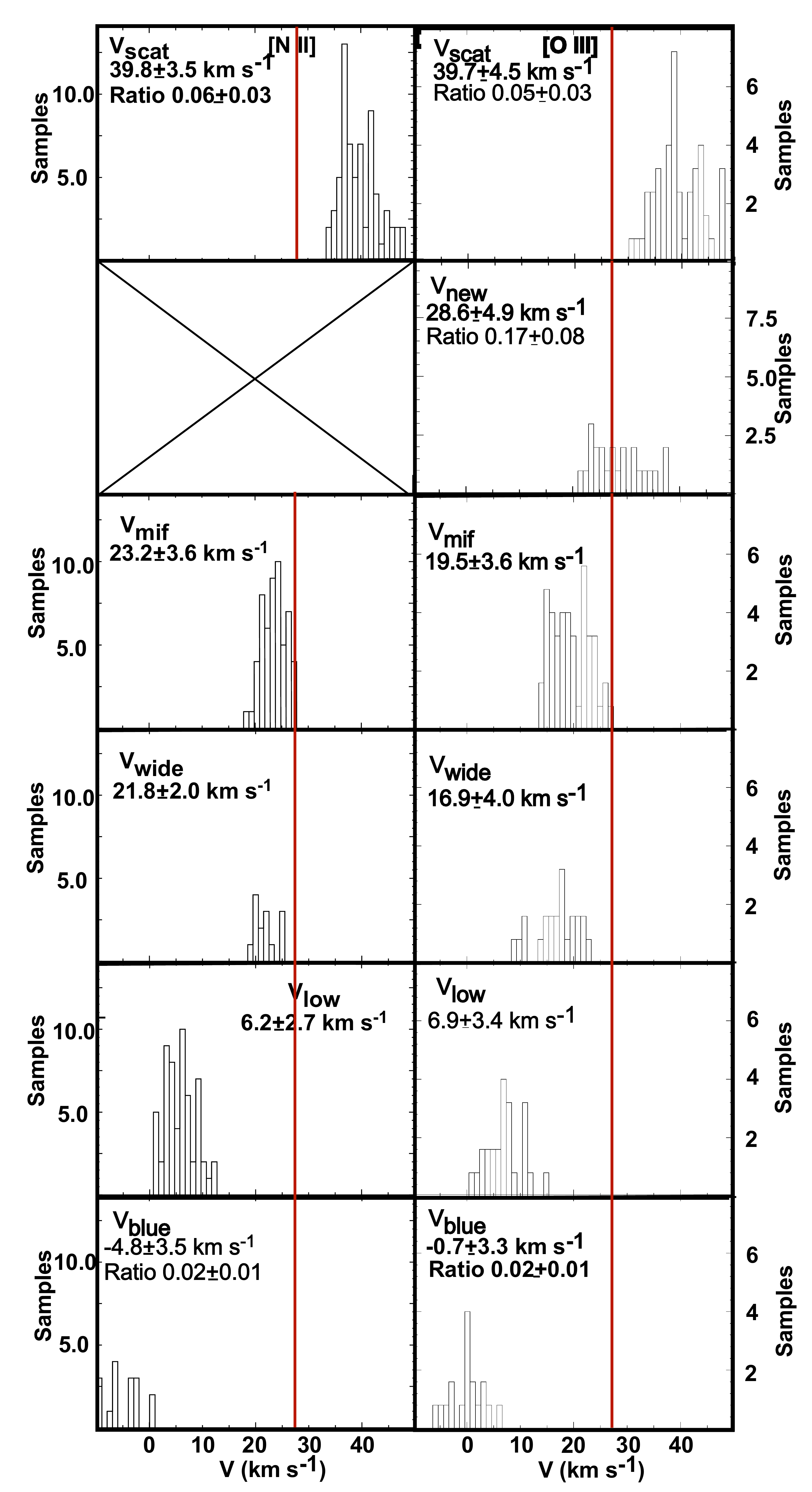}
    \caption{Histograms of the velocity components with the NE-Region are shown and are discussed in Section~\ref{sec:characteristics}.
    The red line indicates the average velocity of the PDR.}
\label{fig:HistogramsBoth}
\end{figure}

Each velocity component in the NE-Region has its own set of characteristics and are never the same for both ions. The frequency of occurrence and the average velocities are shown in Fig.~\ref{fig:HistogramsBoth}. In addition we give the average ratio of the signal of each as compared with the MIF component
(S$\rm_{component}$/S$\rm_{mif}$). This ratio was not calculated where the presumed MIF component has been assigned to the \Vwide\ class. 

{\bf \Vmif\ }average velocities are different, with \oiii\ clearly more blue shifted than \nii. Interpreting the displacement of both from \Vomc~= 27.3$\pm$0.3\kms\ as due to photo-evaporation from the flat face of the PDR, this means that the characteristic photo-evaporation velocity is about 4 \kms\ for \nii\ and 8 \kms\ for \oiii . The relative magnitude of these values is what is expected from gas accelerating away from the MIF, but greater than predicted
in the best theoretical model (Henney, Arthur, \& Garc\'ia-D\'iaz, 2005). The wider distribution of \Vmifoiii\ is consistent with the expectation that it arises from a thicker
emitting layer.  

{\bf \Vwide\ } occurs in a fraction of the Samples. We find that the average velocities are slightly bluer than the \Vmif\ component (-1.4 \kms\ for \nii\ and -2.6 \kms\ for \oiii). This is probably due to the line broadening arising from the MIF component being blended with a lower velocity component. A more quantitative analysis is not possible. The magnitude of the velocity shift and increase of FWHM of a composite (treated as a single line) formed from two separate lines is complex (O17a) as they depend on the relative strength of the two components and their velocity differences.  The different break-point for the two ions 
(18.0 \kms\ for \nii\ and 16.0 \kms\ for \oiii) reflect that the \nii\ lines are generally slightly broader.

{\bf \Vscat\ } is a common feature in both ions with a wide spread of velocities. This is what is expected from the weakness of this component, although this is balanced in part by the large velocity separation from the MIF component. The probable nature of this
component as light back-scattered by the PDR is discussed in Section~\ref{sec:backscat}.  

{\bf \Vnew\ } only occurs in \oiii. Its velocity is about mid-way between that of the much stronger \Vmifoiii\ and the much weaker \Vscatoiii\ components. A similar
strength \nii\ component would be more difficult to detect because of the smaller separation of \Vmifnii\ and \Vscatnii , but we have found no indication of a \Vnew\ component in \nii\ and it is probably simply absent. The nature of the \Vnewoiii\ component is discussed in Section~\ref{sec:DiscQuiVnew}.

{\bf \Vlow\ } occurs in both \nii\ and \oiii , being much more numerous in \nii . Average signal ratios are not shown in Fig.~\ref{fig:HistogramsBoth} because of the wide range of values. In the case of \nii\ 0.61 of the ratios occur between the
lower limit of 0.05 and 0.20, with 0.14 occurring between 0.30 and the maximum of 0.44. In the case of \oiii\ 0.52 of the ratios
occur between the lower limit of 0.07 and 0.20, with 0.41 occurring between 0.30 and the maximum of 1.68. These numbers 
indicate that the \Vlownii\ values mostly clump within the low range of ratios with a small fraction near the highest ratios. This 
contrasts with the \Vlowoiii\ values where the distributions are more nearly equal.

{\bf \Vblue\ } components are rarer than the \Vlow\ components. We have started their identification at -10 \kms, with the assumption
that more negative velocities are the results of outflows from discrete objects. In the case of the intrinsically most common type
of young stellar object (bipolar outflows from sources beyond the PDR), we selectively see the blue shifted components \citep{ode01}. This component is always weaker than the \Vlow\ components. 

The \Vblue\ and \Vlow\ components may be part of a single type of component lying to the blue of the MIF components. 
In Table~\ref{tab:criteria} they have been distinguished by pairs of criteria that overlap. The \Vblue\ velocity range is
-10 \kms -- 1 \kms\ for \nii\ and -10 \kms\ -- 6 \kms\ for \oiii\ with signal ratio criteria maxima of 0.05 for \nii\ 
and 0.07 for \oiii. 
The \Vlow\ velocity range is
0 \kms -- 15 \kms\ for both \nii\ and \oiii\ with signal ratio minima  of 0.05 for \nii\ 
and 0.07 for \oiii. The break-point between the criteria was determined by examining the data and identifying natural divisions.
If there is but a single V~$\leq$~15 \kms\ component for each ion, we used the signal ratio limits.

We consider it most likely that the \Vlow\ and \Vblue\ components are separate systems, with the <\Vlow > for both ions being indistinguishable with a combined average of 6.4$\pm$2.9 \kms, but that there may be different velocities for \nii\ and \oiii.   Further evidence for these being separate systems are presented in Section~\ref{sec:VlowProfiles} and Section~\ref{sec:VblueProfiles}.

\subsection{Examination of possible relations of the Velocity Components}
\label{sec:links}

We have sought to determine if there are relations between the several weaker velocity components with  
the MIF by comparing their velocities. When there are clear correlations between the component and MIF velocities there is likely to be a physical relationship. We will see that there are correlations for \Vscat\ but none for \Vlow, and \Vblue. There may be a correlation between \Vmifoiii\ and \Vnew (Section~\ref{sec:DiscQuiVnew}).

\subsubsection{The \Vscat\ Component}
\label{sec:backscat}

The first study to report the \Vscat\ component (O'Dell, Walter, and Dufour 1992) attributed it to back-scattering from the high density and dusty background PDR. 
That interpretation is strengthened by the fact that the continua in nebular spectra near \tC\ are much stronger than expected from atomic processes and this is probably due to back-scattering of stellar continuum from the Trapezium stars \citep{bal91}. 

The PDR is red shifted with respect to the emitting layers of the MIF (actually, the emitting layers are  physically blue shifted with respect to the nearly stationary PDR) and the scattered light from the PDR appears at about twice the velocity difference between the emitting layers and the PDR.  In a theoretical study \citet{hen98} put this interpretation onto more solid ground when he modeled the case where light  is scattered from a moving layer, covering all the possibilities (red and blue shifted scattering layers in the foreground, red and blue shifted scattering layers in the background). In each case there were upper limits to the displacement of the scattered light, but the flux distribution at velocities less that the maximum varied significantly, depending on the geometric model adopted. A separate paper modeling the predicted polarization of scattered light \cite{hen94} agreed with the limited observations available \citep{llb87}. 

We present here the most complete test of the interpretation of the red component of the emission line profiles as scattered light. This is done by using the high S/N NE-Region spectra. After establishing its origin, the red component can be used to inform the discussion of individual areas within the Huygens Region.

 \begin{figure}
  \includegraphics
[width=3.5in]
{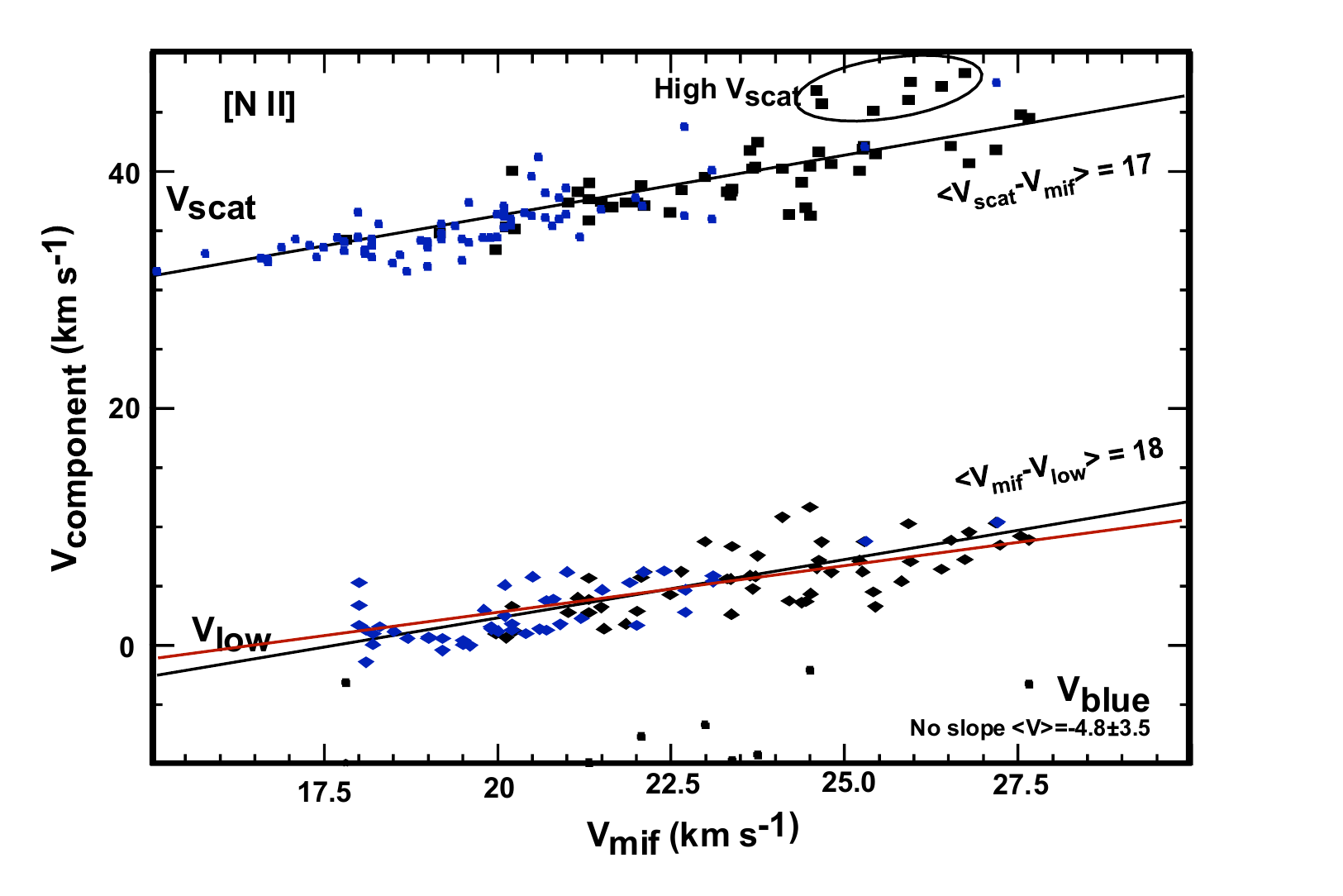}
 \caption{This figure illustrates the relation and lack of relation of \Vmifnii\ with \Vscatnii , \Vlownii, and \Vbluenii . The black symbols are from large Samples and the blue symbols are individual Spectra from outside this region. The black lines indicate
 the best fitting slope 1.0 for \Vscatnii\ and \Vlownii .The red sloped line is the least squares fit of slope 0.8 for \Vlownii . Only <\Vbluenii > values are given as there is no obvious correlation with \Vmifnii. The ellipse surrounds the set of data not included in the fitting of the \Vscatnii\ data, as described in Section~
 \ref{sec:backscat}.  The velocity difference labels refer to the difference in velocity from \Vmifnii\ of the \Vscatnii\ and \Vlownii\ components slope 1.00 fits.}
\label{fig:VmifVallNII}
\end{figure}

 \begin{figure}
  \includegraphics
[width=3.5in]
{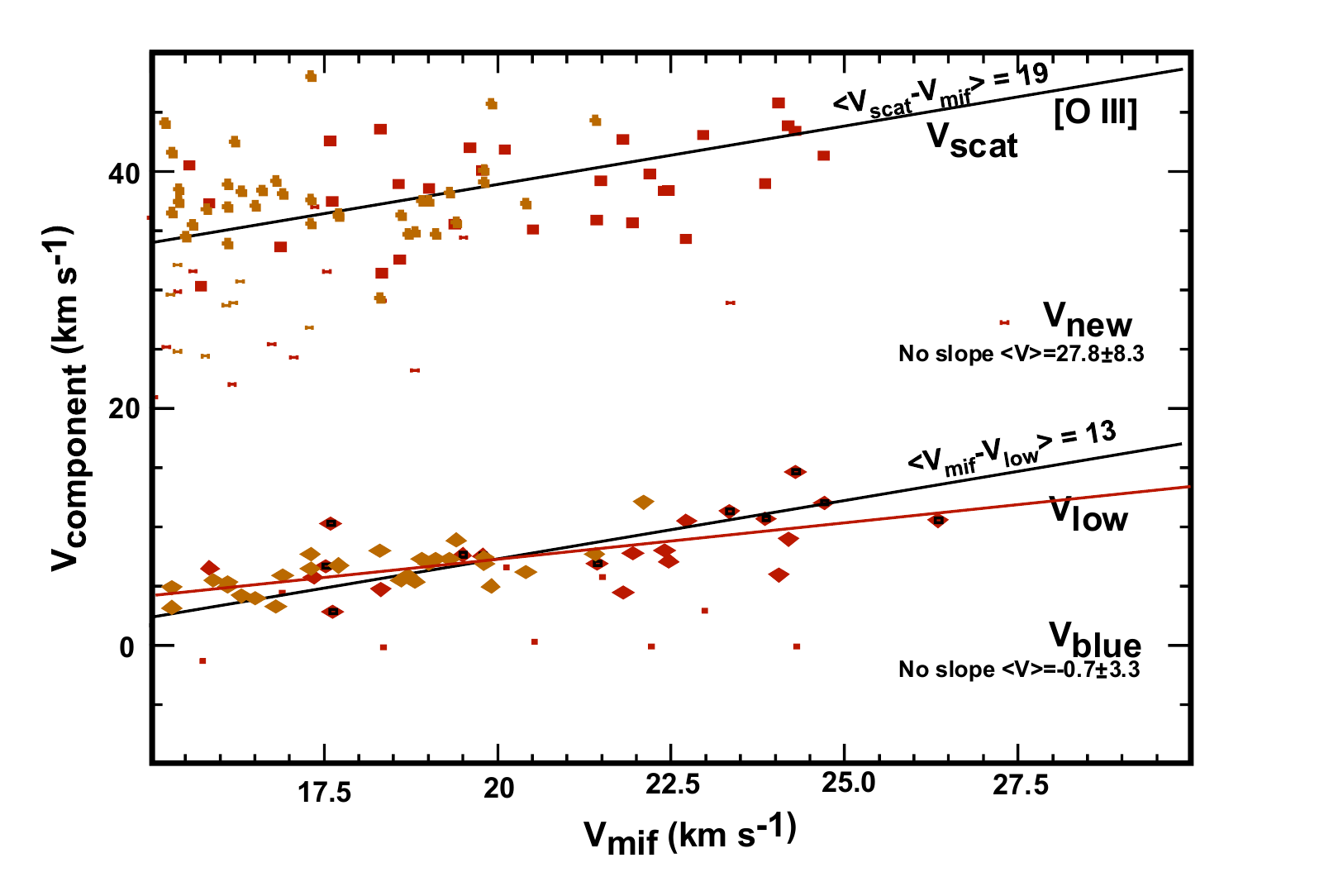}
 \caption{Like Fig.~\ref{fig:VmifVallNII} except for \oiii\ data. It also shows the data for the \Vnew\ components. 
 Unlike Fig.~\ref{fig:VmifVallNII} all of the data points are now used since the wider scatter may hide any peculiar clumping.i
 The red diamonds with central black markings are large Samples with high S(\Vlowoiii)/S(\Vmifoiii) values (0.20 -- 1.68), while the average of the other Samples is 0.10$\pm$0.04. The red symbols are from the large Samples and the orange symbols are from Spectra outside of the NE-Region.
 The black line passing through the \Vscatoiii\ data points is a slope 1.00 line fit to all of the data.
 The red sloped line is the least squares fit of slope 0.6 for \Vlowoiii\ while the black sloped line is a slope 1.00 fit. 
 \Vnew , and \Vblueoiii\ do not show a well defined correlation with \Vmifoiii\ and 
 only their velocities are shown. The velocity difference labels refer to the difference in velocity from \Vmifoiii\ of the \Vscatoiii\ and \Vlowoiii\ components slope 1.00 fits. }
\label{fig:VmifVallOIII}
\end{figure}

In the simplest model (the ionized surface lies in the sky) where the back-scattered component arises from only the column passing from the observer to the PDR, the \Vscat\ component's displacement will be 2$\times$\Vevap, where \Vevap\ is the photoo-evaporation velocity away from the PDR. However, the study of \citet{hen98} demonstrated that the back-scattered light will originate in emission outside the observed column, hence have different relative velocities. The result is that the back-scattered component will have a distorted line profile, whose peak displacement will be no greater than 2\Vevap\ and can be expressed as 2\Ascat $\times$\Vevap, where \Ascat\ is a correction factor for the distorted line profile. \Ascat\ will be one for the simplest geometry and less than one for more realistic cases.  The expected relation of the velocities will be 

\begin{equation}
\rm V_{scat} = V_{MIF}  + 2A_{scat} \times V_{evap} .
\label{eq:Vscat}
\end{equation}

\Vevap\ is different for various ions and is theoretically expected to be lowest for emission occurring near the MIF (e.g.  \nii ) and greater for emission rising further away (e.g. \oiii ) \citep{hen05}. Indeed, this expectation is what gave rise to recognition of the blister
model for the Huygens Region \citep{zuk73}~Ballick, Gammon, and Hjellming (1974).  \Ascat\ can also be different for the two observed ions because of the \oiii\ emission arising further from the MIF.
There may even be sample to sample differences in \Ascat\  within a specific ion because the distance (emitting layer to PDR) can vary across the nebula. However, this variation is expected to be less for \nii\ because that emission is concentrated to near the MIF, whereas the \oiii\ emission arises from throughout the \ozone\ zone. 

These expectations are realized in the upper portions of Fig.~\ref{fig:VmifVallNII} and Fig.~\ref{fig:VmifVallOIII}. There we see that <\Vscat\ - \Vmif >
is 17$\pm$1 \kms\ for \nii\ and 19$\pm$1.5 \kms\ for \oiii. As expected, there is greater spread and uncertainty for \oiii. In the calculation 
of the \nii\ difference we have not included the points within the `High \Vscat ' region because they are obviously anomalous, reflecting basically different conditions than in the other Samples.
These differences would correspond to $\rm  2A_{scat} \times V_{evap}$.

If \Vmif\ varies because of the local tilt, one would expect that the lowest values of \Vmif\ to correspond to where the MIF is nearly in the sky and would have the value \Vmif\ =~\Vpdr\ -~\Vevap\  and the back-scattered component would be
\Vscat\ =~\Vpdr\ + $\rm  2A_{scat} \times V_{evap}$.
The largest value of \Vmif\ would be when one views the MIF edge-on, thus removing the line-of-sight (LOS) component of \Vevap\  and \Vmif\ =~\Vpdr.
At that point the LOS component of the back-scattered light will also have been removed and \Vscat\ will have dropped to \Vpdr.
These considerations mean that the points for the MIF component in Fig.~\ref{fig:VmifVallNII} and Fig.~\ref{fig:VmifVallOIII} should have a negative slope, whereas the actual slopes are positive.

There are several other features to be noted in Fig.~\ref{fig:VmifVallNII} and Fig.~\ref{fig:VmifVallOIII}. In the \nii\ figure we see that the highest \Vmif\ values near 27.5 \kms\ equal that of the PDR, but these occur at Samples  -90\arcsec ,Line15 and -80\arcsec ,Line15 where there is no
change of ionization, as expected when viewing an ionization front edge-on. The strength of \oiii\ in Fig.~\ref{fig:HRtwo}
is due to the high ionization Big Arc East \citep{gar07} which is not a product of tilt of the MIF. 

The positive tilts in the figures and the occurrence of the maximum value of \Vmifnii\ at a region that does not show the
ionization changes of a tilted MIF argue that at large scales 
the variations in \Vmif\ are primarily due to variations
in velocity of the underlying MIF. 

The good agreement of \Vscat~-~\Vmif\ with the expectations for back-scattering from the PDR confirms anew the interpretation commonly applied to the
\Vscat\ velocity component. It also explains the variation in \Vmifnii\ values seen in Fig.~\ref{fig:HistogramsBoth} as being due to variations in \Vpdr ,
since it is established \citep{goi15} that the \cii\ emission that traces the PDR has a FWHM of about 5 \kms.  This agreement means that we can use similar observations of smaller areas of the Huygens Region to evaluate the local conditions. There is one region where the \Vscat~-~\Vmif\ values for \oiii\ are quite different from other regions and is discussed in Section~\ref{sec:DiscQuiVscat}. 

\subsubsection{Are the \Vlow\ components real?}
\label{sec:real}

{\bf Are there deconvolution problems?} The \oiii\ observations are more likely to be affected by deconvolution problems because of the smaller separation from
the MIF components. However, the model calculations discussed in Appendix~\ref{app:Accuracy} indicate that the derived values should be real
for both \nii\ and \oiii .

{\bf Are the \Vlow\ components due to blueshifted shocks?}
The NE-Region contains 75 Samples, 60 of which show \Vlownii\ components and 27 show \Vlowoiii\ components. The \Vlownii\ components are evenly distributed, except that none appear in Line 21. The NE-Region is highly ionized and the broad distribution of the \Vlownii\ components argue that they are not due to low ionization shocks. In contrast, most of the \Vlowoiii\ components fall in the left Samples in the lower half of the NE-Region, with almost all of those with S(\Vlowoiii )/S(\Vmifoiii)~$\geq$~0.2 occurring in 
the leftmost Samples within Lines 10~-- ~14. The Samples showing strong S(\Vlowoiii )/S(\Vmifoiii) occur
where \citet{doi04} detected a blue shifted \oiii\ feature near our \Vlowoiii\ values that they call the `Big Arc East'. We conclude that the strongest \Vlowoiii\ components are due to blueshifted shocks, but that the weaker \Vlowoiii\ and the \Vlownii\ components have another origin.

\subsubsection{The \Vlownii\ and \Vlowoiii\  components as scattered light}
\label{sec:Vlow}
The \Vlownii\ velocity component is probably correlated with the MIF emission, as 
we see in the lower portion of Fig.~\ref{fig:VmifVallNII}. 
\Vlownii\ agrees well with a slope 1.00 line.
The magnitude of the velocity differences is similar to that of the \Vscat\ components,  with <\Vmifnii - \Vlownii > ~=~18 \kms .

In Fig.~\ref{fig:VmifVallOIII} we see that there is only a hint of a correlation of \Vmifoiii\ and \Vlowoiii . This has greater scatter around a slope of one  and the best fitting slope is 0.64. This lack of a good correlation for the \oiii\ velocities may not tell us much since the emission
producing the \Vmifoiii\ layer is more diffuse. 






A blue shifted component can arise when there is a foreground layer of dust sufficiently optically thick in visible light and is blue shifted with respect to the MIF emission. We designate the velocity of the
scattering layer as \Vfront\ and the velocity of its scattered light as \Vlow.  The defining equation will be

\begin{equation}
\rm V_{low} = V_{MIF} -A_{low} \times (V_{MIF} -V_{front}),
\label{eq:VscatBlue}
\end{equation}
where \Alow\ is again a scaling factor related to the maximum velocity shift expected by the simplest version of the scattering model. 

If one assumes that \Alow\ is near the expected value of one, then equation~\ref{eq:VscatBlue} says that the 
scattered component  would appear at the velocity of \Vfront\ along each line of sight. The observed linear relation would argue that the \Vmifnii\ and \Vfront\ are directly related, with the averaged \Vfront\ varying from
about 0~--~10 \kms. The \Vlow\ component would then be expected to show the same spectrum as the
MIF, except without the velocity variation caused by different velocities of photo-evaporation. In the case of the NE-Region Samples, Figure~\ref{fig:HistogramsBoth} shows the average \Vlownii\ (5.7$\pm$2.7 \kms ) and \Vlowoiii\ (8.5$\pm$3.8 \kms ) values are indistinguishably the same.

In summary, we can say that the \Vlow\ velocities are consistent with a foreground scattering layer whose velocity
is closely linked to that of the underlying MIF. However, in the next section we demonstrate that another interpretation is more likely.

\subsubsection{The \Vlow\ layer as ionized gas}
\label{sec:IonizedVlow}
One must consider the possibility that the \Vlow\ layer invoked as a simple scattering layer could also be
ionized. If that is the case, then the \Vlow\ component could be a combination of forward scattered light
from the MIF and locally emitted radiation.

 We can determine if a blue shifted scattering layer is consistent with the \Vlow\ layer being ionized by calculating 
if such a scattering layer is optically thick to the LyC. To produce scattered light at about 10$\%$ the level of the MIF emission requires an optical depth of about 0.1.
 The product of the ratio of the number of dust particles to hydrogen atoms times the ratio of the scattering cross section of a grain to the absorption cross of hydrogen at the start of the LyC is small \citep{agn06}.  This means that a scattering layer sufficiently optically thick to 
 produce the scattered light would be very optically thick to LyC ionizing radiation unless the dust/gas ratio there is much lower than for the general interstellar medium. Under the assumption of a normal gas/dust ratio, the  gas in the scattering layer would
 be ionization bounded and a bright ionization front facing \tC\ would be formed. Another way of saying this conclusion is that a layer of gas and dust in the foreground of the MIF is much more likely to produce locally produced emission lines rather than scattering emission from the MIF.
 
 If there is a low column density of gas and there is a high LyC radiation field, the foreground ionized layer would be completely ionized (the gas bounded case) and some LyC radiation would penetrate and ionize the Veil. This is the model assumed by \citet{abel16}. However, if the column density of gas is high and there is a low LyC radiation field, the foreground material would be only partially ionized (the ionization bounded case)
 and LyC radiation would not penetrate into the Veil.  The presence of \oi\ emission in the \Vlow\ layer (as discovered in this study, Section~\ref{sec:DiscQuiVlow}) argues
 for the ionization bounded case and would necessitate new models to explain the ions and molecules observed in the Veil.

There is other evidence for a blue shifted ionized layer.
 In Section~\ref{sec:DiscOptAbsp} we discuss the results of optical wavelength absorption lines formed
 in the Trapezium stars at velocities similar to the \Vlow\ components. In addition, in Section~\ref{sec:DiscAbel16} we summarize the results from UV absorption lines found in \tB\ at similar 
 velocities. If truly linked to the layer producing the blue shifted \Vlow\ emission, this layer must be
 between the observer and the Trapezium stars. 
 
 In subsequent sections of this paper we try in part to clarify this issue by the study of detailed regions. In any event, as pointed out in \citet{abel16}, the \Vlow\ material is going to be colliding with Veil Component B in 30,000 to 60,000 years.

\section{Study of two highly inclined portions of the MIF}
\label{sec:Inclined}

In Section~\ref{sec:NE} we characterized the properties of a section of the Huygens Region that
is arguably the least complex and added to the diagnostic figures the results from individual Samples, all of the data supporting the model that 
variations in \Vmif\ are primarily due to velocity changes in the PDR. However, there are regions of known or suspected high tilts of the MIF and these are studied in this section.
In the following section we describe the expected variations in the radial velocity across a tilted MIF. This is followed by application of this model to two regions through the use of high spatial resolution composite Spectra. 

The first highly tilted region in the MIF to be recognized was the Bright Bar (sometimes called the Orion-Bar). More recently the NE Orion-S Region was recognized to be similar by \citet{md11} (henceforth M-D). OY-Z pointed out that there are similar, but less pronounced 
features called the East Bright Bar and the E-W Bright Bar. These objects were confirmed in the low ionization radial velocity study of \citet{gar07}, who also discovered another similar object designated as the Near East Bright Bar.

\subsection{Expected variations across a highly tilted MIF}
\label{sec:tilted}
In order to test our basic assumption that variation in \Vmifnii\ values are primarily due to variations in the tilt of the MIF, we have conducted a thorough examination of two regions having strong published evidence for their being locally highly tilted. In order to confirm our interpretation of variations in \Vmifnii\ as due to tilt, one needs supporting data that can be supplied by variations of surface brightness and ionization.

\begin{figure}
	\includegraphics
	[width=3.5in]
	{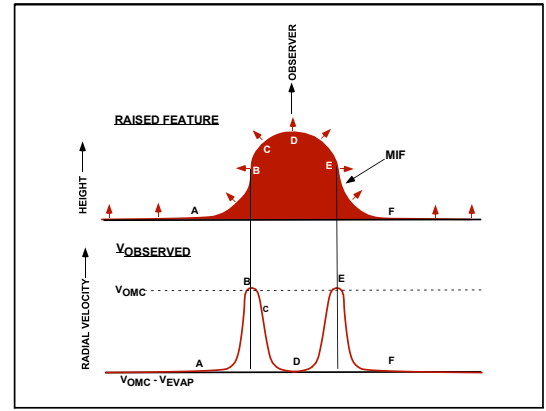}
\caption{This cartoon describes the relation between the angle of the MIF and the observed
radial velocity along lines of sight crossing a raised feature. It is assumed that the photo-evaporation flow (V$\rm_{EVAP}$) is always 
perpendicular to the MIF.  The radial velocity of the Orion Molecular Cloud is V$\rm_{OMC}$ and the significance of the positions A through F are discussed in Section~\ref{sec:tilted}.}
\label{fig:Cartoon}
\end{figure}

In Fig.~\ref{fig:Cartoon} we illustrate the expected variations in the observed radial velocity as the observer's LOS traces across a raised feature in the MIF.  Proceeding from the left to right in this figure the sequence is as follows.
At position A \Vobs\ will be \Vomc\ - \Vevap. Between points A and B as the MIF tilts toward the observer the \Vevap\ component decreases as the cosine of the tilt, thus raising \Vobs. \Vobs\ is a maximum at point B, where the MIF is perpendicular to the observer's LOS. \Vobs\ then decreases to \Vomc\ - \Vevap\ as one reaches the crest of the raised feature at point D. The pattern in \Vobs\ after point D simply mirrors that between A and D. 

If one is observing an escarpment, i.e. a feature that rises from points A through D and then remains flat at the highest level, then one will see only a single velocity peak followed by a drop back to \Vomc\ - \Vevap. If the MIF continues to slope towards the observer beyond the point of maximum tilt, then \Vobs\ will slowly decrease from the maximum value. If \Vobs\ continues near the peak velocity, the MIF surface must continue to rise, albeit slower. 

What the observer can expect to see is determined by the conditions of illumination of the MIF by the LyC photons that 
create the MIF.  If \tC\ were located in the lower left in the figure, then the regions to the right of point D would be shadowed from ionizing photons and there would be no emission. In fact, since \tC\ is the dominant ionizing star in the inner Huygens Region, the physical shape of the MIF will automatically be determined by the interaction of the radiation
from that star and the underlying density variations of the OMC. This is what produces the concavity of the Huygens Region. These considerations mean that in the outer part of the Huygens region we would only expect to see
single-peak velocity features. In contrast, in the regions of the nebula where the MIF is illuminated from above, then 
one can see a double-peak feature. 

A similar velocity pattern would also be seen if the feature of the MIF is a depression, rather than an elevation. The important difference is in the illumination. If \tC\ were located in the upper left in the figure, then the region A and C (and possibly on to D) would be in shadow and one would only see the velocity features arising from the illuminated further portions.  Again, an illumination from above would allow seeing a double-peak feature. The displacement of the peak of
the LOS emission from the direction of the maximum velocity can be very useful in distinguishing whether the velocity
features are caused by local elevations or local depressions.

\subsection{Bright Bar Spectra}
\label{sec:BrtBarSlits}

\begin{figure}
	\includegraphics
	[width=3.5in]
	{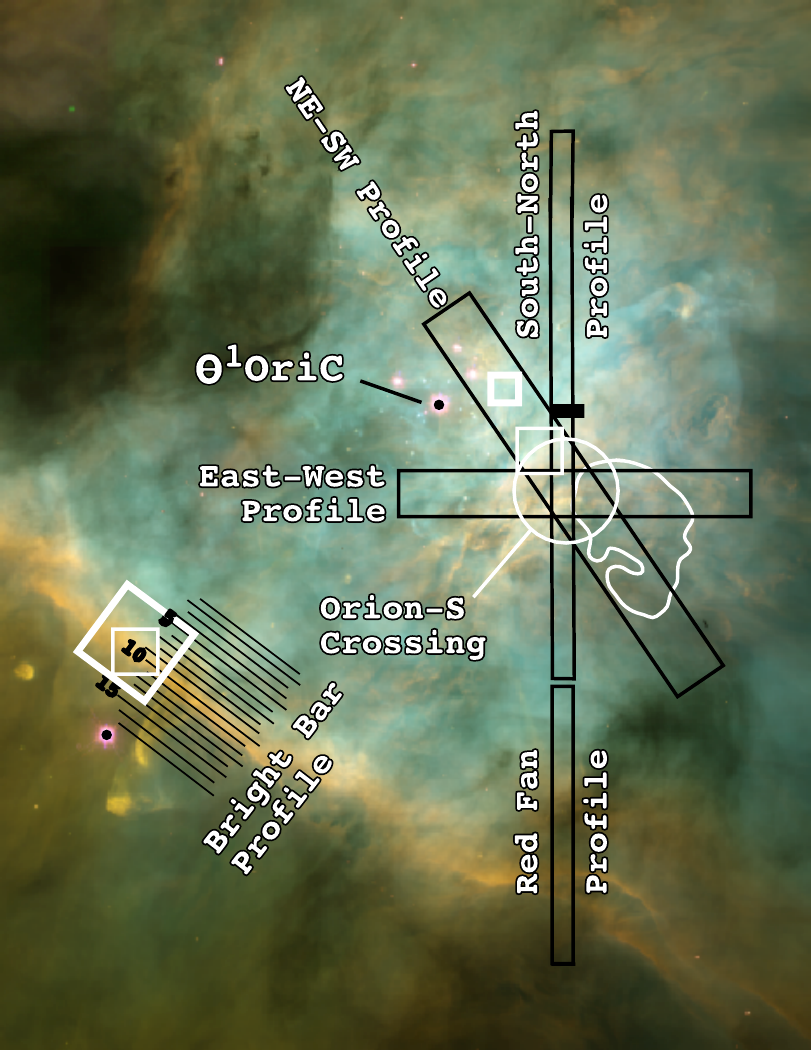}
    \caption{Like Fig.~\ref{fig:HRone} and Fig.~\ref{fig:HRtwo} but excluding some labels and including the individual artificial slits (Bright Bar Profile) crossing the Bright Bar feature and the areas covered in the other profiles that intersect in the Orion-S Crossing (large white line circle). The slit numbers for the profiles intersecting the Orion-S Crossing are given in Fig.~\ref{fig:RatioImage} and those in the Red Fan Profile are
given in Fig.~\ref{fig:SWSF}. 
The black box indicates the most thoroughly studied region in \citet{baldwin93}. The two thin white line boxes show the regions studied  by \citet{md11} while the wide white line boxes show the Bright Bar and Orion-S NE-Regions studied by O'Dell, Ferland, and Peimbert (2017a).
  Much of the region pictured is shown as a ratio image in Fig. ~\ref{fig:RatioImage}, where detailed features of the central slit sequences are shown.}
 \label{fig:HRthree}
\end{figure}

We created Spectra of the Bright Bar along a 53\degr\ position angle (PA) as shown in  Fig.~\ref{fig:HRthree}.
These were made by sampling among the nearest spectra in the Atlas's rectilinear grid, forming Spectra along lines with steps of 2\farcs 7 and average widths of 39\arcsec. 

Henceforth we will use integer numbers in boldface type to indicate  Spectrum numbers unless it is obvious that this is not the case.

\subsection{The Bright Bar}
\label{sec:BrtBarTilted}

Earlier conclusions that the Bright Bar is a highly tilted ionization front \citep{bal74,tie93,pel09,shaw09,md11,oss13,ode17a} or an edge-on view of a curved ionization front \citep{wal00} is confirmed and refined in this study.  Our study is complementary to that of M-D, who used 2-D spectrophotometry in the SE thin line white box shown in Fig.~\ref{fig:HRthree} to determine the electron temperature (\Te) and density (\Ne) and the ionization conditions. A multi-slit study complementary to the present investigation was described by O17a, where they used shorter slits with spacings of 1\arcsec\ across the SE heavy white box region shown in Fig.~\ref{fig:HRthree}. The latter study employed the same set of spectra that we use, in addition to MUSE \citep{wei15} line ratios to derive \Te\ and \Ne , finding similar results as M-D and confirming the edge-on view interpretation of the Bright Bar. Our current study differs from O17a in that it uses longer (37\arcsec\ instead of 24\arcsec ) and wider (2\farcs7 instead of 1\farcs0) slits, thus providing a higher S/N. The higher quality of the slit spectra have allowed study of both \nii\ and the lower signal \oiii , including their red and blue components, at a degraded but acceptable spatial resolution.

\subsubsection{Velocity and Surface Brightness Variations across the Bright Bar}
\label{sec:BarGeneral}

\begin{figure}
	\includegraphics
	[width=3.5in]
	{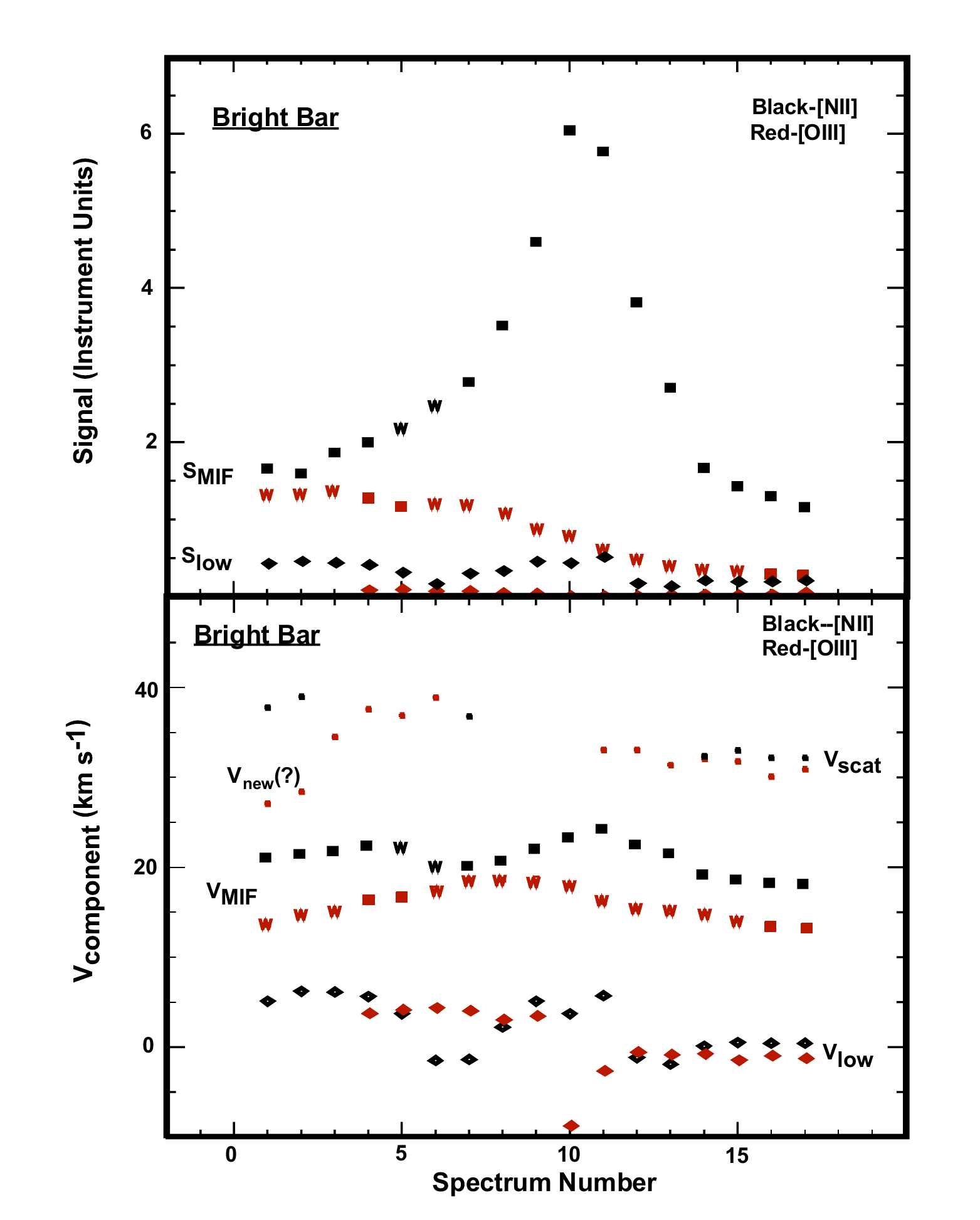}
\caption{The upper panel shows the signal for both the MIF components (heavy box symbol) and the S(\Vlow) components (diamonds) plotted versus the Spectrum number (increasing from NW to SE) for the Bright Bar Spectra shown in Fig.~\ref{fig:HRthree}. The Spectrum spacing is 2\farcs7. The letter W is superimposed on \nii\ MIF components with FWHM > 18 \kms\ and \oiii\ MIF components with FWHM > 16 \kms.
The meaning of these symbols remains the same throughout the remainder of this paper. As new symbols are employed their meanings are presented upon first use, then used without explanation subsequently. The lower panel shows the velocities of the MIF, \Vscat , and \Vlow\ components. The upper small boxes  correspond to \Vscat . 
}
\label{fig:BarDouble}
\end{figure}

We can use the signal and velocity information from our series of slits shown in Fig.~\ref{fig:HRthree}. The results for the individual Spectra are shown in Fig.~\ref{fig:BarDouble} and averages of selected groups of spectra are shown in Table~\ref{tab:BrtBarVelocities} and Table~\ref{tab:BrtBarRatios}. We discuss the individual Spectra in the order of passage across the Bright Bar. 

There is a local maximum in \Vmifnii\ at {\bf 5}, which indicates the direction of a tilted region in the MIF prior to reaching the Bright Bar. In full HST resolution images \citep{ode96} one sees local \nii\ peaks that also indicate tilted structures within the MIF. 

In Fig.~\ref{fig:HRthree} one sees that {\bf 11}
falls onto the bright region usually identified as the Bright Bar. In Fig.~\ref{fig:BarDouble} the peak of the signal is at {\bf 10} while the signal is only slightly less at {\bf 11}. This means that the features near these positions characterize the Bright Bar. In Fig.~\ref{fig:BarDouble}, where the expected accuracy of the determination is about  1 \kms , one sees a clear \Vmifnii\ peak of 24.0 \kms\ for {\bf 11}. 
Following the logic explained in Section~\ref{sec:tilted}, this means that {\bf 11} lies along the LOS where the MIF is 
most tilted.  Beyond {\bf 11} \Vmifnii\ decreases. The average of {\bf 14}~--~{\bf 17} for \nii\ is <\Vobs>~=~18.3$\pm0.5$ \kms\ and the average of 
Samples  that lie beyond {\bf 17}  (-90\arcsec ,Line 7; -80\arcsec ,Line7;-90\arcsec,Line6;-80\arcsec,Line6;-90\arcsec,Line 5;-80\arcsec,Line 5)
is <\Vmifnii >~=~17.5$\pm$1.6 \kms . 
The decrease in \Vmifnii\ outside of {\bf 11} indicates that the MIF flattens in this region. 

We have assigned different velocity components of individual Spectra according to the criteria for the NE-Region Samples as 
described in Table~\ref{tab:criteria}. The results are presented in Table~\ref{tab:BrtBarVelocities} and Table~\ref{tab:BrtBarRatios} (signal ratios).  The sampled regions selected are thought to be representative of the regions immediately before and after the Bright Bar.

The \Vscatoiii\ values for {\bf 1} and {\bf 2} are much lower than the other \Vscatoiii\ values (27.0 \kms\ and 28.3 \kms\ respectively).  These probably belong in the \Vnewoiii\ group in spite of their low
signals compared to the \Vmifoiii\ component (0.06 and 0.048 respectively). 

It is not clear where the \Vlownii\ components for {\bf 6} and {\bf 7} belong within the NE-Region classification. Their velocities
(-1.7 \kms\ and -1.6 \kms) are similiar to those in {\bf 12} and {\bf 13} (-1.4 \kms\ and -2.1 \kms). However, their ratios (0.07 and 0.11) are notably higher than in {\bf 12} and {\bf 13} (0.05 for both). 

The location of the peak signals give us some indication of the thickness of the emitting layers.
The fact that the peak signal for \nii\ lies about one step (2\farcs7) inside the peak velocity indicates that this must be about the thickness of the projected \nii\ zone, which would be larger than the emitting layer if the tilt is not exactly 90\degr.
The \oiii\ signal behaves very differently, monotonically decreasing across the region of maximum tilt and shows a weak rise 
near {\bf 7}, which is just inside of the peak \oiii\ \Vmif\ at {\bf 8}. The displacement of the \oiii\ features of about 
three steps suggest that the \oiii\ emitting layer is about 8\arcsec\ thick.  

\begin{table}
\caption{<V(component)> for Bright Bar Spectra*}
\label{tab:BrtBarVelocities}
\begin{tabular}{cccc}
\hline
Component & Spectrum Range & <V(component)>\\
\Vlownii       & 8~--11                 & 4.0$\pm$1.6  \\
''                   & 14~--~17              & 0.2$\pm$0.02 \\
\Vlowoiii       & 4~--~9                  & 3.8$\pm$0.5\\
''                   & 14~--~17               & -1.2$\pm$0.3\\
\Vmifnii         & 6~--~8                   & 20.0$\pm$0.3\\
    ,,               & 14~--~17               & 18.3$\pm$0.5\\
  \Vmifoiii      &  6~--~8                  & 18.3$\pm$0.5\\
    ,,              &  14~--~17              & 14.0$\pm$0.6\\
\Vscatnii       & 1~--~7                   & 37.8$\pm$1.1\\
''                    & 14~--~17              & 32.4$\pm$0.4\\
\Vscatoiii       & 3~--~6                  & 36.9$\pm$1.8\\
''                     & 11~--~17              & 31.2$\pm$0.8\\
\hline
\end{tabular}\\
*All velocities are in \kms.
\end{table}

\begin{table}
\caption{<S(component)/S(MIF)> for Bright Bar Spectra}
\label{tab:BrtBarRatios}
\begin{tabular}{cccc}
\hline
Component & Spectrum Range & <S(component)/S(MIF)>\\
\Vlownii       & 6~--10                  & 0.09$\pm$0.02  \\
''                   & 14~--~17              & 0.15$\pm$0.03 \\
\Vlowoiii       & 4~--~9                  & 0.06$\pm$0.02\\
''                   & 14~--~17               & 0.10$\pm$0.05\\
\Vscatnii       & 1~--~7                   & 0.04$\pm$0.01\\
''                    & 14~--~17              & 0.18$\pm$0.01\\
\Vscatoiii       & 3~--~6                  & 0.03$\pm$0.01\\
''                     & 11~--~17              & 0.19$\pm$0.04\\
\hline
\end{tabular}\\
\end{table}

These results can be compared with the nearby sample studied in O17a. 
The O17a region does not include the structured features
that produce the local \Vmifnii\ feature at {\bf 6}. However, one sees a similar displacement of the peak signals of \nii\ and \oiii\ and peaks in \Vmif.
Again, the \Vmifnii\ and \Vmifoiii\ drop outside of the Bright Bar, but the region sampled does not go out far enough to establish that one is seeing the
flattening of a previously steep region of the MIF.

Because of the lower S/N, O17a could not measure \Vlowoiii, although \Vlownii\ was measured. In that study the lowest velocity components are called V(blue), but with 
the recognition in the present study of a difference of the blue and low components, they should be considered to be part of the  \Vlow\ components. 
Including all of our new \oiii\ data into the \Vlowoiii\ system requires lowering the lowest velocity of this system to -1.5 \kms\ (from 0 \kms ). For
{\bf 4 --- 9} <S(\Vlowoiii)/S(\Vmifoiii)>~=~0.06$\pm$0.02, thus being slightly below the NE-Region lower limit of 0.07, while the average for {\bf 13 --- 17}
is 0.10$\pm$0.05, well within the NE-Region limit. Although the signal ratio for the \Vlownii\ components fall within the  lower limit of the NE-Region definition of 0.05,  there is a noticeable difference on the two sides of the Bright Bar since <S(\Vlownii)/S(\Vmifnii)> is 0.09$\pm$0.02 for {\bf 6 --- 10} 
and 0.15$\pm0.03$ for {\bf 14 --- 17}. As noted previously (Section~\ref{sec:images}), the region outside the Bright Bar is ionized by \tA , thus it is not surprising that the \Vlownii\ system is different there.

As in O17a we again see that \Vlownii\ drops abruptly when crossing the Bright Bar feature and now we note that the same is true for \Vlowoiii. The significantly different low values of \Vlownii\ at {\bf 6} and {\bf 7}, are probably associated with 
the secondary regions of tilt associated with the drop there in \Vmifnii , that follows the velocity peak at {\bf 5}. This region was not included in the O17a study.

We can summarize this section by saying that in this section we see strong evidence for variations in \Vmif ,
especially \Vmifnii , as being due to structure (varying tilt) within the MIF. The scale of these changes are of
a few slit spacings (three would the 8\farcs1, slightly less than the size of the NE-Region Samples)

\subsection{The M-D Region}
\label{sec:OriSTilted}

\begin{figure}
	\includegraphics
	[width=3.5in]
	{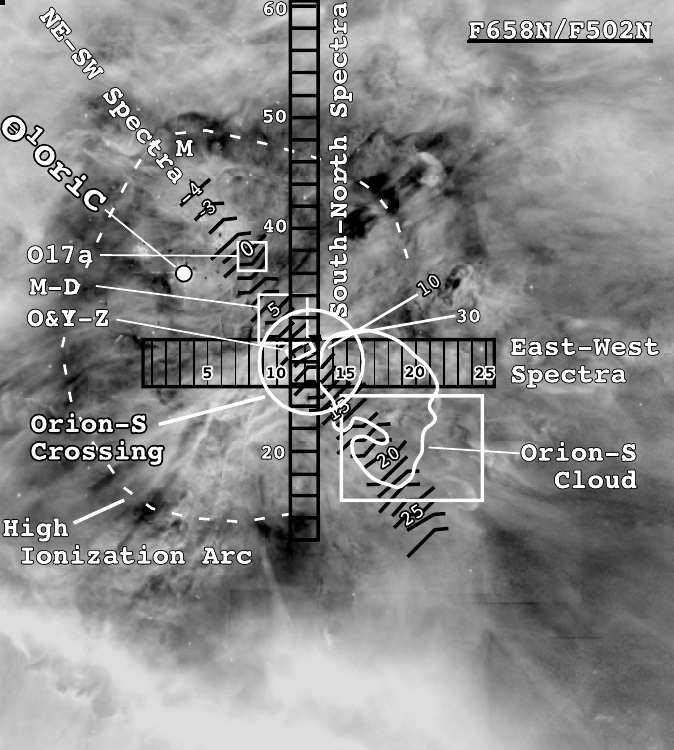}
\caption{This 194\arcsec $\times$216\arcsec\ image of the central Huygens Region depicts a full resolution HST image in the F658N (\nii ) filter divided by the image in the F502N (\oiii) filter. The smaller white boxes indicate regions studied in detail in O\&Y-Z, M-D, and O17a. The large white box indicates the region to the SW of the Orion-S Cross that is discussed in Section~\ref{sec:SWofCrossing}. The position of three sequences of Spectra are shown, together with their slit numbers. . 
The Orion-S Cloud is indicated to the SW of \tC\ (marked by a white dot) by the outermost 21-cm line absorption profile from the study of \citet{vdw13}. The approximate outline of the High Ionization Arc \citep{ode09b} is shown with the dashed white line.}
\label{fig:RatioImage}
\end{figure}

\begin{figure}
	\includegraphics
	[width=3.5in]
	{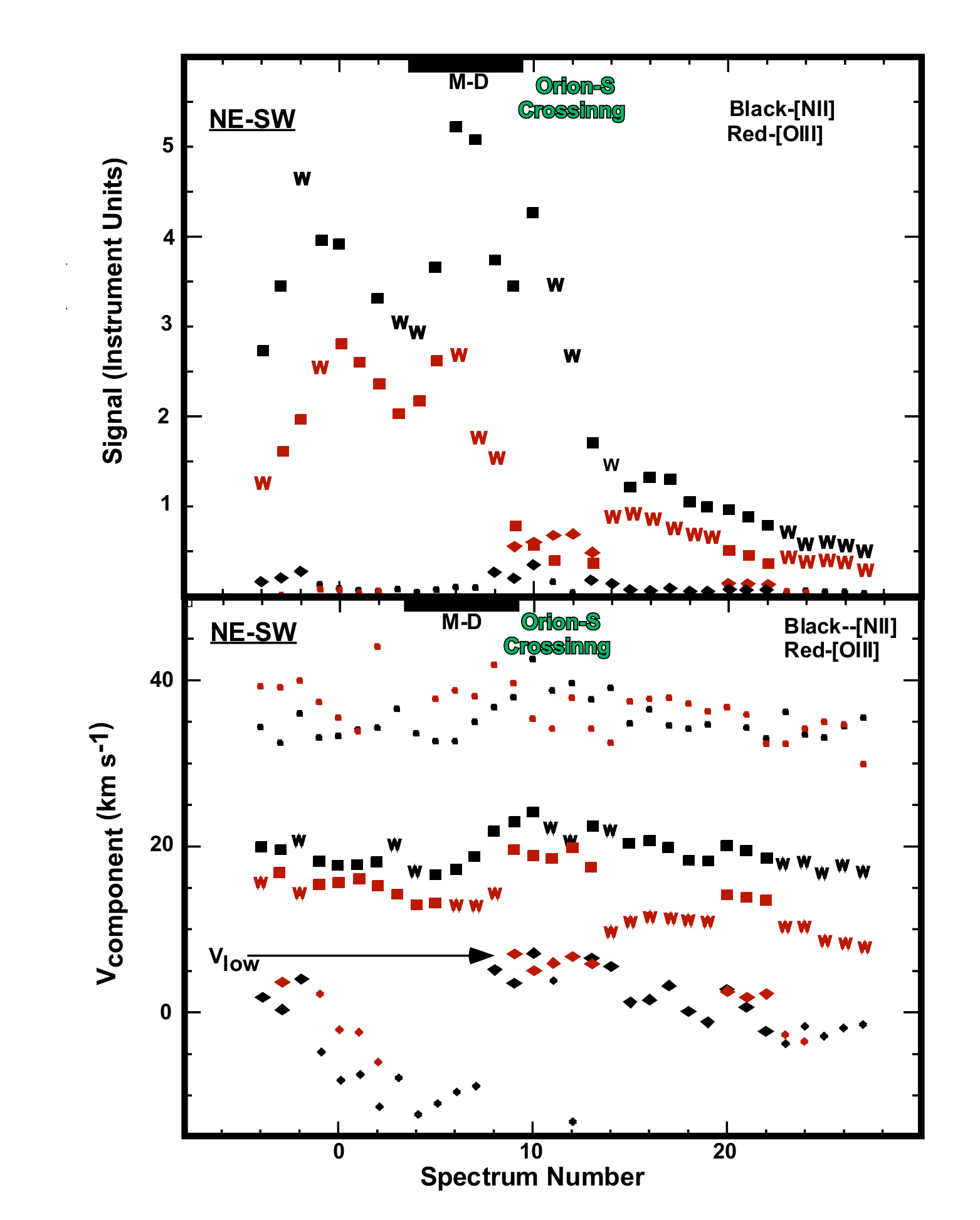}
\caption{
Like Fig.~\ref{fig:BarDouble} except now for the  NE-SW Spectra shown in Fig.~\ref{fig:RatioImage}. These have
an average spacing of 3\farcs6 (however there are narrower or wider spacings as necessary for resolution of features).  
The symbols have the same meaning as before except now the lower velocity systems are broken into
two components, as in Table~\ref{tab:criteria} established for the NE-Region Samples. 
Large diamonds indicate the  low velocity components discussed in Section~\ref{sec:Vlow} and the small diamonds the blue velocity components discussed in Section~\ref{sec:VblueProfiles}.  Orion-S Crossing refers to the  Orion-S Region centred on the
Dark Arc feature marked in Fig.~\ref{fig:RatioImage}. The heavy line indicates the range of Spectra that diagonally cross the M-D region studied spectrophotometrically at low spectral resolution.} 
\label{fig:OriSDouble}
\end{figure}

The NE part of the Orion-S Region is the second region with a well established highly tilted structure (M-D). It is the brightest part of the Huygens Region in \Ha\ radiation. This is due to a combination of two facts: the region has a high LyC flux due to the proximity to \tC\ and there is limb-brightening due to viewing an ionized layer nearly edge-on. Its structure was very accurately evaluated using monochromatic images of multiple ions in the M-D study; however, they did not consider variations in \Vobs\ and was over a limited range of positions (16\arcsec $\times$ 16\arcsec\ Field of View (FOV)).  In this section we will
expand on the M-D study of this region. 

We made a series of Spectra described in detail in Section~\ref{sec:OrionSslits} passing through the M-D FOV. The features in their study were collectively called `NE Orion-S'. 
The positions of our Spectra crossing the NE Orion-S feature are shown in Fig.~\ref{fig:HRthree}. Our average Spectrum spacing was 3\farcs6, being slightly larger in the negative
value slits and closer where conditions were changing rapidly. The resulting spatial resolution is poorer than the 1\arcsec\ spaxels of M-D, although they over-sampled their poor resolution images. The results from our Spectra are shown 
in Fig.~\ref{fig:OriSDouble}, with the slits crossing the M-D FOV indicated by heavy dashed lines. These are drawn slightly wider than the M-D FOV because of the effects of spatial resolution.

Within the area of overlap of our Spectra and the M-D FOV, there is excellent agreement of our S(\nii) and S(\oiii) distributions with those of M-D. {\bf 4}~--~{\bf 9} cross the M-D FOV. Our velocity information for \nii\ indicates that the MIF tilts up beginning at {\bf 4} and reaches a maximum
    tilt at {\bf 10}, then slowly decreases in tilt to the SE, with a temporary increase in tilt at {\bf 13}.  
The change of velocity (7.5 \kms) is much greater than in the Bright Bar (4.2 \kms). The similarity continues
in that the two peaks in \nii\ Signal ({\bf 6}~and {\bf 10}) are displaced towards \tC\  from the maximum tilt Spectra, although
in this case by larger amounts. The greater change in velocities indicates that either \Vevapnii\ is much greater in
this part of the nebula (entirely reasonable because of the higher illumination by LyC photons) or the NE Region
is tilted almost edge-on.

In the case of \oiii , we see evidence for a maximum tilt at {\bf 9} and the corresponding peak in Signal occurs
    at {\bf 6}. After the peak at {\bf 9}, the \Vobsoiii\ slowly decreases, with a slight rise at {\bf 12}.
The peculiar nature of the \Soiii\ and \Vobsoiii\ values in {\bf 9}~--~{\bf 13} are discussed in Section~\ref{sec:DiscCrossing}.

These results are similar to those concluded in M-D, where they demonstrated that this region has many of the ionization features associated with seeing an edge-on ionization front. These included an elongated peak in \Ne\ inside an elongated increase in \Te , in addition to a progression of increasing ionization towards \tC\ (both are features seen in the Bright Bar). 
The difference in our two studies is that M-D drew only on spectrophotometric data and placed the region of maximum tilt in the SW corner of their FOV,
whereas we use velocity data to present evidence that the peak occurs about 6\arcsec\ further to the SW.

For the purposes of this study, the important result here is that the NE Orion-S feature velocity changes over scales of about 25\arcsec\ can be primarily attributed to variations in the tilt of the MIF.

 \section{Profiles Crossing the Orion-S Crossing}
 \label{sec:OriS}
 
 In Section~\ref{sec:appearance} we summarized the properties of the Orion-S Cloud and the evidence 
 for its being a high molecular density region with active star formation. Through the presence of neutral hydrogen and molecular absorption lines seen in the radio thermal continuum, we know that it has
 ionized gas both on the nearer side facing the observer and on the far side.  
 
 Some comments here on  the adopted nomenclature are appropriate.
The Orion-S Cloud will often be called the ''Cloud'' (not to be confused with the background Orion Molecular Cloud-the OMC). The larger region near the Cloud and most visible in optical
radiation we will call the ''Orion-S Region''. The smaller area within the Orion-S Region where there is an intersection of the axis of a series of Spectra will be called the ''Orion-S Crossing''. The sequences of Spectra will be called ''Profiles''. 

It is debatable if the emission arising from the observer's side of the optically thick Cloud
should be called the MIF. More properly, the layer that produces the continuum against which the radio absorption lines should be called the MIF. For this reason we designate the ionized layer on the observer's side of the Cloud as the Orion-S Ionization Front (OriS-IF).  

There are many regions of small-scale structures in the Orion-S Region and this has driven the decision
 to use fairly narrow spectra, which in turns means that the S/N ratios of the Spectra are less than in the larger Samples used in the study of the NE-Region. The exceptions to this are the very bright regions in the NE corner of the Orion-S Cloud. This probably accounts for our not seeing some features present in the NE-Region Samples.

At the Orion-S Cloud we know that the observed radiation will be dominated by the OriS-IF and clockwise from the SE through the north of the cloud the radiation will be from the MIF. 
Beyond the Cloud the observed radiation can continue to 
 be from the OriS-IF, can be dominated by the MIF, or can be a combination of both.  We will
 first summarize the information from each of the profiles, then combine this information with that from a large Sample SW of the crossing of the profiles (Section~\ref{sec:SWofCrossing}).    
 
 The Northeast-Southwest (NE-SW), East-West (E-W), and South-North (S-N) Profiles naturally divide into three sections. In NE-SW and E-W the first section represents what resembles
 MIF emission, the second clearly dominated by OriS-IF emission, and a third within which it is more ambiguous about the type of emission that dominates. 
 
Division of the S-N Profile is less clear. In the North section both the \nii\ and \oiii\ lines are a product of the MIF and then transitions to the OriS-IF area.  In the South section of this profile \nii\ emission appears to be from the MIF, but \Vmifoiii\ drops to velocities associated with the \Vlow\ system. 

 \subsection{Orion-S Spectra}
\label{sec:OrionSslits}
In order to fully map the Spectra of the Orion-S Region, we created three sequences of Spectra, all passing through or near 
the Dark Arc feature that lies near the NE boundary of the Orion-S Cloud.  One sequence was perpendicular to the multiple linear features on the NE boundary of the Orion-S cloud (PA~=~124\degr\ , O15) discussed in Section~\ref{sec:OriSTilted}. The other two sequences are north-south and east-west in orientation.

{\bf NE-SW Spectra} were created in a fashion similar to those for the Bright Bar. However, in this case the average spacing of the slits was close to 3\farcs6 and the average width was 13\arcsec .The spacing between slits -3 and -4 was 5\farcs0 and there are a total of 32 spectra. As shown in Fig.~\ref{fig:RatioImage}, the axis of the sequence was along 214\degr , which is perpendicular to the orientation of the local highly tilted ionization front (O15). 
The width was selected to match the 16\arcsec $\times$16\arcsec\ region studied with a multi-aperture spectrograph by \citet{md11}.

The axis  of this series of slits lies in the direction of the opening of the C-shaped high ionization arc that surrounds 
 \tC\ \citep{ode09b,ode17a}. As discussed in \citet{ode09b}, this well defined feature is best seen in images that are the ratio of high and low ionization emission lines. Therefore, we show in Fig.~\ref{fig:RatioImage} a ratio image in \nii\ over \oiii . 
 It appears that the Orion-S Cloud either lies in front of the high ionization shell or interrupts it. 

{\bf E-W Spectra} were created from a sequence of 25 slits with a spacing of 4\farcs0. The easternmost slit is 9\farcs4
east of \tC\ and the centre of the sequence is 25\farcs7 south of \tC. The slit heights are 12\farcs8. 

{\bf S-N Spectra} were created with a width of 8\farcs0, centred 35\farcs0 west of \tC. The top slit ({\bf 61})  is 81\farcs4  north of \tC\ and the bottom slit ({\bf 12}) is 75\farcs6 south of \tC. 

{\bf Red Fan Spectra} were made from the same north-south Atlas spectra as the S-N spectra, except now the slits run from {\bf -14} at 122\farcs2 south of \tC\ to {\bf 11} at 74\farcs1 south of \tC.  

The results of the NE-SW Spectra are shown in Fig.~\ref{fig:OriSDouble}. To these we add the results for the E-W Spectra
 in Fig.~\ref{fig:EWDouble} and for the S-N Spectra in Fig.~\ref{fig:ShortDouble}.  In this section we will first analyze the results from \nii , as it arises from close to an ionization front, and then \oiii\  which
 can arise from further from an ionization front or even in fully ionized regions not immediately associated with an ionization front. We will see that
 there are \oiii\ components not seen in \nii. We first discuss the results for sections of the Profiles lying outside the Orion-S Crossing and Samples  near the Profiles (Section~\ref{sec:OutsideTheCrossing}).
 We then discuss the Profile sections within the Orion-S Crossing (Section~\ref{sec:Crossing}). 
  
 \subsection{Profiles and nearby features outside of the Orion-S Crossing}
 \label{sec:OutsideTheCrossing}
 
\subsubsection{Analysis of the strongest component of \nii\ in the profiles}
\label{sec:NIIprofiles}

The \nii\ emission must arise from a thin layer that is close to either the MIF or the OriS-IF. This makes it
useful in developing a picture of the 3-D structure of the Huygens Region.

\begin{figure}
	\includegraphics
	[width=3.5in]
	{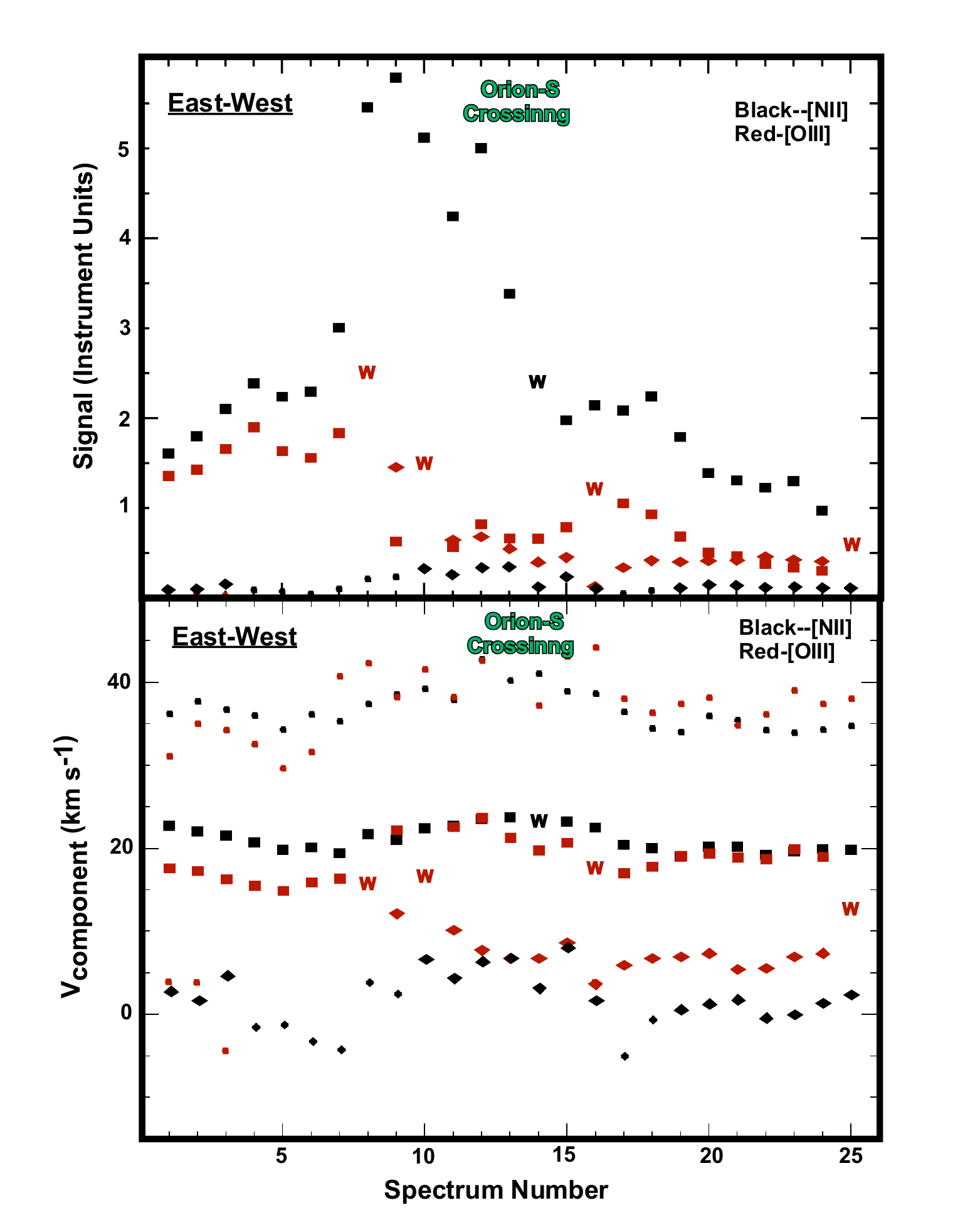}
\caption{The upper panel gives the Signal results from the series of Spectra in the E-W sequence shown in Fig.~\ref{fig:RatioImage}. The lower panel gives the velocities. The spacing of the Spectra is constant at 4\farcs0. Again the Orion-S crossing is shown.  
}
\label{fig:EWDouble}
\end{figure}

\begin{figure}
	\includegraphics
	[width=3.5in]
	{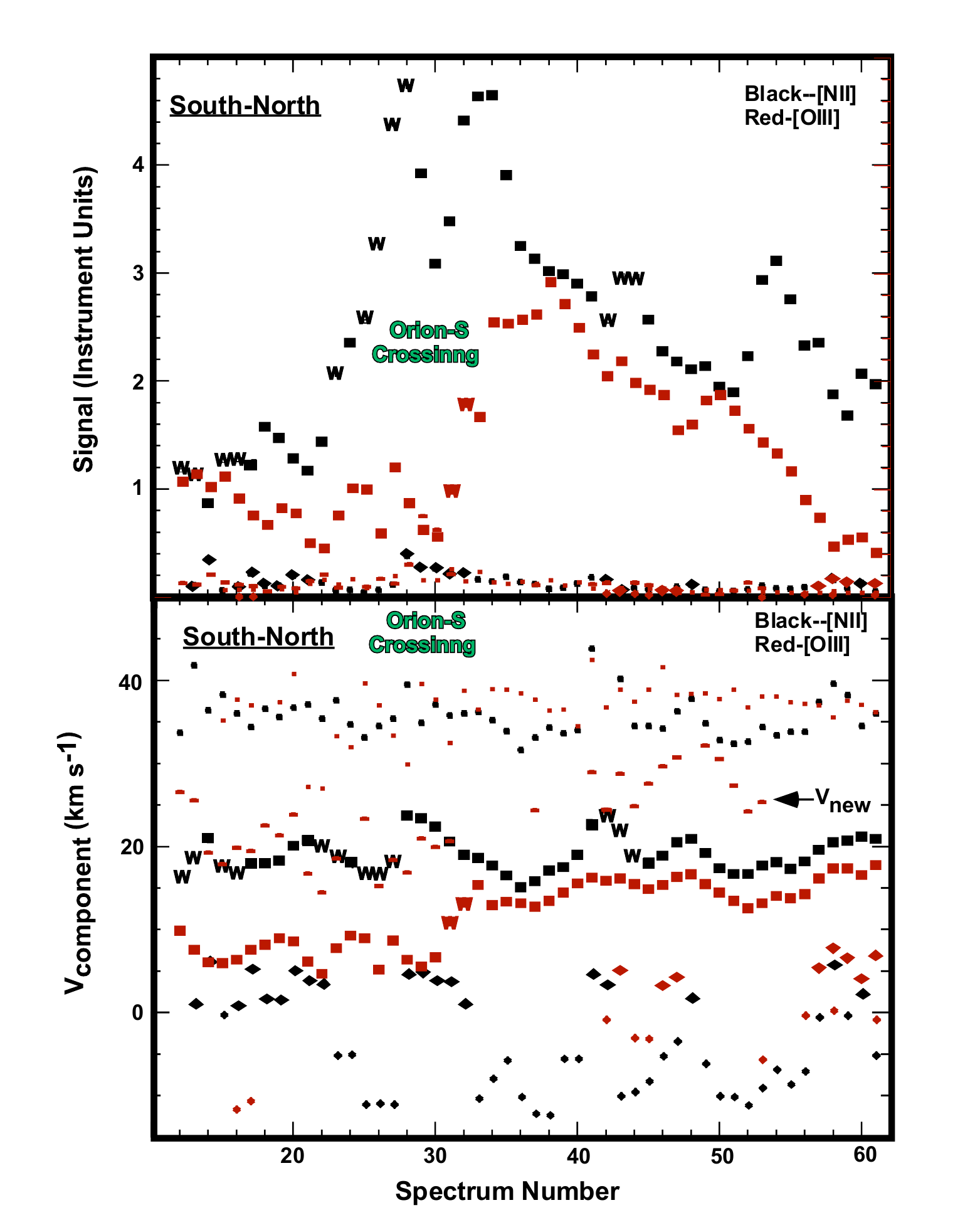}
\caption{The upper panel gives the Signal results from the series of Spectra in the S-N sequence shown in Fig.~\ref{fig:RatioImage} are shown. The spacing of the Spectra is constant at 3\farcs2. Again the Orion-S crossing is shown. The lower panel gives the Velocities. The horizontal bar symbol depicts the data for the \Vnew\ velocity system seen in this Profile and the more crowded upper panel shows S(\Vnew ) as horizontal bars.
}
\label{fig:ShortDouble}
\end{figure}

We see in Fig.~\ref{fig:OriSDouble}, Fig.~\ref{fig:EWDouble}, and Fig.~\ref{fig:ShortDouble}
that near the end of the profiles \Vmifnii\ values are generally below those for the NE-Region (23.2$\pm$3.6 \kms) with the exception of the East end of the E-W Profile. This argues that the extreme
regions, which are well beyond the Orion-S Crossing are similar to the NE-Region, but are either flatter or have a lower \Vpdr.  The NE-SW Profile shows a S(\Vmifnii) peak and a \Vmifnii\ peak of 24.1 \kms\ at {\bf 10}, which is located just beyond the M-D FOV and on the northeast edge of the Dark Arc. These peaks are almost certainly associated with the OriS-IF, while the slow decrease in \Vmifnii\ to the SW may be due to an increase of tilt of the OriS-IF, a 
transition to a mix with MIF emission, or a change in the underlying \Vpdr\ (or a mix of these factors).  

The E-W Profile shows its strongest signal peak at {\bf 9} where it is established on the rise of the MIF and possible transition to the OriS-IF. A second signal peak is at {\bf 12}, where it crosses the series
of high velocity features associated with HH~269. The smooth decrease of \Vmifnii\ to the West is
subject to the same multiple possibilities as in the SW portion of the NE-SW Profile.

The S-N Profile is the most structured of this series. Its first sharp signal and velocity peak at 
 {\bf 28} occurs within the Orion-S Crossing at the passage of the HH~269 associated features. The second signal peak at {\bf 33}-{\bf 34} within a region where \Vmifnii\ is decreasing but where this decrease is almost stopped. The signal and velocity peaks at {\bf 42}~--~{\bf 44} occur immediately before where the profile crosses the inner boundary of the High Ionization Arc. The signal peak at {\bf 54} is in the low ionization portion of the ionization stratified High Ionization Arc.  
 
 Well to the NE of {\bf 10} in the NE-SW Profile the observed radiation must originate in the MIF. This group is the closest to the dominant ionizing star \tC\ and thus probably subjected to the greatest radiation field and stellar wind. It provides a useful region of reference to the others. The distance between the MIB and \tC\ in the foreground has been determined from the extinction corrected \Hb\ surface brightness \citep{bal91} to be 0.23$\pm$0.03 pc, corresponding to 122\arcsec $\pm$16\arcsec .

These profiles all support the idea that the Orion-S Crossing region is in a region closer to the observer
than the MIF in the NE-Region Samples but do not indicate quantitatively their relation to the MIF. Unfortunately, the data do not indicate if the OriS-IF is dominant
to the SW.

\subsubsection{Analysis of the strongest component of  \oiii\ in the profiles}
 \label{sec:OIIIprofiles}

The characteristics of the Cloud region are quite different in \oiii . This is not surprising because in photo-evaporation flow the \oiii\ emission arises further from the MIB than the \nii\ emission. In addition,  gas not directly linked to photo-evaporation flow is expected to be fully ionized because of proximity to \tC\ and there will be no accompanying \nii\ features. 
 This means that very different radial velocities can be encountered.
In this section we restrict our discussion to the regions outside the Orion-S Crossing area because the \oiii\ emission is quite different. A separate section (\ref{sec:Crossing}) will discuss the Orion-S Crossing in both \nii\ and \oiii .

Outside of the Orion-S Crossing area all of the strongest \oiii\ components clearly belong to the \Vmifoiii\ system as defined in Table~\ref{tab:criteria}. However, there are revealing features. In  Fig.~\ref{fig:OriSDouble} we see that SW of the Orion-S Crossing the strongest component has \Vobsoiii ~<~15.0 \kms, but in the case of the lowest values the line is unusually broad (FWHM~>~18 \kms)
and it is likely to be a blend with a \Vlowoiii\ component. In Fig.~\ref{fig:EWDouble} we see that the \Vmifoiii\  values in the East fit into the pattern established in the NE-Region, but in the West the values are slightly higher and almost the same as \Vmifnii . 
We see in Fig.~\ref{fig:EWDouble} that S(\Vmifoiii) has dropped to the level of the almost constant S(\Vlowoiii) values. 

Taken together, these results for outside the Orion-S Crossing indicate that the \Vmifoiii\ components are
very similar to those in the NE-Region, with the exception of the S-N South Region,  and probably reflect similar conditions. In addition, in the west area of the E-W Profile where the strength of the \Vlowoiii\ and \Vmifoiii\ components are about the same, \Vmifnii\ and \Vmifoiii\ are about the same, indicating that the photo-evaporation model does not apply in this region. The behavior of the \Vmifoiii\ wide components in the SW portion of the NE-SW Profile also indicates that S(\Vmifoiii ) and S(\Vlowoiii) are about the same there also. 
 A region to the SW of the Orion-S Crossing composed of low resolution Samples is discussed in Section~\ref{sec:SWofCrossing}.

\subsubsection{Analysis of the \Vscat\ components in the line profiles}
\label{sec:VscatProfiles}

We see in Fig.~\ref{fig:OriSDouble}, Fig.~\ref{fig:EWDouble}, and Fig.~\ref{fig:ShortDouble}
that in the profiles the difference in velocities between the \Vscat\ and \Vmif\ components are very similar in both ions
to those in the NE-Region. 

This is not true in both ions for the S-N Profile South Region ({\bf 12}~--~{\bf 22}). 
The separation there in \nii\ (<\Vscatnii\ -\Vmifnii >) is 17.8$\pm$2.4 \kms, comparable to 17$\pm$4 \kms\ in the NE-Region. 
The \oiii\ velocities are more complex. For \oiii\ <\Vscatoiii\ -\Vmifoiii >~=~31$\pm$5 \kms and <\Vnewoiii\ -\Vmifoiii >~=~17$\pm$4 \kms , while \Vscatoiii\ - \Vmifoiii\ = 20$\pm$4 \kms\ for the NE-Region. While the \Vnewoiii\ separation more closely agrees the NE-Region results, the signal
of \Vnewoiii\ is not strong enough to produce the observed S(\Vscatoiii). In addition, the most likely interpretation of the \Vnew\ component
places it far away from the MIF (Section~\ref{sec:DiscQuiVnew}. Because of the peculiar shift in \Vmifoiii , it is probably not wise
to draw conclusions based on the \Vscatoiii\-\Vmifoiii\ differences in the S-N Profile South Region.

\subsubsection{Further Evidence for a new velocity system in \oiii }
\label{sec:VnewProfiles}
The new \oiii\ velocity system, designated as \Vnew , seen in the NE-Region is also present in the S-N Profile.
The average values for the S-N Profile are 27.3$\pm$2.9 \kms\  and the signal ratios are low in both the S-N South and North regions, about 0.01.
 It is seen in ten Spectra within the Orion-S Crossing, being unusually strong in {\bf 29}. 

\subsubsection{The \Vlow\ components in the line profiles}
\label{sec:VlowProfiles}

Many of the spectral components in the Profiles clearly fall into the \Vlow\ and \Vblue\ categories
as defined in Table~\ref{tab:criteria}. Since these criteria depend upon both the velocity range and the
strength of the signal relative to the MIF component, the boundaries are sometimes uncertain, especially when the MIF component is wide. However, some clear patterns emerge. 

We find the \Vlownii\ component in all of the profiles and also consider it here only where it appears 
outside the Orion-S Crossing area. In the NE-SW Profile it is at {\bf -4}~--~{\bf -2}(<\Vlownii >~=~1.8$\pm$1.8 \kms ) then appears again
after the Orion-S Crossing in {\bf 15}~--~{\bf 22} (<\Vlownii >~=~1.6$\pm$2.8 \kms ). 
In the E-W Profile it is at {\bf 1}~--~{\bf 3} (<\Vlownii >~=~3.1$\pm$1.5 \kms ), then reappears after the
Orion-S Crossing at  {\bf 19}~--~{\bf 25} (<\Vlownii >~=~1.0$\pm$1.0 \kms ). 
In the S-N Profile it is at {\bf 13}~--~{\bf 22} (<\Vlownii >~=~3.3$\pm$2.0 \kms ), except for {\bf 15}, then reappears north of the Orion-S crossing at {\bf 41}~--~{\bf 42}, {\bf 48}, {\bf 58}, and {\bf 60} (<\Vlownii >~=~3.6$\pm$1.7 \kms ). All of these averages fall within the dispersion in the NE-Region 
\Vlownii\ values,
where the average is 6.2$\pm$2.7 \kms.  

We also find the \Vlowoiii\ component in all of the profiles and we consider it in this section only where it appears outside the Orion-S Crossing area. In the NE-SW profile it is at {\bf -3} and {\bf 20}~--~{\bf 22}, with <\Vlowoiii >~=~2.3$\pm$0.8 \kms. In the E-W profile it is in {\bf 16}~--~{\bf 24}, with <\Vlowoiii >~=~6.3$\pm$1.2 \kms.  In the S-N profile it appears in {\bf 43},  {\bf 46}~--~{\bf 47}, and {\bf 57}~--~{\bf 61}, with <\Vlowoiii >~=~5.4$\pm$1.6\kms. Except from the NE-SW Profile (only four occurrences), all of these averages fall within the dispersion in the NE-Region 
\Vlowoiii\ values,
where the average is 6.9$\pm$3.4 \kms.  

In summary, we can say that \Vlowoiii\ is clearly present to the SW, west, and north of the Orion-S Crossing area and is present in the NE and north. It is comparable in signal to the MIF component to the west. \Vlownii\ is clearly present to the NE, SW, west, and north of the Orion-S Crossing. When both \Vlow\ components fall in the same region, <\Vlowoiii > is always larger, by 0.7 to 5.3 \kms.  As an overall group the Profile \Vlow\ values are similar to \Vlow\ values NW of the Bright Bar.

\subsubsection{The \Vblue\ components in the line Profiles}
\label{sec:VblueProfiles}

In the Profiles, the \Vblue\ components are less frequent than \Vlow\ components, as was the case
for the NE-Region. They are almost never seen in the Orion-S Crossing region and outside of there the \Vbluenii\ components (37) are more frequent than the \Vblueoiii\ components (19).  Their average velocities  give weak evidence that the  \nii\ component (<\Vbluenii >~=~-5.2$\pm$3.7 \kms >) is more blueshifted than the \oiii\ component (<\Vblueoiii >~=~-0.9$\pm$3.0 \kms). These properties are discussed in Section~\ref{sec:DiscQuiVblue}.

\subsubsection{Characteristics of a Region SW of the Orion-S Crossing}
\label{sec:SWofCrossing}

In order to verify the trends in the Spectra from the Profiles in the region to the SW, South, and West of the Orion-S Crossing
we employed a group of our Samples (40\arcsec--~80\arcsec , Lines 10~--~13) . The boundary is shown in Fig.~\ref{fig:RatioImage} and was selected to avoid contamination by
the high velocity flows associated with HH~269 and to cover the area of the Orion-S Cloud. The results of the deconvolutions are given in Table~\ref{tab:SWOriS}, where the criteria for assignment
of velocity components established in the NE-Region (given in Table~\ref{tab:criteria}) were used. 

A comparison of the results in Table~\ref{tab:SWOriS} with the Profiles bracketing and crossing this grouping of Samples 
indicates that their properties extend across the Orion-S Cloud. The \Vlowoiii\ component agrees with the 
values at {\bf 20}~--~{\bf 22} in the NE-SW Profile, indicating that our interpretation of the highest slits as blends of \Vmifoiii\ and 
\Vlowoiii\ is correct. A notable difference between the results of the Profiles and these Samples is that \Vbluenii\ is much less 
frequent across the Orion-S Cloud and the \Vblueoiii\ is not detected.  These properties indicate that the Orion-S Cloud either
intercepts the layer producing the blue components or is preventing its ionization. 

The FWHM of the MIF components were different, with all of the \Vmifnii\ components having FWHM~$\leq$~18.0 \kms ,
while 11 of the 12 \Vmifoiii\ components had FWHM~$\geq$~16.0 \kms . Therefore, the \Vmifoiii\ lines would be considered
wide in the Profiles.  

A previously unknown \nii\ velocity component was found in all four of the Line 11 Samples. It has an average velocity of  31.4$\pm$2.3 \kms\ and is strong (<S(component)/S(MIF)>~=0.16$\pm$0.11). As shown in Fig.~\ref{fig:RatioImage} it lies
along the inner edge of the High Ionization Arc and probably indicates the motion of this feature.  At its western end it includes HH 1148. 

In the 80\arcsec ,Line 13~Sample a strong (S(\oiii)/S(MIF)>~=0.43) component was seen at 3.7 \kms. 
The strength and location indicates that this component is associated with HH 1131 .    

 \begin{table}
\caption{Properties of Region SW of the Orion-S Crossing*}
\label{tab:SWOriS}
\begin{tabular}{lcc}
\hline
Component &<Velocity>        &S(component)/S(MIF)\\
\hline
\Vscatnii      &  35.7$\pm$3.3 &  0.10$\pm$0.05\\
\Vmifnii        &  19.2$\pm$1.5 &        1.0 \\     
\Vlownii        &  0.5$\pm$2.1   &  0.08$\pm$0.03\\
\Vbluenii **     &  -12.5$\pm$0.8 &  0.02$\pm$0.01 \\
\Vscatoiii      &  35.7$\pm$ 1.8 & 0.10$\pm$0.03 \\
\Vmifoiii        &  11.9$\pm$1.6  &    1.0   \\
\Vlowoiii       &  -1.0$\pm$0.6   &  0.08$\pm$0.02\\
\hline
\end{tabular}\\
*All velocities are in \kms. Samples: 50\arcsec~--~80\arcsec ,in Lines 11~--~13. ** Four Samples only.
\end{table}

\subsubsection{The Dark Arc Feature.}
\label{sec:DarkArcRegion}

The Orion-S Crossing also contains the curious feature designated (OY-Z) as the Dark Arc. 
It is highly visible,
even in the reduced resolution image in Fig.~\ref{fig:HRone} and more clearly in the full HST resolution
F658N over F502N images shown as Fig.~\ref{fig:RatioImage} and in various color depictions such as Fig. 20 in Bally, O'Dell, and McCaughrean (2000).
Although in single filter images it appears
as a dark feature, it has little if any extinction as determined by OY-Z from the ratio of surface brightness in the \Ha\ optical line and the 20.46 cm radio continuum.
 The surface brightness in \oiii\ drops the least of the strong optical emission lines,
 probably due to the \oiii\ emission arising from an ionized layer larger that the physical size of the Dark Arc feature itself. This is what makes it least visible in this ion. 

This feature is best
evaluated using the NE-SW Profile because it crosses almost perpendicular to the east side of the feature. 
It's NE boundary falls between {\bf 9} and  {\bf 10}, while the maximum height along the
NE-SW Profile occurs at {\bf 12}.  This means that it falls on the rapidly descending OriS-IF in the 
direction of \tC.The SE portion of {\bf 13} includes a portion of the peculiar feature we designate here as the Dark Box (5:35:14.0 -5:23:59, 4\farcs8 $\times$1\farcs6, orientation PA = 107.5\degr), which has many of
the characteristics of the Dark Arc.

 OY-Z present profiles of the Dark Arc
along the short-axis of the small Box in Fig.~\ref{fig:RatioImage}. The HST images are more than 20$\times$ better resolution than our Spectra, so they become the primary source of information about the nebula near the Dark Arc. The NE portion of the HST
sample falls near {\bf 9}. In the NE of the HST sample there is a peak in \nii\ adjacent to a narrow peak of \oi\ emission away from \tC\ and immediately on the NW edge of the Dark Arc.  

Within 
the dark edge \oi\ and \nii\ emission drop sharply (\nii\ about 50\%), whereas \Ha\ only drops about 30\%\ and \oiii\ only about 20\%.  All have
recovered from the effects of the Dark Arc by about 3\arcsec\ from the dark edge (about {\bf 11}).  

O15 point out that the series of shocks designated as HH~1127 arise from either of the  high extinction sources  MAX 46 or COUP 602, which lie north of the Dark Arc. These shocks appear at the NW edge of the  Dark Arc. This strongly argues that the space south of the
Dark Arc's edge is open, rather than being dark because it is beyond an ionization front. The MIB continues to the SW, but there is a 
brief space where it is only illuminated by scattered LyC photons, rather than directly by \tC. The region illuminated by scattered LyC photons will be only about one-tenth as bright as an adjacent directly illuminated region and this is what produces the Dark Arc. The feature is a curved ridge on the descending OriS-IF. The cause of this ridge is uncertain. Near its centre of symmetry there is an unidentified source producing a rapidly moving series of \oiii\ shocks directed towards the Dark Arc (O15) and the momentum of this flow of material may produce the local  curved ridge.  

\subsection{Velocity and Signal Changes in the Orion-S Crossing}
\label{sec:Crossing}

The Orion-S Crossing is arguably the most complex area within the Huygens Region.
It physically lies above the MIF level of the sub-\tC\ region and contains the high density Orion-S Cloud. There is
a centre of imbedded young stars near the peak of the rise. In this section we will draw heavily
upon Fig.~\ref{fig:OriSDouble}, Fig.~\ref{fig:EWDouble}, and Fig.~\ref{fig:ShortDouble}. 
In this area we see that both \Vmif\  and \Vlow\ systems are present for both \nii\ and \oiii .

\nii\ is the less complex ion. We see an increase in the strongest (MIF) component by 4~--~5 \kms\ above the values in the adjacent regions outside of the Crossing. S(\Vmifnii) has a double peak in each profile. One of the peaks occurs within the Crossing
in the NE-SW Profile and the E-W Profile, while the peaks in the S-N Profile straddle the two sides of the Crossing. The coincidence 
of velocity and signal maxima at NE-SW {\bf 10}, E-W {\bf 12}~--~{\bf 13}, and S-N {\bf 28} indicate that these are highly tilted regions, as
well they should be as one sees in Fig.~\ref{fig:RatioImage} that these Spectra coincide and lie on the SW side of the M-D region. NE-SW {\bf 6}, E-W {\bf 9} and the almost coincident  S-N {\bf 33}~--~{\bf 34}) have lower velocities, 
indicating that these are not as highly tilted.

The \Vlownii\ values appear in all of the Crossing Profiles. They appear at velocities about 4.4$\pm$1.5 \kms\ and 4.3$\pm$1.8 \kms higher than the regions outside the Crossing
in the NE-SW and E-W Profiles respectively, while the S-N Crossing \Vlownii\ values are about the same as the adjacent regions. However, if one 
only considers the four \Vlownii values nearest the S-N {\bf 28} signal maximum and the two regions closest to the Crossing {\bf 13}~--~{\bf
 22} and {\bf 41}~--~{\bf 48}, there is a small velocity difference (about 1 \kms) of the Crossing and background values.
 
 Better defined than
 the small \Vlownii\ velocity differences (well defined at about 4 \kms\ for \Vmifnii\ and less well defined as somewhat less for \Vlownii),
 there is a big difference in the S(\Vlownii) as one enters the Orion-S Crossing. In each Profile, the crossing values are about 3 or more times
 stronger than the adjacent regions.  
 
 The \oiii\ properties are more complex. The \Vmifoiii\ components within the Crossing are higher for both the NE-SW (4.5$\pm$0.3 \kms ) and E-W 
 (4.0$\pm$1.6 \kms) Profiles. In both of these Profiles, the S(\Vmifoiii) values drop well below an interpolation across the nearby
 regions and the \Vlowoiii\ components have become comparable in signal. In the NE-SW Crossing \Vlowoiii\ is 2.6$\pm$0.8 \kms\ greater than
 the adjacent values and in the E-W Profile it is 0.9$\pm$0.9 \kms\ greater.
 
 The S-N Profile in \oiii\ is more difficult to determine (Fig.~\ref{fig:ShortDouble}) because of the ambiguity in assignment to different velocity
 systems. South of the Crossing the strongest signal has been assigned to \Vmifoiii , which has only a slightly higher velocity than the much weaker
 \Vlowoiii\ north of the Crossing. The \Vnew\ components south of the Crossing have similar velocities to the \Vmifoiii\ components north of the Crossing, but they are much weaker than the S-N Profile South Region \Vmif\ signals. The velocity difference above the local values is impossible to determine  because this is where the average
 \Vmifoiii\ drops from 13.5$\pm$2.8 \kms\ north of the Crossing to 7.2$\pm$1.5 \kms\ south of the Crossing. The complexity of the core of the Crossing 
 {\bf 26}~--~{\bf 31} is peculiar in its \oiii\ signal.  Average S(\Vmifoiii) has decreased to 0.81$\pm$0.26, while the average S(\Vnew) has increased to 0.36$\pm$0.26. Another peculiarity in this region occurs at {\bf 26}, where the \Vnew\ and \Vmifoiii\ velocites are 5.1 \kms\ and 15.2 \kms and the
 signal ratio S(\Vnew)/S(\Vmifoiii)~=~0.27. Both components are unusually narrow, with FWHM(\Vmifoiii)~=~8.5 \kms\ and FWHM(\Vnew )~=~9.2  \kms, much narrower that the usual value of about 14 \kms.
 
 Proceeding from north to south we can say that in the core of the Crossing that the \Vlowoiii\ component has disappeared,
 not to be seen again to the south. We can also say that the S(\Vmifoiii) drops markedly in the Core region never to be as strong as it was on the \tC (north) side of the Cross. The fact that S(\Vnew) temporarily increases in the core of the Crossing (it is seen in ten Spectra in the Crossing, being unusually strong in {\bf 29}. indicates some interaction with the layer producing \Vnew.
 
\section{Additional Profiles and Samples}
\label{sec:Additional}
In order to more completely investigate regions outside the Orion-S Crossing, we also studied a region to the south of the S-N Profile, a high
extinction feature to the SW, and a profile along the eastern boundary of the Atlas.

\subsection{A Profile Across the Red Fan Feature}
\label{sec:RedFan}

\begin{figure}
	\includegraphics
	[width=3.5in]
	{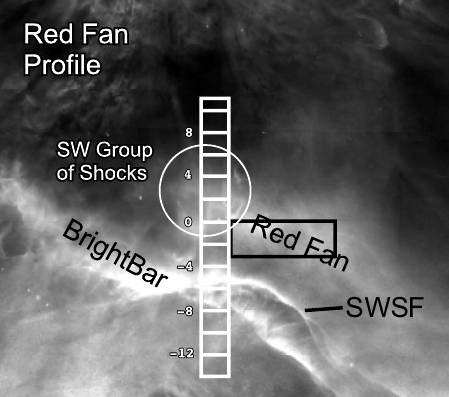}
\caption{
Like Fig.~\ref{fig:RatioImage} this is a ratio image except now of a 129\arcsec $\times$114\arcsec\ FOV centred 37\farcs7 west and 111\farcs5 south of \tC . The white sequence shows the slit locations of the Red Fan Profile as shown in Fig.~\ref{fig:HRthree} The black box indicates the three low spatial resolution Samples used to create high S/N spectra of the Red Fan. The SW Group of shocks are discussed in O15. }
\label{fig:SWSF}
\end{figure}

\begin{figure}
	\includegraphics
	[width=3.5in]
	{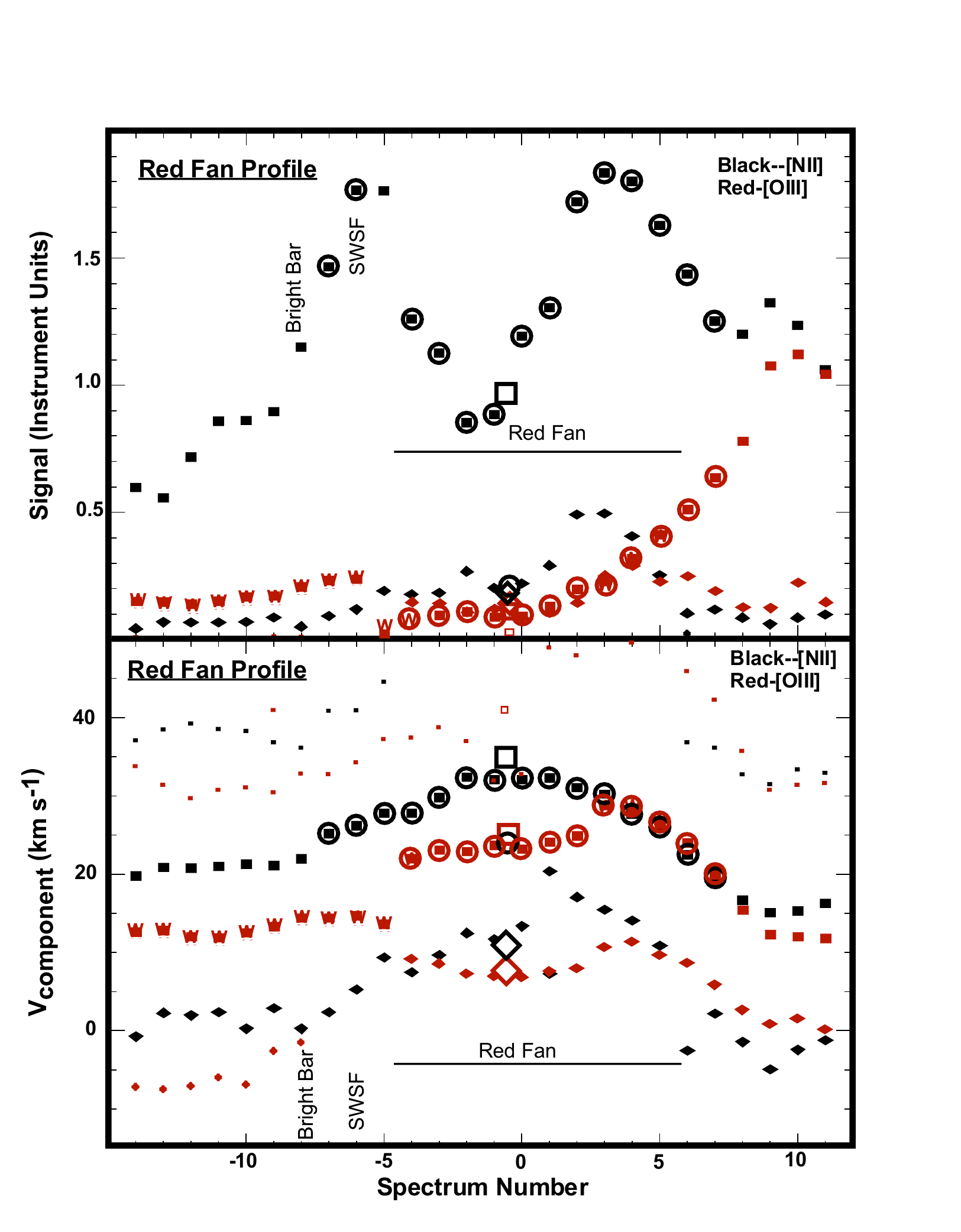}
\caption{Like Fig.~\ref{fig:ShortDouble} except showing the results for the Red Fan Profile and with the strongest components that are dominated by the Red Fan Cloud are shown as circled squares (Section~\ref{sec:RedFan}). The results for the high S/N sample using the results in the large box sampling the Red Fan, as shown in Fig.~\ref{fig:SWSF}, are presented. The large Sample results are depicted with open symbols, with the circle indicating the results for the \nii\ component that falls at the position where \Vlownii\ changes abruptly.}
\label{fig:RedFan2}
\end{figure}

Near the Bright Bar region south of the Orion-S Cloud lies a unique structure. The Southwest Spoked Feature (SWSF) 
is the leading edge of an incomplete parabolic  arc of bright gas pointed to the NNW. Within this arc are multiple radial spokes. The feature has
been most extensively described in O15 and is well illustrated in Fig.~24 of that publication and in Fig.~\ref{fig:SWSF}.  A thin section
of the Bright Bar crosses the arc. To the NW of the SWSF lies a broad low ionization feature first identified because of its high velocity by
 \citet{gar07} and was designated there as the Red Fan. We consider in this section a series of slit Spectra (the Red Fan Profile) that lie south
 of the S-N Profile, having the same widths and spacings. Their location is shown in the SW portion of Fig.~\ref{fig:HRthree} and Fig.~\ref{fig:SWSF}.

The greatest variations of velocities in the present study occur in the Red Fan Profile. In addition, we see features not found in other profiles. Fig.~\ref{fig:RedFan2} gives the
results for this profile, using the same symbols as before, i.e. an assignment to the previously recognized velocity systems 
has been made. This is straightforward for \nii , where we see \Vmifnii , \Vlownii, and \Vscatnii\ and all these components are continuations from the southern portions of the S-N Profile.  However, no \Vbluenii\ components were seen, nor were any \Vscatnii\ components across the Red Fan feature. In addition to the Red Fan Profile, Fig.~\ref{fig:RedFan2} also shows the results from a region of grouped low resolution Samples at 40\arcsec\ --~60\arcsec ,Line 5 and 40\arcsec\ --~50\arcsec ,Line 4 (these cover the brightest part of the Red Fan). 

With a maximum \nii\ velocity of 32 \kms, it appears that the Red Fan feature is the result of the MIF moving about 17 \kms\ greater velocity in \nii\ than the MIF immediately to its north and 9 \kms\ greater than typical for the NE-Region and is rapidly moving into the host OMC.

The strongest \nii\ component reaches an unprecedented high of 35.0$\pm$1.8 \kms\ at the centre of the Red Fan feature.
This indicates a motion of 8$\pm$2 \kms\ towards the OMC at 27.3$\pm$0.3 \kms. This variation in velocity is almost certainly due to motion in
a discrete cloud, rather than a changing tilt because the velocity maximum occurs where there is a S(\Vmifnii) minimum. 

In both the North and South Regions of the Red Fan Profile, the  \Vlownii\ component has a separation from \Vmifnii\  similar  (about 18~--~19  \kms) to that in the other profiles , even as \Vmifnii\ changes. The separation narrows as one moves towards the brightest part of the Red Fan. This smooth change breaks down as one crosses the brightest part of the Red Fan. There one sees that the \Vlownii\ component abruptly jumps to a lower velocity, which then continues to mimic  \Vmifnii\  with the original separation. The high S/N large Sample reveals two \Vlownii\ components of very similar strengths, one at the projection of the north to south pattern and the other at the projection of the south to north pattern.

The broad rise in \Vmifnii\ over {\bf -7}~--~{\bf +7} probably does not actually arise from the MIF, rather, from a discrete cloud moving towards the OMC that we now designate as the Red Fan Cloud (RFC).  The large change in the \Vlownii\ values that occur at {\bf 1}~--~{\bf 2} probably indicates that the RFC interferes with the layer that usually produces the \Vlownii\ system. The peaks in S({\Vmifnii}) at {\bf -6}~--~{\bf -5} and {\bf 3}~--~{\bf 4} would indicate where we are observing along
the edge of the RFC. The argument for a discrete cloud is strengthened by the lack of a \Vscatnii\ component across the RFC, indicating that there
is not a nearby back-scattering layer.

Patterns in the \oiii\ velocities are similar to those in \nii . However, in the middle of the RFC, the \oiii\ components are about  9 \kms\ lower velocity.
The \Vlowoiii\ components are about 16 \kms\ lower than the \Vmifoiii\ components, only slightly smaller than in other regions within the Huygens Region.

The sense of the RFC velocities (\Vmifoiii\ less than \Vmifnii) indicates that we are seeing flow towards the observer from a cloud moving even
faster than the central \Vmifnii\ value. The RFC would be ionized by \tC. As indicated by the low S(\Vmifoiii )/S(\Vmifnii) values in the centre direction, the RFC is very low ionization.  As encountered previously, in the RFC centre the S(\Vmifoiii) and S(\Vlowoiii) are about the same,
whereas the S(\Vmifnii) component remains much larger than S(\Vlownii). Fig.~27 of \citet{wei15} shows that the \sii\ derived electron density rises to 
about 2000 \cmq\ along the NW edge of the SWSF and to about 1600 \cmq\ at the centre of the RFC.

\subsection{The High Extinction Southwest Cloud}
\label{sec:SWcloud}

In order to further clarify the effects of obscuring foreground features, we isolated a Sample on the high extinction Southwest Cloud (O-YZ) and compared it with a nearby region that is not heavily obscured. The expectation is that the \oiii\ components will be more affected than \nii\ components by the selective
reddening of extinction.  Our Southwest Cloud group was made 
from Samples: 60\arcsec ,Line 8; 70\arcsec ,Line 8; 80\arcsec ,Line 8. The comparison group was: 30\arcsec ,Line 8; 40\arcsec ,Line 8; 30\arcsec,Line 9;40\arcsec ,Line 9. The results
are shown in Table~\ref{tab:SW}, where the classification criteria of Table~\ref{tab:criteria} were used. No \Vblue\ components were seen.

\begin{table}
\caption{Average Characteristics in the Southwest Cloud and a nearby comparison region*}
\label{tab:SW}
\begin{tabular}{lcc}
\hline
Component & Comparison  & Southwest Cloud \\
\Vlownii              & -1.7$\pm$2.9    &     -1.2$\pm$2.9\\
S(\Vlownii)/S(MIF,\nii) & 0.07$\pm$0.04 &     0.09$\pm$0.03 \\
\Vmifnii                     & 16.8$\pm$1.5   &   17.6$\pm$1.7  \\
\Vscatnii                   &   33.7$\pm$2.4  &   33.8$\pm$1.4   \\
S(\Vscatnii)/S(MIF,\nii)  & 0.14$\pm$0.04 &     0.19$\pm$0.03\\
\Vlowoiii              & -1.0$\pm$0.5    &     0.3$\pm$5.6\\
S(\Vlowoiii)/S(MIF,\oiii) & 0.07$\pm$0.03 &     0.06$\pm$0.03 \\
\Vmifoiii                     & 11.0$\pm$2.8   &   12.0$\pm$0.5  \\
\Vscatoiii                   &   27.7$\pm$3.5  &   29.6$\pm$3.7   \\
S(\Vscatoiii)/S(MIF,\oiii)  & 0.09$\pm$0.02 &     0.13$\pm$0.03\\
S(\Vmifnii)/S(\Vmifoiii )   & 1.07$\pm$0.16     & 1.72$\pm$0.11\\
S(\Vlownii)/S(\Vlowoiii) & 0.93$\pm$0.01~(2)       & 3.01$\pm$1.50 \\
\hline
\end{tabular}\\
*Values in parentheses indicate the number of samples if less than 4.
\end{table}

We see that the components fall into the usual systems of \Vscat , \Vmif, and \Vlow.
The small differences in velocity don't reveal anything about the effects of the Southwest Cloud.
However, the big difference of S(\Vmifnii)/S(\Vmifoiii ) (1.07$\pm$0.16 in the comparison region and 1.72$\pm$0.11 in the Southwest Cloud)
indicate that the emission from both \nii\ and \oiii\ predominantly arise from beyond the obscuring Southwest Cloud.
The difference in S(\Vlownii)/S(\Vlowoiii) indicates that the Southwest Cloud also lies in front of the layer producing both \Vlow\ 
components. The fact that the difference in this ratio is even greater than for the MIF components suggests that proportionately
more of the \Vlowoiii\ component is being obscured by the Southwest Cloud.
\subsection{A Low Resolution Profile along 90\arcsec\ east}
\label{sec:Minus90}

\begin{figure}
	\includegraphics
	[width=3.5in]
	{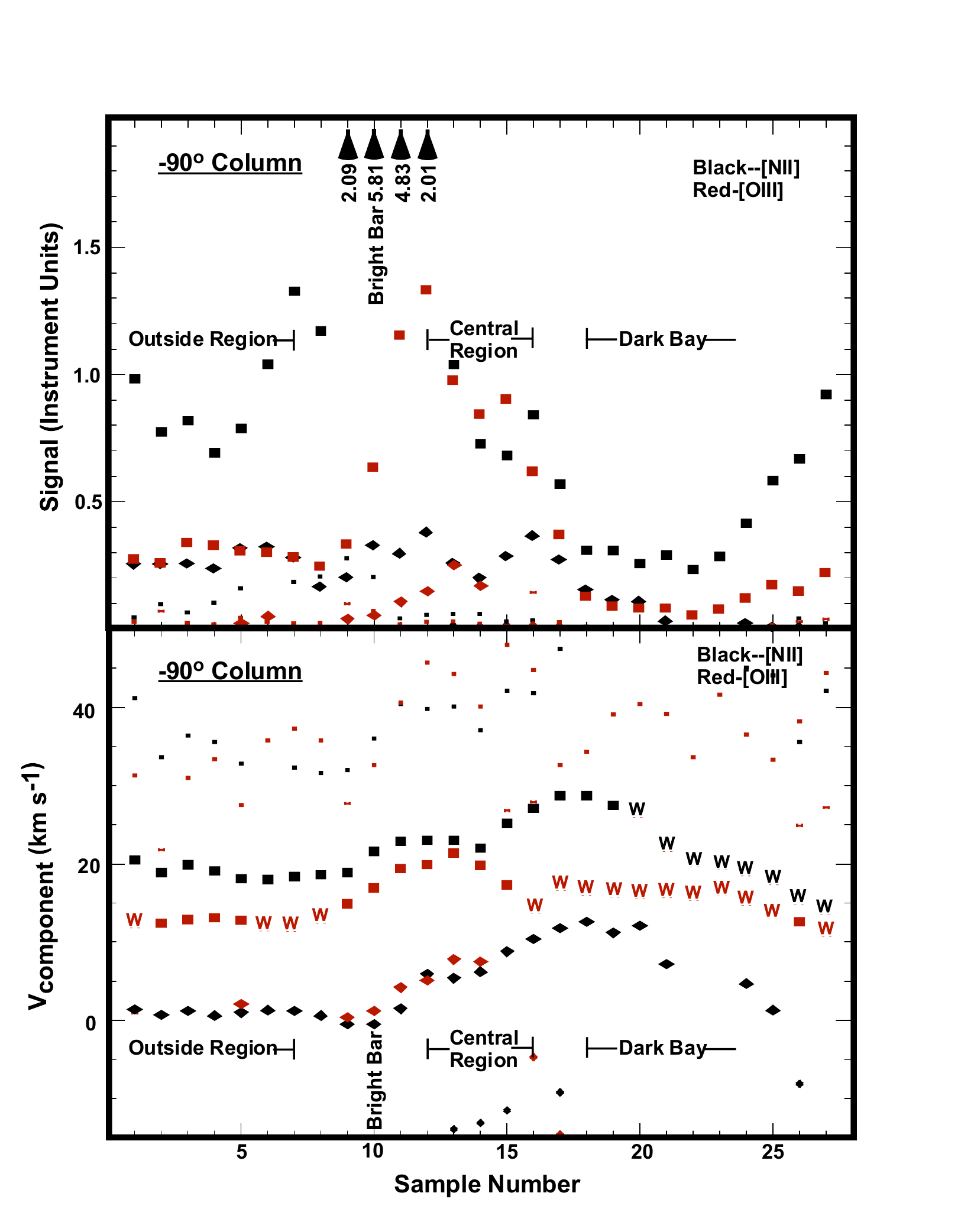}
\caption{Like Fig.~\ref{fig:ShortDouble} except now for the series of Samples at -90\arcsec\ as shown in Fig.~\ref{fig:HRtwo}. The details are described 
in Section~\ref{sec:Minus90}.  Like the earlier figures, W indicates the velocity of a MIF component with FWHM(\nii) larger than 18.0 \kms\ or FWHM(\oiii) larger than 16.0 \kms , these being Samples that may be a blend with another component. For clarity in this figure, these Samples are not denoted with a W in the upper panel. The highest S(\nii) values lie off-scale and their values are indicated. The locations of the Bright Bar, the Central Region, and the Dark Bay are indicated.
}
\label{fig:Minus90}
\end{figure}

In order to further illuminate the large scale variations in properties of the velocity systems we created a profile from the series of
low spatial resolution Samples all with centres 90\arcsec\ east of \tC\ (Fig.~\ref{fig:HRtwo}). The results are shown in Fig.~\ref{fig:Minus90}. Lines  {\bf 5}~--~{\bf 8} include portions of HH~203 and HH~204 but the velocities of those shocks are so large that
they did not interfere with identification of the components arising from the nebula. 

The Samples to the south of the Bright Bar are labelled as the ''Outside Region'' in Fig.~\ref{fig:Minus90} and represent regions
primarily ionized by \tA. There we confirm the results found in Section~\ref{sec:BarGeneral} that the \Vmifnii , \Vmifoiii , \Vscatnii , \Vscatoiii ,
\Vlownii , and \Vlowoiii~components are present (although \Vlowoiii\ is seen only in Line 5) and no \Vblue components are found.
\nii~emission dominates, with S(\Vmifnii)/S(\Vmifoiii)~=~3.1$\pm$0.9, in contrast with 1.61$\pm$0.63 for the NE-Region. The velocity separation
of the two components (6.4$\pm$0.8) is slightly larger than in the NE-Region (3.1$\pm$2.6 \kms ).  These differences probably reflect the different radiation
field seen in the two regions.  

The Central Region (Lines {\bf 12}~--~{\bf 16}) shows the same velocity systems as in the Outside Region, but now one commonly sees the
\Vbluenii\ component and also the \Vblueoiii\ component at the Line (16) closest to the Big Arc East \citep{gar07}. The velocities (<\Vmifnii >~=~24.2$\pm$2.1 \kms\ and <\Vmifoiii >~=~18.6$\pm$2.6 \kms ) and signals (S(\Vmifnii)/S(\Vmifoiii)~=~1.1$\pm$0.3) are all
similar to those in the NE-Region. 

In the highly obscured (O-YZ) Dark Bay region we again (like the Southwest Cloud Feature described in Section~\ref{sec:SWcloud}) see the effects of greater
extinction at the shorter wavelength of the \oiii\ emission (signal ratio changes from 2.4 to 4.5). The separation velocity of \nii\ and \oiii\ appears to be 
different from that in the Central Region and changes from 11.9 \kms\ to 2.9 \kms\ over Lines {\bf 18}~--~{\bf 27}, but the significance of this is 
not clear since one or both components may be the results of blends of the \Vmif\ and \Vlow\ components. Certainly the \Vlownii\ component
is present in the direction of the Dark Bay, at a much reduced signal, which places its source between the distance of the Dark Bay material (a part of the Veil) and \tC.

\section{Discussion}
\label{sec:discussion}

We see a variety of features in each of the regions investigated. Many of the regions share properties and here we summarize and discuss the regions separately. Appendix~\ref{app:Accuracy}  has established 
the validity of the identified velocity systems and gives the relation of the intrinsic relative signals to what is derived from using IRAF task splot. Assignment to a velocity system was based on the velocity relative to the strongest (outside the Orion-S Crossing) MIF component and the signal strength relative to the MIF component. There are of course regions where 
assignments are not clear, near the borders of the velocity and relative strength criteria. 

\subsection{A summary of earlier studies}
\label{sec:DiscEarlierStudies}

This study is only the most recent of high velocity resolution studies that cover much or most of the Huygens Region. Their techniques have varied, but we
have been able to draw on them in the discussion of our results. Unfortunately the much more numerous lower velocity resolution and single slit setting studies are of little value for this investigation.

\subsubsection{Photographic Spectroscopic Multi-slit mapping by Wilson et al. (1959)}
\label{sec:DiscOCW59}
The multi-slit observations of \citet{ocw59} was the first complete survey of radial velocities, covering the inner 
portions of the Huygens Region in \oii , \oiii , and \Hb\ at a velocity resolution of about 10 \kms. Since the detector was a photographic emulsion,
it was not possible to get precise relative component strengths, but it could give good values of \Vmif\ and the blue component, where seen, 
that we now describe as the \Vlow\ component. These were seen in both the 372.6 nm \oii\ and the \oiii\ line. The average velocity
of the \oii\ \Vmif\ component was 23.3$\pm$4.3 \kms\ and for the \oii\ \Vlow\ component 11.1$\pm$9.3 \kms , both calculated for the
202 most certain samples. The \oii\ emission should arise from the same \nzone\ zone as the \nii\ emission. The average \Vmif\ for \oii\ agrees well with \Vmifnii~=~23.2$\pm$3.6 \kms\ for the NE-Region, and the great uncertainty in the average \Vlow\ for \oii\ means that it
falls into the range of the NE-Region (\Vlownii ~=~6.2$\pm$3.4 \kms).

\subsubsection{Photographic Fabry-Perot mapping by Deharveng (1973)}
\label{sec:DiscDEH73}
The Fabry-Perot study of \citet{deh73} covered a larger area than the \citet{ocw59} study. Her area was located in the northern part of the EON and used a similar velocity resolution in \nii. An absolute velocity calibration was not made and the paper is limited to reporting line splitting in three areas bordering the Huygens Region. This line splitting was not seen in the inner, much more densely exposed regions, and it is not clear that their absence there is
physical or due to emulsion saturation. The three regions of detected line splitting were:
\newline
{\bf A}, a trapezoid starting at the south side of the Dark Bay and proceeding south to the FOV limit at about 159\arcsec\ south. The top is 107\arcsec\ wide and lies immediately south of the  Dark Bay, the bottom is 196\arcsec\ wide. The average split is 14.2$\pm$3.3 \kms. This area runs along the inside of the Rim Feature (SW of \tC ) designated in \citet{ode10} that extends from the SSW to the NNE of \tC\ and bounds this portion of the EON.\newline
 {\bf B}, an elliptical form (200\arcsec $\times$ 73\arcsec) with the long axis oriented towards PA~=~258\degr , centred 45\arcsec W,187\arcsec N. 
 This runs along the inside of the north portion of the Rim Feature. The average split is 12.1$\pm$2.8 \kms .
 \newline 
{\bf C}, an L shaped form varying from 73\arcsec --104\arcsec\ wide. It starts at 160\arcsec W,53\arcsec N , goes to 375\arcsec W,0\arcsec N then ends at 643\arcsec W,324\arcsec N. The average split is 14.6$\pm$7.2 \kms .
This starts within the central cavity of the Huygens Region west of the Orion-S Crossing and extends in the direction of a large concave form defining the limit of the EON to the NW.

Adopting the NE-Region \Vmifnii\ of 23.2 \kms, the blue components are at {\bf A} 9$\pm$4 \kms , {\bf B} 11$\pm$3 \kms, {\bf C} 9$\pm$7 \kms. These are to be 
compared with the NE-Region <\Vlownii >~=~6.2$\pm$2.7 \kms.

\subsubsection{CCD Spectroscopic mapping at high velocity resolution near \tC\ and \tA\ by Casta\~neda (1988)}
\label{sec:DiscHOC88}
\citet{hoc88} mapped the regions close to \tC\ and \tA\ in \oiii\ in \oiii\ at twice the velocity resolution of the Atlas. He used a 3\arcmin\ long slit centred on
the stars or using combinations of Trapezium stars, producing good mapping to distances of about 90\arcsec\ from the alignment stars. One of these
slits lies within the Orion-S Crossing Region. The lines were submitted to deconvolution into multiple components, much as done in the present study. Multiple
velocity systems were identified, including what we now designate as \Vmifoiii~(his system A), \Vlowoiii~(his system B), and \Vscatoiii\ (his system C).
There are also components that are probably attributable to \Vblueoiii\ and \Vnewoiii .
Looking through his many tables of data it appears that \Vlowoiii\ is on the average several \kms\ higher than for the NE-Region Samples. The advantages of using CCD detectors with their linearity of response and large dynamic range were employed in all subsequent optical and UV studies.

\subsubsection{21-cm H~I absorption line study of the Veil by van der Werf \&\ Goss (1989)}
\label{sec:DiscHIorig}
In their original study \citet{vdw89} mapped the entire Huygens Region in the 21-cm line seen in absorption against the radio continuum emission
from the background MIF.  The amount of LOS \Hi\ generally increased towards the NE and the line saturated in the vicinity of the Dark Bay.
Three velocity systems were identified, very clear systems at 24 \kms (designated as A) and 21 \kms\ (designated as B) and a less well defined
system C at 16 \kms. The values for A and B were later refined to 23.2 \kms\ and 19.5 \kms\  by \citet{tom16}. 

\subsubsection{[O~I] study of O'Dell \&\ Wen (1992)}
\label{sec:DiscOWOI}
\citet{ode92} mapped the vicinity of \tC\ in \oi\ in the 630.0 nm line, much in the manner of \citet{hoc88}. In studies of \oi\ the location of the 
foreground night sky line can be a serious problem. Most of their observations were made about 1988 November 14, when the heliocentric velocity correction
was +12.8 \kms, which means that the night sky component was at that velocity, while the observed nebular component was around 26.7$\pm$1.2 \kms. As shown in Fig. 2a and Fig. 2b of their paper, the proximity of the night sky line (at a separation of about 14 \kms) only allows measurements of the MIF component, although the presence of a \Vscat\ component is obvious. Later observations (1991 January 1) at a heliocentric velocity 
correction of -8.8 \kms\ clearly shows (their Fig. 2c) \Vlow , \Vmif , and \Vscat\ components. 

\subsubsection{Optical absorption line study of O'Dell et al. (1993)}
\label{sec:DiscOptAbsp}
In this high resolution (FWHM~=~3.3 \kms) \citet{ode93} measured the \Hei\ line at 388.9 nm and the \Caii\ line at 393.4 nm in the four brightest
Trapzezium stars and in \tA. The \Hei\ line arises from a state populated by recombination of ionized helium that has no permitted transitions to lower energy levels. Thus it is a measure of the column density of He$^{+}$. In contrast, the ground state \Caii\ line arises from a trace state of ionization, 
most of the calcium being \Caiii\ since ionization of \Caii\ requires only 11.9 eV and the next stage of ionization requires 47.3 eV photons and those are infrequent in \tC\ radiation.  They found an average \Hei\ velocity of 2.1$\pm$0.6 \kms\ for the Trapezium stars, which is most likely associated
with the \Vlow\ system. Multiple components were seen in \Caii . The strongest were seen in all of the Trapezium stars and were at
7.5$\pm$1.9 \kms , 22.0$\pm$1.2 \kms , and 30.9$\pm$1.0 \kms . A separate study by \citet{hobbs} showed strong \Nai\ components at
-16.8 \kms , 6.0 \kms , 17.4 \kms , 22.3 \kms , and 31.1 \kms\ in \tC . Similar to \Caii , \Nai\ is a minor ionization stage since it requires only 5.1 eV photons to ionize it to \Naii\ and the next stage of ionization requires 50.9 eV photons, which are again infrequent in \tC\  radiation.
\subsubsection{Multiple high ionization study of Doi et al. (2004)}
\label{sec:DiscDoiVel}
\citet{doi04} mapped a 3\arcmin $\times$5\arcmin\ region with north-south slits in \Ha , \oiii , and \nii . These data were then absolute velocity corrected  later and incorporated in the Atlas 
prepared by \citet{gar07}.  Seeing limited samples in declination were submitted to deconvolution using splot and from this a map of velocities was prepared, being presented in their Fig.~2 in velocity bands. It should be noted that their velocities are presented in terms of displacement from a systemic velocity of 18$\pm$2 \kms.  The 
purpose of the study was to detect new features of high relative velocities and many were found, including the Big Arc East, the Big Arc South, HH~512, HH~725, and HH~726. With the exception of the Big Arc components, all of these new features were attributed to shocks arising from
outflows. 

\subsubsection{Multiple low ionization lines study of Garc\'ia-D\'iaz et al. (2007)}
\label{sec:DiscGar07}

In their study of the low-ionization lines from the Atlas \citet{gar07} worked with the \oi\ 630.0 nm line, the red \sii\ 671.7 nm and 673.1 nm doublet, and the \siii\ 631.2 nm line, smoothing the data to give a uniform FWHM of 12 \kms. Rather than deconvolution of the spectra, their analysis was based on
identifying velocity bands of about 10 \kms\ that sampled different portions of the typical line profiles. This technique was well suited for the identification of multiple
velocity features throughout the Huygens Region. These are summarized nicely in their Fig.~7. Of particular interest to the present study is the region 
NW of the Bright Bar near \tA\ that they designate as the Southeast Diffuse Blue Layer. There the \sii\ lines are split, with the longer component (which we associate with \Vmif )
at about 20~--~25 \kms and the weaker and shorter component at about -2 \kms (which we associate with \Vlow ). They were able to determine the
density of this feature as about 400 \cmq.

A closer examination of a selected set of \oi\ spectra in the Atlas indicates that \oi\ is present in nearby regions, in contrast with its being absent in the Southeast Diffuse Blue Layer , as reported in \citet{gar07}. The problem is not one of presence in one area and not the other, rather, some of
the Atlas data are more suitable for looking for an \oi\ \Vlow\ component.
As presented in Section~\ref{sec:DiscOWOI}, a \Vlow\ nebular component is seen in the \citet{ode92} spectrum with a heliocentric velocity correction
of -8.8 \kms .
The bulk of the Atlas observations in \oi\ were made on 2003 December 14 when the heliocentric velocity correction of was -0.4. In processing
the data they identified the strong night sky component and subtracted it before inclusion in the Atlas. The separation 
would have been 25 \kms\ and with a FWHM of 12 \kms\ they should have been able to see the blue component, however the S/N in the region of the \Vlow\ 
component is quite low, therefore \Vlow\ in \oi\ was reported as not present. 

In the search for an \oi\ \Vlow\ component three observations (-28\farcs8, -11\farcs4, -4\farcs3 East) are more useful.  These
were made on 2003 January 19 when the heliocentric velocity correction was -15.3 \kms. In order to isolate a region free of the effects of either the Big Arc or the Bright Bar we chose to group the slit spectra over the range of 45\arcsec\ north and south from \tC . In order to avoid the effects of systematic correction for the night sky lines, we worked with calibrated spectra before the night sky line correction was made. These were kindly provided by W. H. Henney, a coauthor of the Atlas. 
Deconvolution of the averaged spectra revealed \Vmif , \Vscat , and \Vlow\ components, in addition to the night sky component. The \Vscat\ component is very weak 
and the velocity is quite uncertain. Although the relative signal of the \Vlow\ component is weak, it is still at a believable level. This
averaged spectrum is very similar to Fig. 2c in \citet{ode92}, except that their figure includes the night sky component. \oi\ is discussed further
in Section~\ref{sec:DiscQuiVlow}.

\subsubsection{21-cm H~I high velocity absorption line features of van der Werf et al. (2013)}
\label{sec:DiscHIhigh}
In their higher spatial resolution (7\farcs2$\times$5\farcs7) and higher S/N study \citet{vdw13} refined their earlier study \citep{vdw90} of high negative 
velocity 21-cm \Hi\ absorption line features. They established that several regions are associated with emission line features, in particular the region NW of \tA\ (3.5~--~7.3 \kms ) that overlaps with Southeast Diffuse Blue Layer of \citet{gar07} and a region to the NW of \tC\  (-3.0~--~+7.3 \kms ) that they associate with HH~202. Our N-S Profile Spectra that cross this feature {\bf 41}~--~{\bf 48} have an average of  both \nii\ and \oiii\ of 3.8$\pm$1.2  \kms .

\subsubsection{UV absorption line and optical emission line study of Abel et al. (2016)}
\label{sec:DiscAbel16}
This most recent \citep{abel16} of a series of four papers is primarily concerned with conditions in the partially ionized Veil as determined by
a host of absorption lines, extending the list to include \htwo . The UV absorption lines include \Piii\ (4.8$\pm$3.0 \kms) and
\Siii\ (4.5$\pm$0.9 \kms), both of which are expected to arise in a \nzone\ zone. Their velocities suggest an association with 
the \Vlow\ layer as they are unlikely to be part of the higher velocity Veil. They also present emission line velocities for the MIF
over a 40 square arc second area and for \Vlow\ components of \nii\ (1.8$\pm$1.9 \kms) and \oiii\ (0.9$\pm$2.8 \kms). All of the available information
was used in creating models of the physical conditions and locations of Veil components A and B.

\subsection{Conditions in the Quiescent Regions }
\label{sec:DiscQuiescent}
Each of the velocity components have their own characteristics in the quiescent regions, which we define here
to mean those outside of the Orion-S Crossing, the Red Fan, the Bright Bar and the -90\degr\ Profile. The quiescent regions are expected to show features related to most of the Huygens Region.

\subsubsection{The \Vmif\ Components}
\label{sec:DiscQuiVmif}
The profiles NE-SW (Northeast and Southwest Regions), E-W(East Region), and S-N (North Region) all resemble the NE-Region in that \Vmifnii~-~\Vmifoiii\ is about four to five \kms.  This is what is expected from models of a photo-evaporation front, although the observed magnitudes of the velocities are more than those expected from the models \citep{hen05}. The \Vmifnii\ components in the NE-Region are on the average a few \kms\ higher than in the other regions. The most anomalous region is in the South portion of the S-N Profile. There
we see that \Vmifnii\ is similar to the North portion of that profile and the non-NE-Region Samples. However, \Vmifoiii\ drops to velocities characteristic of the \Vlowoiii\ components. It is as if few
photons of greater than 24.6 eV are reaching the layer associated with the MIF.

As established in Section~\ref{sec:backscat}, the comparison of \Vscat\ and \Vmif\ components indicate that the usual variations in \Vmif\ are caused
by changes of velocity of the underlying PDR just beyond the MIB. Only small scale variations (<25\arcsec ) can be explained by changes in the tilt of the ionized layers near the MIB.  

\subsubsection{The \Vscat\ Components}
\label{sec:DiscQuiVscat}
The \Vscat\ components satisfy the predictions of this being due to back-scattering from dust in the PDR. Within the NE-Region the 
\Vscat\ components are about the same average velocities, but when one examines Fig.~\ref{fig:VmifVallNII} and Fig.~\ref{fig:VmifVallOIII} one sees the velocity differences 17 \kms\ and 19 \kms\ respectively support the idea of back-scattering from 
ionized layers with different photo-evaporation velocities.

 We also see this difference in the non-central portions of the other Profiles with the exception of the S-N South region),
where the velocity difference is 17.8$\pm$2.4 \kms\ for \nii\ and 27.8$\pm$4.3 \kms\ for \oiii . This is the region where \Vmifoiii\ is about the same 
as the usual values for \Vlowoiii. The velocity difference between \Vmifoiii\ and \Vscatoiii\ is larger than normal
(20$\pm$5 \kms\ in the NE-Region) 
This is expected if the PDR region has the usual velocity (as indicated by the \Vmifnii\ velocities), but the 
\Vmifoiii\ velocity is lower. However if the usual backscattering model applies, the velocity difference should be about  47 \kms . This may indicate that the backscattering from the \Vmifoiii\ component arises from emission
farther from the LOS passing through the observed point. This is what would be expected from the \Vmifoiii\ 
component originated further from the PDR. This is also consistent with the S(\Vscatoiii)/S(\Vmifoiii) ratio
being 1.5~--~2 times larger than normal for quiescent regions.

\subsubsection{The \Vlow\ Components}
\label{sec:DiscQuiVlow}
 \begin{table*}
\caption{Averaged Properties of the \Vlow\ components outside the Orion-S Crossing Area*}
\label{tab:DiscVlow}
\begin{tabular}{lcccccc}
\hline
Region        &<\Vmifnii >     &<\Vlownii >   &<\Vmifnii -\Vlownii& <\Vmifoiii >    &<\Vlowoiii > &\Vmifoiii -\Vlowoiii >\\
\hline
NE-Region  &23.7$\pm$2.0&5.7$\pm$2.7&21.5$\pm$3.3(52)&20.9$\pm$3.6&8.5$\pm$3.8&12.4$\pm$2.3(12)\\
NE-SW(NE)&20.0$\pm$0.6&1.8$\pm$1.8&18.2$\pm$1.3(3) & 16.8(1)   &3.4~~(1)                   &13.4~~(1)     \\
NE-SW(SW)&19.4$\pm$1.0&0.5$\pm$1.9&18.9$\pm$1.3(8)& 13.7$\pm$0.4  &  2.0$\pm$0.4&11.8$\pm$0.4(3)\\  
E-W(East)   &22.1$\pm$0.6&3.1$\pm$1.5&19.0$\pm$1.9(3)& ---                     & ---                 & ---\\
E-W(West) &19.7$\pm$0.5&1.0$\pm$1.0&18.7$\pm$0.8(7)& 18.6$\pm$0.9   & 6.6$\pm$0.8 &12.0$\pm$0.9(8)\\
S-N (North)&21.9$\pm$1.5&3.6$\pm$1.6&18.3$\pm$2.2(5)& 16.6$\pm$0.7 &5.4$\pm$1.6    &11.1$\pm$0.9(8)\\
Minus90\degr (Outside)&19.1$\pm$0.9&1.0$\pm$0.3&18.1$\pm$0.9(8)&12.8~~(1)&2.1~~(1)&10.7~~(1)\\
Minus90\degr (Central)&24.2$\pm$2.1&7.3$\pm$2.2&16.6$\pm$0.8(5)&20.1$\pm$0.9&6.2$\pm$1.8&14.0$\pm$1.3(4)\\
Weighted Average**&  --- & 3.9$\pm$2.5 & --- & ---& 5.9$\pm$2.3 &---\\
\hline
\end{tabular}\\
*All velocities are in \kms. Parentheses indicate the number of samples. Only samples with both \Vmif\ and \Vlow\ values were used.
** Weighted by number of samples and without Minus90\degr (Outside).
\end{table*}
Although velocity components have previously been reported in the \Vlow\ range, generally described as \Vblue , this velocity system is best defined in this study. 
The \Vlownii\ average of 3.9$\pm$2.5 \kms\ and \Vlowoiii\ average of 5.9$\pm$2.3 \kms\ are indistinguishable, although Fig.~\ref{fig:VmifVallNII}
and Fig.~\ref{fig:VmifVallOIII} suggest that there is a small difference of about the amount in Table~\ref{tab:DiscVlow}, \oiii\ being higher by about 2 \kms . 

Addressing the important question of the presence of \oi\ in the \Vlow\ system, we have averaged and analyzed the Atlas spectra for the same three slits as described in Section~\ref{sec:DiscGar07} over a declination range of 45\arcsec\  north and south from \tC. \Vlow\ components were seen in all the lines.  The \Vmif\ (\kms), \Vlow\ (kms), and relative signal strengths S(\Vlow)/S(\Vmif)  values are: \oiii , 18.1, 3.2, 0.03; \nii , 22.7, 5.7, 0.08; \sii , 22.8, 4.1, 0.02; \oi , 27.2, 8.0, 0.04. The velocity determinations are all very close (average 5.3$\pm$2.1 \kms) and within the measurement spread of these weak components. With cautions about their uncertainties, the ratios of the \Vlow\ components may be useful in models that compare the conditions in the \Vlow\ layer with that of the MIF. 

 Given the agreement of the \nii\ and \oiii\ velocities and those in \oi\ (8 \kms) and \sii\ (-2 \kms) with those of
the ionized absorption lines seen in the Trapezium stars (\Hei\ 2.1$\pm$0.6 \kms , \Nai\ 6.0 \kms , \Caii\ 7.5 \kms ,\Piii\ 4.8$\pm$3.0 \kms , \Siii\ 4.5$\pm$0.9 \kms) 
it is likely that all of these components arise from a common layer that is stratified in ionization. \oii\ emission should arise in the same ionization zone (\nzone) as \nii\ and the \citet{ocw59} average of the blue components of 11.1$\pm$9.3 \kms , strengthens the interpretation of \nii\ and \oii\ as part of the \Vlow\ system. As noted in Section~\ref{sec:DiscDEH73}, the three outlying blue \nii\ areas have velocities of 9$\pm$4 \kms , 11$\pm$3 \kms, and 9$\pm$7 \kms , which argues that the lowest ionization portion of the \Vlow\ layer extends beyond the Huygens Region.

The best fit slope in the \nii\ data (Fig.~\ref{fig:VmifVallNII}) is 0.8 while that for \oiii\  (Fig.~\ref{fig:VmifVallOIII}) is 0.6. This argues
that there is a good but not excellent correlation of \Vlownii\ and \Vmifnii\ and a poorer correlation of \Vlowoiii\ and \Vmifoiii. These are consistent with the general trends that result from the \Vmifoiii\ components arising at a greater distance and thicker layer than \nii .
Certainly the \Vlownii\ layer knows about the velocity of the \Vmif\ in each LOS. This presents a dynamics problem
since one is in the foreground and the other the background of \tC .

Because of the fact that an \oi\  component is
now recognized, this layer is probably ionization bounded, unlike the assumption in \citet{abel16} that it is mass-bounded. This 
presents a problem in explaining the radiation field that reaches the two primary layers of the Veil. This problem will be addressed
in a future publication. Certainly the layer lies between \tC\ and the Veil and the high relative velocity of the \Vlow\ layer towards the Veil 
indicates that a collision of the two is imminent, about 30,000~--~60,000 yrs \citep{abel16}.

It is likely that the highly blue shifted \Hi\ absorption lines 
seen by \citet{vdw13} at the end of the features driving HH~202 (their Sample F at -3.0~--~+7.3 \kms ) are 
the products of shocks causing high compression densities that have led to de-ionization in those areas of the \Vlow\ layer.

The fact that S(\Vlowoiii) and S(\Vmifoiii) become about the same in the E-W Profile West Region while S(\Vlownii) remains weak compared with S(\Vmifnii) argues that some form of selective shadowing of the \Vmifoiii\ region is occurring, as discussed in Section~\ref{sec:DiscCrossing}.

\subsubsection{The \Vblue\ Components}
\label{sec:DiscQuiVblue}

The \Vblue\ components are difficult to isolate and are the rarest of the components in the quiescent regions. When present, their inherent weakness makes
determination of their velocities uncertain. Since the shocks from young star outflows that we see in the Huygens Region are selectively blue shifted, it is possible that the components do not form a system. In this case the \Vblue\ components would
be attributed to low velocity blue shifted shocks.

\begin{table*}
\caption{Averaged Properties of the \Vnew\ components outside the Orion-S Crossing Area*}
\label{tab:DiscVnew}
\begin{tabular}{lcccccc}
\hline
Region  &S(\Vnew)/S(\Vmifoiii) &<\Vlowoiii >    & <\Vmifoiii>       &<\Vnew>       &\Vnew -\Vmif\ & \Vnew -\Vlow\\
\hline
NE-Region    (18)        & 0.17$\pm$0.08        &6.9$\pm$3.4   &19.5$\pm$3.6 &27.6$\pm$4.5 &8$\pm$4       & 21$\pm$4    \\
S-N (South)      (12)     & 0.15$\pm$0.10        &   ---                 & 7.2$\pm$1.6 &18.5$\pm$6.2  & 12$\pm$6    &   ---\\             
S-N (North)     (13)     &  0.04$\pm$0.02        & 5.4$\pm$1.5  &14.9$\pm$1.6 &27.6$\pm$2.8 &13$\pm$3    &   22$\pm$3\\
Minus90\degr  (4)        &0.14$\pm$0.06          &5.3$\pm$2.4   &13.5$\pm$4.5 &25.5$\pm$3.3 & 8$\pm$5     & 20$\pm$3\\
\hline
\end{tabular}\\
*All velocities are in \kms. Parentheses indicate the number of samples.
\end{table*}

\subsubsection{The \Vnew\ Components}
\label{sec:DiscQuiVnew}

The \Vnew\ components have no counterparts in \nii . They are present in the high velocity resolution \oiii\ study of \citet{hoc88} although
not recognized there as a common velocity system. Because of lying between \Vmifoiii\ and \Vscatoiii\ they are difficult to identify
except when they are strong relative to a clearly defined \Vscatoiii\ feature.  

The \Vnew\ component appears in three areas and the results are shown in Table~\ref{tab:DiscVnew}. There we see that there are well
defined groupings of the velocity differences. <\Vnew -\Vmifoiii > is about 11 \kms , much less than the 21 \kms\ for <\Vnew -\Vlowoiii >. 
The S(\Vnew )/S(\Vlowoiii) ratio is usually similar, which makes it unlikely that \Vlowoiii\ and \Vnewoiii\ are related. The
fact that the \Vnew\ component is characteristically strong everywhere (except in S-N Profile North) yet is absent in \nii , that the velocity difference from \Vmif\ is nearly constant, and \Vnew\ is too large to be the result of back-scattering all argue that it must
arise from a fully ionized layer. In Section 6.3.1 of \citet{abel16} those authors discuss
outflow away from the observer from the ionized side from Component B of the Veil (the Veil component argued to be closer to \tC). They 
predict that this \oiii\ should occur at 28.1 \kms. This is indistinguishably the same in all the sampled regions in Table~\ref{tab:DiscVnew} except
for the peculiar S-N Profile South Region, which is obviously very different from other regions of the MIF. \citet{abel16} also predicted the outflowing \nii\ emission to be at 27.3 \kms. This lower velocity component would be lost in the much stronger higher velocity \Vmifnii\ component. How this outflowing emission
from the Veil that we see in \oiii\ can be dynamically linked to the \Vmifoiii\ layer is an open question. Nevertheless, the most likely explanation of the \Vnew\ 
component is that it is the outflow from the Veil predicted by \citet{abel16}.

\subsection{Conditions in the Orion-S Crossing}
\label{sec:DiscCrossing}

The most dramatic changes in velocity occur in the Orion-S Crossing and those in the NE-SW and E-W Profiles are easiest to understand.
In these profiles \Vmif\ and \Vlow\ are greater than the adjacent regions by about 4~--~5 \kms\ in both \nii\ and \oiii . This indicates that
the underlying PDR is moving away with respect to the nearby MIF.  

The changes of the signal levels in the NE-SW and E-W Profiles behave very differently than the velocities.
S(\Vmifnii) progresses across the Crossing as if it were a simple ionized layer. In contrast, the S(\Vlownii) \ values increase significantly within
the Crossing. S(\Vmifoiii ) drops dramatically in the core of the Crossing, while the S(\Vlowoiii ) values increase to about the same signal level as the \Vmifoiii\ signals. The combined signals of the \Vmifoiii\ and the \Vlowoiii\  within the Crossing are about equal to an interpolation of nearby values.  It is as if both the \Vmifoiii\ and the \Vlowoiii\ layers are receiving similar amounts of ionizing photons above 24.6 eV that are necessary to
produce an \ozone\ zone. 

In the S-N Profile there is a familiar increase in \Vmifnii\ at the crossing. However, the \Vmifoiii\  transitions from background values south of the Crossing to the higher values to the north in the middle of the Crossing. Although the S(\Vmifnii) varies smoothly across the Crossing, S(\Vlownii)
temporarily increases there.  Proceeding from north to south, we see that the S(\Vmifoiii ) drops abruptly at the Crossing, never to fully recover, 
probably because of increasing distance from \tC .

\Vlownii\ appears a few times in the S-N Profile North Region, then becomes common in the Crossing and the region to its south. Again \Vlowoiii\ appears a few times in the North region, then does not reappear within the Crossing or to its south. The strength of the low velocity \oiii\ in and south 
of the Orion-S Crossing is what cause its identification as the \Vmifoiii\ component, even though its velocities are characteristic of the \Vlowoiii\ 
component in the rest of the Spectra. The most straight-forward interpretation is that
the higher energy photons necessary to create an \ozone\ zone do not reach the S-N Profile South Region, probably because it is shadowed
by the Orion-S Cloud. But, the lower energy photons necessary to create the \nzone\ zone do reach the surface of the PDR. Since the \oiii\ emitting
zone is further from the PDR, the contrasting behavior of \nii\ and \oiii\ can be explained by the Orion-S Crossing being a cloud high enough to
block the high energy photons that usually create the \oiii\ emitting layer and the lower energy photons pass under the cloud to continue
the MIF \nii\ emission.

The \Vnew\ velocity system, whose properties are presented in Table~\ref{tab:DiscVnew} appears eight times in the Orion-S Crossing six of which in the core region ({\bf 26}~--~{\bf 31}) have unusually large S(\Vnew )/S(\Vmif ) values  0.53$\pm$.48  although their velocities (18.6$\pm$2.3 \kms) are lower than in other regions in Table~\ref{tab:DiscVnew} except for the S-N Profile South Region (18.5$\pm$6.2 \kms). The S(\Vnew )/S(\Vmif ) ratios
are very different than the other regions, with this ratio being greater than 1.0 for two ({\bf 29} and {\bf 30}). These two Spectra are where
the S-N Profile crosses the Dark Arc feature.

\subsection{Samples from outside the Orion-S Crossing}
\label{sec:DiscOutsiders}
\subsubsection{The Red Fan Cloud}
\label{sec:DiscRedFanCloud}
In Section~\ref{sec:RedFan} we found evidence that the Red Fan Cloud is a low ionization discrete cloud moving into the OMC at about 8$\pm$2 
\kms. In the middle of the RFC the \Vlownii\ value changes abruptly indicating that the RFC physically interferes with the layer producing \Vlownii .
As first noted in the discussion of the Orion-S Crossing we again see that the S(\Vlowoiii) and S(\Vmifoiii) components are nearly the same 
in the RFC and that S(\Vlownii) is higher within the Red Fan Cloud than in the adjacent background regions.  These many similarities of
the Orion-S Crossing and the RFC (velocities and velocity changes, S(\Vlow) and S(\Vmif) changes) indicate that the two regions share
several physical characteristics. 

However, it is not the case in either region that we are seeing the ionized layer on a foreground cloud. This
would require that the cloud is located far enough from the MIB that only the far side of the cloud is photo-ionized by \tC. In such a model one 
would expect the evaporative flow to be towards the OMC, which could explain the high value of \Vmifnii , but in this model the \Vmifoiii\ velocity should be more positive than the \Vmifnii\ velocity and it is not.

\subsubsection{Changes produced by heavy local extinction}
\label{sec:DiscSWandDarkBay}
In Section~\ref{sec:SWcloud} we saw that there were no statistically relevant differences in the velocities of the different \nii\ and \oiii\ components.
Changes in the signal ratios behaved in the fashion expected, except that greater changes in the signal ratios for the \Vlow\ components suggest
that the \Vlow\ emitting layer is more strongly affected by the SW Cloud than the \Vmif\ emitting layer. This produces a dilemma because this
would be the only evidence that the \Vlow\ emitting layer is further than the \Vmif\ emitting layer. This is in contrast with the Dark Bay area presented in Section~\ref{sec:Minus90}

\subsection{Questions raised by this study or remaining after it.}
\label{sec:DiscQuestions}
In this intensive presentation of observational data and patterns, many questions of interpretation remain open.

{\bf \Vlow .} There seems to be a correlation of the \Vlow\ and \Vmif\  components. Given that this component produces absorption lines in the Trapezium stars and is distant from the MIF, what gives rise to this correlation? 

{\bf The S-N Profile South Region.} What occurs in this region causing the \Vmifoiii\ component to have
velocities normally associated with \Vlowoiii , while \Vmifnii\ does not?

{\bf Ionization in the \Vlow\ Layer.} Given that one sees ionized states in the absorption lines of the Veil, how is this reconciled with the low
ionization \oi\ and \sii\ features in the intervening \Vlow\ system?

{\bf The regions with the most positive \Vmifnii.} These are the Orion-S Crossing and Red Fan Cloud features (the other high velocity region at 
S-N Profile {\bf 42} is apparently linked to the High Ionization Arc). These velocities are not caused by highly tilted regions because there is no
localized peak in S(\Vmifnii). This leaves the uncomfortable answer that the Orion-S Crossing and Red Fan Cloud are rapidly approaching the OMC.

{\bf \Vnew .} Is this component an outflow away from Component B of the foreground Veil? How is the layer producing high ionization emission dynamically linked to the \oiii\ emission coming from near the MIF?

{\bf Is the Orion-S Cloud a finger-tip or a cloud?} Does the selective shadowing of the MIF \oiii\ producing emission outside the Orion-S Crossing indicate that the Cloud is an isolated cloud and not the fingertip  of a column of material pointed towards \tC?
 
\subsection{Conclusions}
\label{sec:Conclusions}

~~~~~$\bullet$ Deconvolution of spectra across the Huygens Region of the Orion Nebula establish that there are several distinct velocity systems,
each having their own origin.

$\bullet $The faint red velocity component (\Vscat) is confirmed as back-scattering of MIF emission by grains in the dense background PDR that 
lies just beyond the MIB.

$\bullet$ \Vmif\  velocities vary at small scales due to tilted regions in the MIF, but at scales of >25\arcsec, the variations are caused by local
variations in motion of OMC gas near the PDR.

$\bullet$ A low velocity (\Vlow) component is seen in many areas of the Huygens Region. The layer producing \Vlow\  must be beyond (away from the observer) with respect to the foreground 
Veil (because the Veil is very low ionization) but must lie closer to the observer than the Trapezium stars (because the absorption line velocities seen in the stars coincide with those in \Vlow).

$\bullet$ In the Orion-S Crossing both \Vmifnii\ and \Vmifoiii\ move about 4 \kms\ away from the observer with respect to the adjacent regions. 
In this region the S(\Vmifoiii) drops sharply while the S(\Vlowoiii ) components increase enough that the sum of these two components
is about what is expected from interpolating across adjacent regions.

$\bullet$  In the region south of the Orion-S Crossing the brightest \oiii\ component has velocities characteristic of the \Vlow\ component and \Vlowoiii\ is not seen. In this same region the \Vmifnii\ component remains at velocities typical of the MIF and a \Vlownii\ component is seen.

$\bullet$ In addition to the Orion-S Crossing, regions of similar signals of the \Vmifoiii\ and \Vlowoiii\ components are found in an extended region to the west
of the Crossing.

$\bullet$ The above points  indicate that the Orion-S Cloud is selectively shadowing the higher layers where the \Vmifoiii components are formed but that \tC \ lower energy photons that create the \nii\ emitting layer pass below the Orion-S Cloud.This shadowing could also explain the drop in 
S(\Vmifoiii) and the increase in S(\Vlowoiii) in the Orion-S Crossing.

$\bullet$ The Red Fan feature is identified as an isolated cloud moving rapidly towards the OMC and has many of the features of the Orion-S Cloud Region.

$\bullet$ A new velocity system (\Vnew) is found in many regions. The velocity of this \oiii\ component and the lack of a detected
\nii\ component is interpreted as flow away from the observer of the Veil's ionized layer that faces \tC.

\section*{acknowledgements}

The author is grateful to Will Henney of the Institute of Radio Astronomy and Astrophysics for digital copies of the spectral Atlas, including \oi\ without the sky line subtracted, and for clarification of the properties of scattered emission lines. Gary Ferland of the University of Kentucky extracted and shared the information from the Cloudy model for the central nebula that allowed estimating the thickness of the \nii\ and \oiii\ emitting layers and was a constructive reviewer of the penultimate draft manuscript.

In this study we have made extensive use of the SIMBAD database, operated at CDS, Strasbourg, France and its mirror site at Harvard University and to NASA's Astrophysics Data System Bibliographic Services. We have used IRAF, which is distributed by the National Optical Astronomy Observatories, which is operated by the Association of Universities for Research in Astronomy, Inc.\ under cooperative agreement with the National Science foundation. 

The observational data were obtained from observations with the NASA/ESA Hubble Space Telescope,
obtained at the Space Telescope Science Institute, which is operated by
the Association of Universities for Research in Astronomy, Inc., under
NASA Contract No. NAS 5-26555; the Kitt Peak National Observatory and the Cerro Tololo Interamerican Observatory operated by the Association of Universities for Research in Astronomy, Inc., under cooperative agreement with the National Science Foundation; and the San Pedro Martir Observatory operated by the  Universidad Nacional Aut\'onoma de M\'exico.

Since this is likely to be the last lead-author paper on the Orion Nebula by the author, it is appropriate to acknowledge the colleagues who
have enabled this series (the first was \citep{ode65}) of studies. These begin with Arthur D. Code, Rudolph L. Minkowski and Donald E. Osterbrock, all then 
at the University of Wisconsin; Ira S. Bowen, Guido M\"unch, and Olin C. Wilson of the (then) Mt. Wilson and Palomar Observatories; Manuel Peimbert then a graduate student at the University of California at Berkeley, now at the Institute for Astronomy-Mexico City; my former graduate students at Rice University, Hector O. Casta\~neda, Zheng Wen, Michael R. Jones, Xihai Hu, and Takao Doi; and my long time collaborators Nicholas P. Abel of the University of Cincinnati, Gary J. Ferland of the University of Kentucky, and William J. Henney of the Institute for Radio Astronomy and Astrophysics-Morelia, Mexico.


\appendix
\section{Accuracy of the Deconvolutions}
\label{app:Accuracy}

\begin{figure}
	\includegraphics
	[width=3.5in]
	{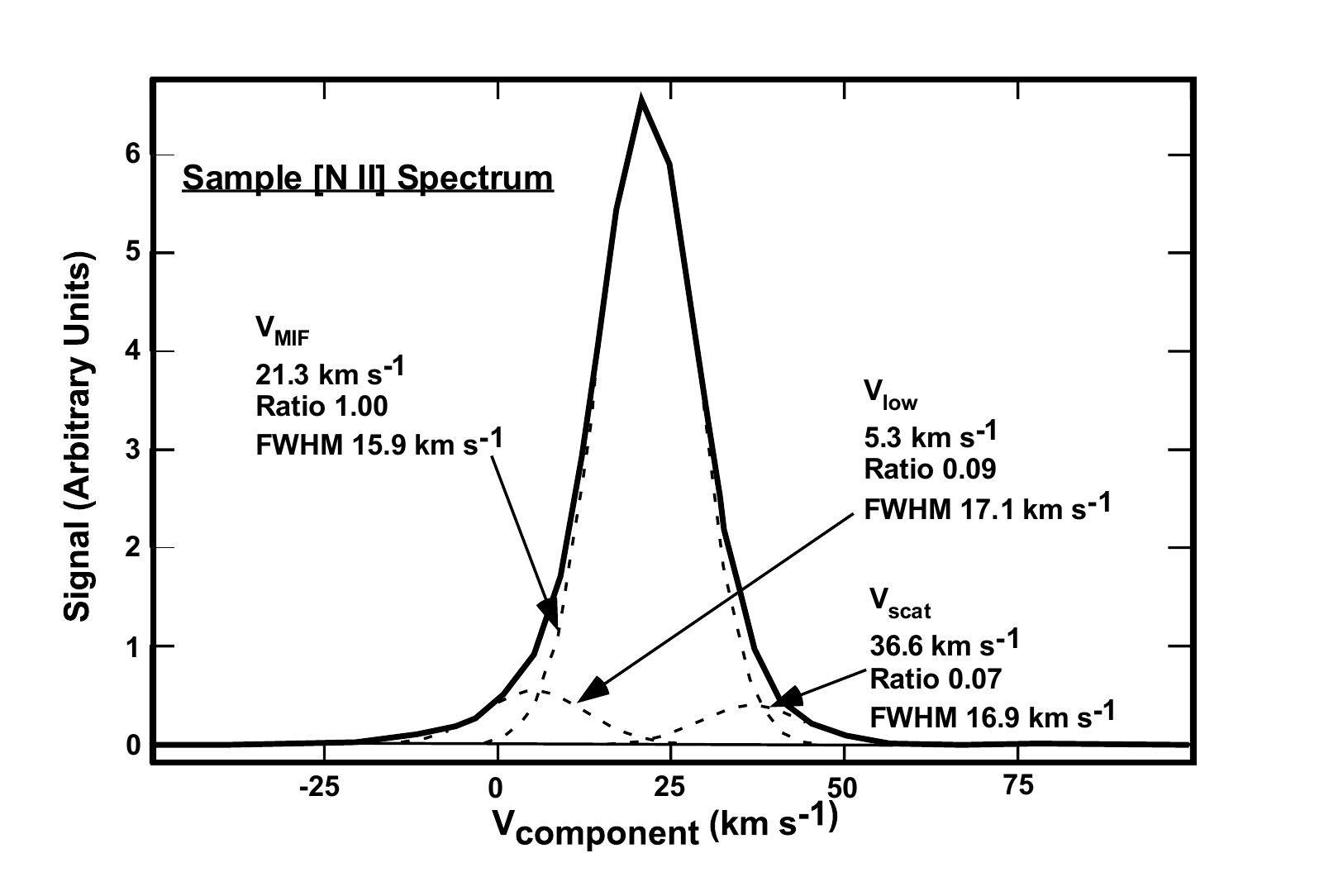}
\caption{The \nii\ spectrum of Sample -30\arcsec ,Line 17 is presented for comparison with the following artificial spectrum, as described in this section. The heavy line depicts the observed profile, the dashed lines the components given from deconvolution. For each component the velocity, the total signal relative to the strongest (MIF) component, and the FWHM are given. The sum of the three fitted 
components falls within the width of the line showing the observed profile.}
\label{fig:Sample2}
\end{figure}

The emission line deconvolution task `splot' takes assumed approximate velocities of line components and produces best fits. Our method of measurement was to first do a fit to a single assumed line. From the region of a poor fit a second line was added to the solution and if there was a third region of poor fit a third line was added. When well separated high velocity lines were seen, they were treated separately.   An illustration of our observed spectra is shown in Fig.~\ref{fig:Sample2}. It is from Sample -30\arcsec ,Line 17.

\begin{figure}
	\includegraphics
	[width=3.5in]
	{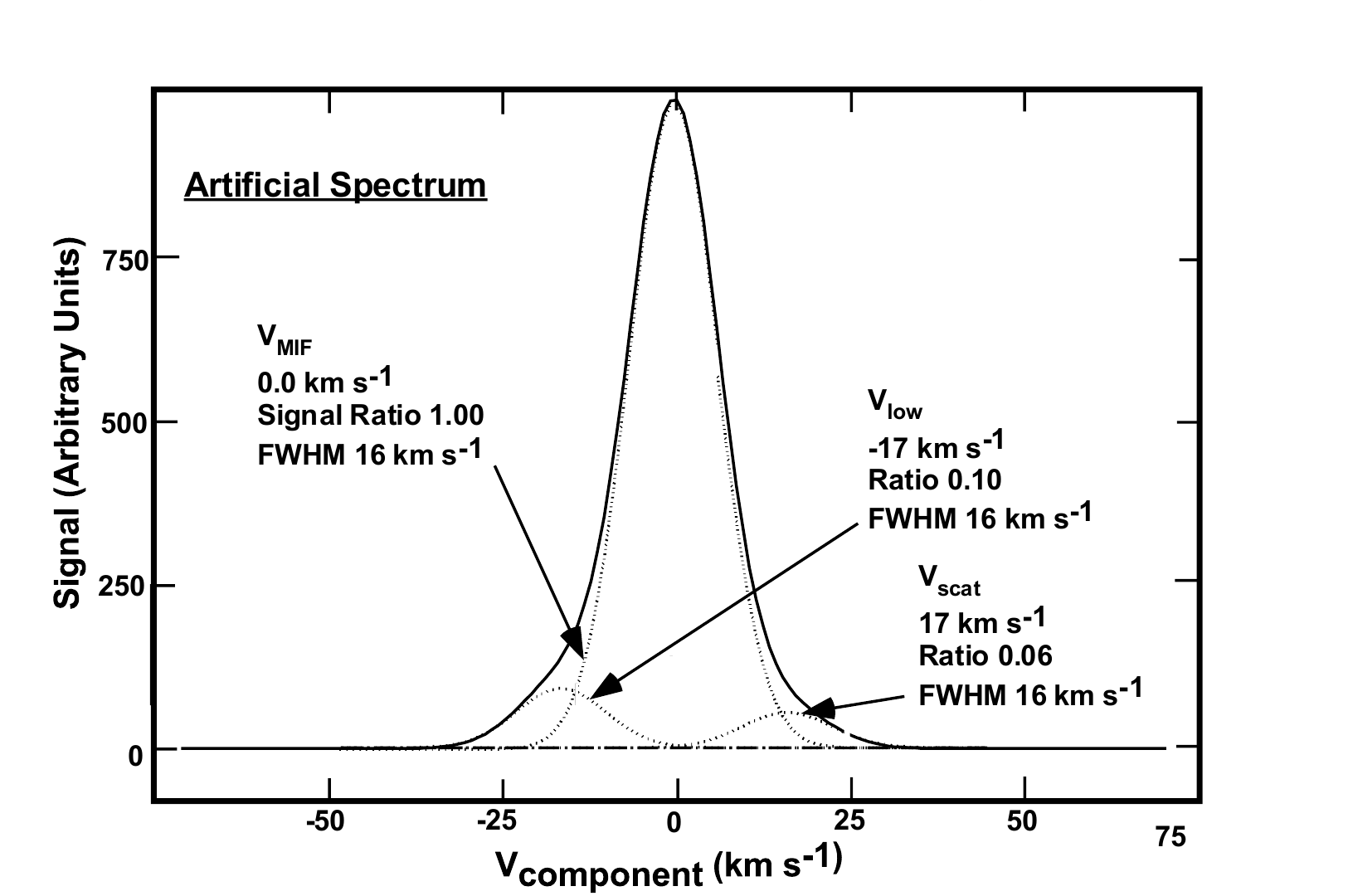}
\caption{An artificial spectrum used in the determination of accuracies of the results of using task `splot' as described in Section~\ref{app:Accuracy}. It is model Std+Low 0.100 from Table~\ref{tab:Composite}
The fit of the sum of the individual components is within the breadth of the line tracing the artificial line profile. The sum of the three fitted 
components falls within the width of the line showing the calculated profile}
\label{fig:Fake}
\end{figure}

This procedure was tested  by creating a series of simulated spectra, then deconvolution of them in the above manner. In order to best mimic
the nebular spectra we assumed three components for the simulations. Since the separations and typical FWHM are similar for both \nii\ and \oiii , it was not necessary to make a line-specific series of calculations. We adopted velocity shifts of $\pm$17 \kms\ and a FWHM of 16 \kms. The
typical value for the S(\Vscat)/S(\Vmif) ratio of 0.06, was employed. This left only the more highly variable S(\Vlow)/S(\Vmif)  ratio as the quantity changing within the set of models. The parameters of the models are given in the left column in Table~\ref{tab:Composite} and the results of the splot deconvolution are in the remaining columns. The model most closely agreeing with the representative spectrum (Fig.~\ref{fig:Sample2}) is shown
in Fig.~\ref{fig:Fake}.
\begin{table*}
\centering
\caption{Deconvolution of Artificial Spectra*}
\label{tab:Composite}
\begin{tabular}{lcccccccc}
\hline
Spectrum          &    \Vmif   &FWHM(MIF)& \Vscat\ &S(\Vscat)/S(\Vmif) &FWHM(\Vscat)& \Vlow\ & S(\Vlow)/S(\Vmif)  &FWHM(\Vlow)\\
\hline
Assumed Value   &    0.0 &   16.0 &  17.0 & 0.06  &   16.0 &    -17.0 & ---  &   16.0 \\
\hline
MIF                      &  0.0   &  16.0&    ---     &       ---    &   ---       & ---       & ---       &---\\
MIF+Scat          &  0.0     &   16.0  &  17.0   &   0.060   &    15.9  &  ---       &   ---     &   ---\\
Std**+Low 0.01  & 0.0   &      16.0 &   16.8  &    0.063 &    16.4  &   -15.3  & 0.014  & 18.3\\
Std+Low 0.025   & 0.0    &     15.8 &   16.8  &    0.067  &   14.5  &  -17.1  & 0.036 &  17.7\\
Std+Low 0.050   & 0.0    &      16.0  &  17.4  &    0.059   &  16.0  &   -17.2  & 0.049 &  16.0\\
Std+Low 0.075  &  0.0    &     16.1  &  17.2  &    0.059  &   16.1  &   -17.4  & 0.072  & 15.9\\
Std+Low 0.100 &  0.0    &     16.1  &  17.3  &    0.059  &  16.0   &  -17.1  & 0.099   &16.0\\
Std+Low 0.150  & 0.0   &      15.8  &  17.0   &   0.061  &   16.4  &  -17.1 &  0.145  & 16.0   \\
Std+Low 0.200  & 0.0   &      16.0  &  17.0  &    0.064  &   16.4  &   -17.0  & 0.201 &  16.0   \\ 
\hline
\end{tabular}\\
*All velocities are in \kms\ with respect to the assumed MIF component at 0.0 \kms.
**Std means the MIF+Scat spectrum
\end{table*}

Examination of Table~\ref{tab:Composite} shows that the derived characteristics of the \Vmif\ and \Vscat\ components are always close to the 
assumed values, even over a wide range of S(\Vlow)/S(\Vmif) values. The derived values of \Vlow\ show no systematic trends with S(\Vlow)/S(\Vmif) although the deviation is -1.7 \kms\ when there is only a one percent contribution of S(\Vlow). The derived S(\Vlow)/S(\Vmif) values
agree with the assumptions down to a five percent contribution but the derived S(\Vlow)/S(\Vmif) values are 1.4 times too large for the 2.5 and 1.0 percent models. 

The above comparison indicates that the velocity data for \Vlow\ is accurate down through a true contribution of 2.5 percent and the relative 
signals are accurate down to five percent, but at assumed values of 2.5 percent and 1.0 percent, the derived values are about 40 percent too large.
This means that almost all of the data used in our discussions are close to the real values, with the caveat that the lower S/N individual slit spectra
must have more uncertainty.

\clearpage

\bsp
\label{lastpage}

\newpage
\begin{thebibliography}{99}

 \bibitem[Abel et al.(2016)]{abel16} Abel, N. P., Ferland, G. J., O'Dell, C. R., Troland, T. H., \ 2016, \apj, 819, 136
 \bibitem[Arthur et al.(2016)]{art16} Arthur, S. J., Medina, S.-N. X., Henney, W. J., \ 2016, \mnras, 423, 2864
 \bibitem[Baldwin et al.(1991)]{bal91} Baldwin, J. A., Ferland, G. J., Martin, P. G., Corbin, M. R., Cota, S. A., Peterson, B. M., Sletteback, A., \ 1991, \apj, 374, 580
 \bibitem[Balick et al.(1974)]{bal74} Balick, B., Gammon, R. H., Hjellming, R. M., \ 1974, \pasp, 86, 616
\bibitem[Bally et al.(2000)]{bom} Bally, J., O'Dell, C. R., McCaughrean, M. J., \ 2000, \aj, 119, 2119
 \bibitem[Casta\~neda (1988)]{hoc88} Casta\~neda, H. O., 1988, \apss\ 67, 93
 \bibitem[Deharveng (1973)]{deh73} Deharveng, L., 1973, \aap , 29, 341
 \bibitem[Doi et al.(2004)]{doi04} Doi, T., O'Dell, C. R., Hartigan, P., 2004, \aj, 127, 3456
 \bibitem[F\'ur\'esz et al.(2008)]{fur08} F\'ur\'esz, G., Hartmann, L W., Megeath, S.T., Szentgyorgyi, A. H., Hamden, E. T.,  2008, \apj, 676, 1109
 \bibitem[Garc\'ia-D\'iaz \& Henney (2007)]{gar07} Garc\'ia-D\'iaz, Ma. T., Henney, W. J.,  2007, \aj, 133, 952
 \bibitem[Garc\'{\i}a-D\'{\i}az et al.(2008)]{gar08} Garc\'{\i}a-D\'{\i}az , Ma.-T., Henney, W. J., L\'opez, J. A., Doi, T., 2008, \rmxaa, 44, 181
 \bibitem[Goicoechea et al.(2015)]{goi15} Goicoechea, J. R., et al.  2015, \apj, 812, 75
 \bibitem[Goicoechea et al.(2016)]{goi16} Goicoechea, J. R., Pety, J.,  et al.  2016, Nature, 537, 207
 \bibitem[Goudis (1982)]{gou82} Goudis, C. 1982, The Orion complex: A case study of interstellar matter (Dordrecht, Netherlands, D. Reidel Publishing Co.), Astrophysics and Space Science Library,Volume 90
 \bibitem[G\"udel et al.(2008)]{gud08} G\"udel, M., Briggs, K. R., Montmerle, T., Audard, M., Rebull, L., Skinner, S. L., 2008, Science, 319, 309
 \bibitem[Henney (1994)]{hen94} Henney, W. J., 1994, \rmxaa, 29, 192
 \bibitem[Henney (1998)]{hen98} Henney, W. J., 1998, \apj, 503, 760
 \bibitem[Henney et al.(2005)]{hen05} Henney, W. J., Arthur, S. J., Garc\'ia-D\'iaz, Ma.-T., 2005, \apj, 627, 813
 \bibitem[Hobbs (1978)]{hobbs} Hobbs, L. M.,1978, \apjs, 38, 129
 \bibitem[Kounkel et al.(2017)]{mk17} Kounkel, M., et al. 2016, \apj, 834, 142
 \bibitem[Leroy \& Le Borgne (1987)]{llb87} Leroy, J. L., Le Borgne, J. F., 1987, \aap, 186, 322
 \bibitem[Menten et al.(2007)]{men07} Menten, K. M. Reid, M. J., Forbrich, J., Brunthaler, A., 2007, \aap, 474, 515
 \bibitem[Mesa-Delgado et al.(2011)]{md11} Mesa-Delgado, A., N\'u\~nez-D\'iaz, M., Esteban, C., L\'opez-Mart\'in, L., Garc\'ia-Rojas, J., 2011, \mnras, 417, 420 (M-D),
 \bibitem[Muench et al.(2008)]{mue08} Muench, A., Getman, K., Hillenbrand, L, Preibisch, T.,  2008, in Handbook of Star Forming Regions, Vol. 1:The Northern Sky, ASP Monograph Publications, Vol. 4, ed. B. Reipurth, p. 483
 \bibitem[O'Dell (2001)]{ode01} O'Dell, C. R., 2001, ARAA, 39, 9
 \bibitem[O'Dell (2009a)]{ode09a} O'Dell, C. R., 2009, \pasp , 121, 428
 \bibitem[O'Dell et al.(2015)]{ode15} O'Dell, C. R., Ferland, G. J., Henney, W. J., Peimbert, M., Garc\'ia-D\'iaz, Ma. T., 2015, \aj, 150, 108 (O15)
 \bibitem[O'Dell et al.(2017a)]{ode17a} O'Dell, C. R., Ferland, G. J., Peimbert, M., \ 2017, \mnras, 464, 4835 (O17a)
 \bibitem[O'Dell \& Harris (2010)]{ode10} O'Dell, C. R., Harris, J. A., 2010, \aj, 140, 985
 \bibitem[O'Dell et al.(2009b)]{ode09b} O'Dell, C. R., Henney, W. J., Abel, N. P., Ferland, G. J., Arthur, S. J.,  \ 2009, \aj, 137, 367
 \bibitem[O'Dell \& Hubbard(1965)]{ode65} O'Dell, C. R., Hubbard, W. B., 1965, \apj, 142, 591
 \bibitem[O'Dell et al.(2017b)]{ode17b} O'Dell, C. R., Kollatschny, W., Ferland, G. J., 2017, \apj, 837, 151
 \bibitem[O'Dell et al.(2008)]{ode08} O'Dell, C. R., Muench, A., Smith, N., Zapata, L., 2008, in Handbook of Star Forming Regions, Vol. 1:The  Northern Sky, ASP Monograph Publications, Vol. 4, ed. B. Reipurth, p. 544
 \bibitem[O'Dell et al.(1993)]{ode93} O'Dell, C. R., Valk, J. H., Wen, Z., Meyer, D. M., 1993a, \apj, 403, 478
 \bibitem[O'Dell et al.(1992)]{ode92} O'Dell, C. R., Walter, D. K., Dufour, R. J., 1992, \apj, 399, L67
 \bibitem[O'Dell \& Wen (1992)]{ode92} O'Dell, C. R., Wen, Z., 1992, \apj, 387, 229
 \bibitem[O'Dell \& Wong (1996)]{ode96} O'Dell, C. R., Wong, S. K., 1996, \aj, 111, 846
 \bibitem[O'Dell \& Yusef-Zadeh (2000)]{ode00} O'Dell, C. R., Yusef-Zadeh, F., 2000, \apj, 120, 382 (OY-Z)
 \bibitem[Ossenkopf et al.(2013)]{oss13} Ossenkopf, V., R\"ollig, M., Neufeld, D. A., Pilleri, P., Lis, D. C., Fuente, A., van der Tak, F. F. S., Bergin, E., 2013, \aap, 550, 57
 \bibitem[Osterbrock \& Ferland (2006)]{agn06} Osterbrock, D. E., Ferland, G. J., 2006, Astrophysics of Gaseous Nebulae and Active Galactic Nuclei (second edition), University Science Books (Mill Valley)
 \bibitem[Pelligrini et al.(2009)]{pel09} Pelligrini, E. W., Baldwin, J. A., Ferland, G. J., Shaw, G., Heathcote, S., 2009, \apj, 693, 285
 \bibitem[Shaw, et al.(2009)]{shaw09} Shaw, G., Ferland, G. J., Henney, W. J., Stancil, P. C., Abel, N. P., Pellegrini, E. W., Baldwin, J. A., van Hoof, P. A. M., 2009, \apj, 701, 677
 \bibitem[Sicilia-Aguilar et al.(2005)]{sic05} Sicilia-Aguilar, A., et al., 2005, \aj, 129, 363                   
 \bibitem[Tielens et al.(1993)]{tie93} Tielens, A. G. G. M., Meixner, M. M., van der Werf, P. P., Bregman, J., Tauber, J. A., Stutzki, J. Rank, D., 1993, Science, 262, 86
 \bibitem[Troland et al.(2016)]{tom16} Troland, T. H., Goss, W. M., Brogan, C. L., Crutcher, R. M., Roberts, D. A., \ 2016, \apj, 825, 2 \bibitem[van der Werf \& Goss(1989)]{vdw89} van der Werf, P. P., Goss, W. M., 1989, \aap, 224, 229
 \bibitem[van der Werf \& Goss(1990)]{vdw90} van der Werf, P. P., Goss, W. M., 1990, \aap, 364, 157
 \bibitem[van der Werf et al.(2013)]{vdw13} van der Werf, P. P., Goss, W. M., O'Dell, C. R., 2013, \apj, 762, 101
 \bibitem[Walmsley et al.(2000)]{wal00} Walmsley, C. M., Natta, A., Oliva, E., Testi, L., 2000, \aap, 364, 301
 \bibitem[Weilbacher et al.(2015)]{wei15} Weilbacher, P. M., et al. 2015, \aap, 582, A114
 \bibitem[Wilson et al.(1959)]{ocw59} Wilson, O. C., M\"unch, G., Flather, E. M., Coffeen, M. F., 1959, \apjs , 4, 199
 \bibitem[Zuckerman (1973)]{zuk73} Zuckerman, B.,1973, \apj, 183, 163
\end{thebibliography}
\end{document}